\RequirePackage{fix-cm}
\documentclass[a4paper]{article}                     
\usepackage{graphicx}
\usepackage{color}
\usepackage{amsmath}
\usepackage{amsfonts}
\usepackage{caption}
  \usepackage{subcaption}
\usepackage{a4wide}
\newcommand{\cS}{\mathcal{S}}
\newcommand{\scS}{{\scriptscriptstyle \cS}}

\newcommand{\Ub}{{\bf U}}
\newcommand{\E}{{\bf E}}
\newcommand{\vb}{{\bf v}}
\newcommand{\hv}{\hat{\bf v}}

\newcommand{\x}{{\bf x}}

\newcommand{\ub}{{\bf u}}
\newcommand{\ie}{{\it i.e.}}
\newcommand{\eg}{{\it e.g.}}

\title{Multi-cue kinetic model with non-local sensing for cell migration on a fibers network with chemotaxis}

\author{Martina Conte \thanks{BCAM - Basque Center for Applied Mathematics, Alameda de Mazarredo, 14, 48009 Bilbao, Spain (\texttt{mconte@bcamath.org})}
\and
	Nadia Loy \thanks{Department of Mathematical Sciences ``G. L. Lagrange'', Politecnico di Torino, Corso Duca degli Abruzzi 24, 10129 Torino, Italy, and Department of Mathematics ``G. Peano'', Via Carlo Alberto 10, 10123 Torino, Italy
                (\texttt{nadia.loy@polito.it})}
                \thanks{Corresponding author: \texttt{nadia.loy@polito.it}}}

\providecommand{\keywords}[1]{\textbf{Keyword.}#1}
\providecommand{\AMS}[1]{\textbf{AMS subject classifications.}#1}

\begin{document}

\maketitle

\begin{abstract}
Cells perform directed motion in response to external stimuli that they detect by sensing the environment with their membrane protrusions. In particular, several biochemical and biophysical cues give rise to tactic migration in the direction of their specific targets. This defines a multi-cue environment in which cells have to sort and combine different, and potentially competitive, stimuli. We propose a non-local kinetic model for cell migration in presence of two external factors both influencing cell polarization: contact guidance and chemotaxis. We propose two different sensing strategies and we analyze the two resulting models by recovering the appropriate macroscopic limit in different regimes, in order to see how the size of the cell, with respect to the variation of both external fields, influences the overall behavior. Moreover, we integrate numerically the kinetic transport equation in a two-dimensional setting in order to investigate qualitatively various scenarios.
\end{abstract}
              
\\

\AMS{ 35Q20, 35Q92, 92B05, 45K05, 92C17}
              
\section{Introduction}
Cell migration is a fundamental mechanism in a huge variety of processes, such as embryogenesis, wound healing, angiogenesis, immune response and tumor stroma formation and metastasis.

During such processes, cells sense the environment and respond to external factors that induce a certain direction of motion towards specific targets (\textit{taxis}): this results in a persistent migration in a certain preferential direction. The guidance cues leading to directed migration may be biochemical or biophysical. 
Biochemical cues can be, for example, soluble factors or growth factors that give rise to {\it chemotaxis}, which involves a \textit{mono-directional} stimulus. Other cues generating mono-directional stimuli include, for instance, bound ligands to the substratum that induce \textit{haptotaxis}, \textit{durotaxis}, that involves migration towards regions with an increasing stiffness of the ECM, {\it electrotaxis}, also known as \textit{galvanotaxis}, that prescribes a directed motion guided by an electric field or current, or {\it phototaxis}, referring to the movement oriented by a stimulus of light \cite{Lara2013IB}.
Important biophysical cues are some of the properties of the extracellular matrix (ECM), first among all the alignment of collagen fibers and its stiffness. 
In particular, the fiber alignment is shown to stimulate contact guidance \cite{Friedl_Brocker.00, Friedl.04}. Contact guidance is a key mechanism in a number of in vivo situations in which cells tend to migrate crawling on the fibers, thus following the directions imposed by the network structure of the ECM. This is a \textit{bi-directional} cue, as, if the fibers network is not polarized, there is no preferential \textit{sense} of migration along them. For example, during wound healing fibroblasts migrate efficiently along collagen or fibronectin fibers in connective tissues; in cancer spread and metastasis formation, cancer cells migrate through the stromal tissue and are thus facilitated to reach blood and lymphatic vessels \cite{Steeg2016NRC, Provenzano2006BMCMed, Provenzano2009TCB}.

In many processes there are several directional cues that may induce different simultaneous stimuli. While the cell response to each of them has been largely studied, from both an intracellular and a migrative point of view, cell responses to a multi-cue environment are much less understood. The fundamental issue is the way cells rank, integrate or hierarchize multiple cues, in particular when these give conflicting stimuli, because, for example, they are not co-aligned \cite{RAJNICEK2007DB}. Some studies have shown that there may be competition or cooperation between different stimuli in the directional response of a cell in a multi-cue environment. Considering the angle between the relative orientation of the directional cues, in the mono-directional case they compete when this angle is $\pi$, whereas they collaborate when this angle is $0$. Bi-directional cues, such as contact guidance, compete when the angle is $\pi/2$. Then, many intermediate scenarios may happen and guidance stimuli submit or prevail according to other factors, among all their average concentration and intensity, that relates to the steepness of the gradient for taxis processes and to the degree of alignment for contact guidance. 
In particular, regarding the external environment, the average value of the directional cue (fiber density, molecule concentration, etc.) and the steepness of the gradient, or the degree of fiber alignment, are fundamental parameters that can be quantified. 
While, for cell migration, the angle between the polarization direction and the preferential direction imposed by the guidance cue
can be measured, as well as the displacement, the mean squared displacement and the persistence time \cite{Dickinson_Tranquillo.93}.
However, in general, when cues are aligned, a simple additive mechanism is not what governs multi-cue migration \cite{Lara2013IB}, even if it is weighted by the average cue concentrations or intensities.

In the framework of kinetic models, in the present paper we will focus on how the environmental sensing of two different stimuli over a finite radius can influence the choice of the direction of motion of a cell. In particular, we combine chemotaxis, a mono-directional biochemical cue, with contact guidance, defining the new orientation of the cells as a result of the sensing of the two cues over a finite neighborhood, that gives a non-local character to the model.
In particular, the combination of chemotaxis and contact-guidance happens in vivo in a variety of situations, for example in wound healing and in breast cancer.
In wound healing, fibers guide cells towards the provisional clot, whilst in breast cancer cells follow the aligned fibers at the tumor-stroma interface for migrating out of the primary tumor. Chemotaxis accelerates and enhances these processes \cite{Lara2013IB, BROMBEREK2002ECR,Provenzano2006BMCMed,Provenzano2009TCB}. 
Therefore, a deep understanding of multi-cue migrational responses is a key step for the comprehension of both physiologic and pathologic processes, but also for building engineered tissues, as their structure is realized for guiding cell migration in a focused way \cite{Lara2013IB}.

There are not many experimental studies concerning chemotaxis and contact guidance, as well as other combinations of directional guidances cues \cite{Lara2013IB}. One of the main reasons is the difficulty in designing environments for controlling multiple directional cues, in particular soluble factors and aligned fibers and fibrous materials.
For example, in one of the first works studying in vitro contact guidance of neutrophil leukocytes on fibrils of collagen \cite{WILKINSON1983ECR}, it is shown that migration is more efficient in the direction of alignment, instead of in the perpendicular direction; in the presence of chemotaxis, obtained by adding a chemoattractant, they observe that these cues cooperate or compete in dependence on their relative orientation. In particular, the chemotactic response is lower for cells trying to cross fibers in the perpendicular direction. In \cite{BROMBEREK2002ECR}, it is shown that alignment along the fibers is greater in presence of a co-aligned chemoattractant. 
In \cite{MAHESHWARI1999BioP}, the authors study how multiple uniformly distributed cues quantitatively regulate random cell migration. 
One of the latest works concerning the competition between chemotaxis and
contact guidance shows that less contractile cells are dominated by chemotaxis,
while contact guidance might dominate in more contractile cells \cite{Woo2005LoC}.
This suggests that, as amoeboid cells are less contractile, while mesenchymal cells are more contractile, and there may be a switching between amoeboid and mesenchymal migration, perhaps there can also be a switching between the dominance of
chemotaxis (amoeboid migration) and contact guidance (mesenchymal migration) \cite{Wolf2003JCB}. 
\noindent One of the most interesting 2D platforms, allowing to study contact guidance and chemotaxis, was proposed in \cite{Sunda2013BB}, in which the authors 
demonstrated an additive effect of chemical gradients and fiber alignment by measuring the persistence time; they also observed that cells were directed by fiber alignment and there was no effect of the chemical gradient when fibers were aligned perpendicular to it. A similar setting was also used for studying the dependence of contact guidance on the  cell cycle \cite{Pourf2018APLB}. However, In the case of different multi-directional cues, totally different scenarios may happen, $\eg$ in \cite{RAJNICEK2007DB} it is shown that for contact guidance and electrotaxis in the cornea, electrotaxis wins when competing with the direction of alignment of the fibers.


There is a huge variety of mathematical models concerning cell migration. They range from microscopic models (also called individuals based models), that describe migration at the cell level, up to macroscopic ones, that describe collective cell-migration at a tissue level. There are many examples of individual based models regarding chemotaxis (\cite{Menci,Gininait2019ModellingCC} and references therein) and migration on the ECM \cite{Col_Sci_Prez.17, Scianna_Preziosi.13.2, Schlute2012BP}.  Concerning macroscopic models, first among all the famous Keller and Segel model is a drift-diffusion model postulated at the macroscopic level \cite{Keller_Segel}. Many efforts were made in order to encompass the defects of the Keller and Segel model, as well as for deriving it from lower scale models (see \cite{PAINTER2019JTB, Hillen_Painter.08, Othmer_Hillen.02, Othmer_Stevens.97,bellomo2015} and references therein). 
Between microscopic and macroscopic models there are mesoscopic models that are an intermediate representative scale, as they include microscopic dynamics and describe the statistical distribution of the individuals. They also allow, for instance in the case of kinetic theory, to recover the appropriate macroscopic regime which inherit some details of the microscopic dynamics, thus giving more significance to some of the parameters \cite{Othmer_Hillen.02}. Some examples are \cite{Col_Sci_Tos.15, Chalub_Markowich_Perthame_Schmeiser.04, Eftimie2}.
The two major models for contact guidance at the mesoscopic level were proposed in \cite{Hillen.05} and \cite{Dickinson}, both local models in the physical space. Concerning multiple cues, not many models exist. In \cite{Painter1999}, the authors propose a macroscopic drift-diffusion model derived from a space jump process in which they include the response to multiple chemicals. A recent review for macroscopic PDEs including multiple-taxis has been proposed in \cite{kolbe2020modeling}. 
In \cite{wagle2000}, the authors propose one of the first models for both contact guidance and chemotaxis, derived from a microscopic dynamics description. In a recent work \cite{Azimzade2019PRE}, the authors propose a microscopic stochastic model for studying contact guidance and add chemotaxis in order to study migration at the tumor-stroma interface for classifying TACS (tumor associated collagen signature). In \cite{Chauviere_Hillen_Preziosi.07}, a kinetic model for cell-cell interactions on a fibers network in presence of a tactic cue is considered.
In \cite{loy2019JMB, Loy_Preziosi2}, the authors propose a non-local kinetic model with a double biasing cue: the first one affecting the choice of the direction and the second one affecting the speed, including, through the non-locality, the sensing of macroscopic quantities performed by the cell, that depends on the cell size, $\ie$, on its maximum protrusion length.

As already stated, in this paper we want to include chemotaxis and contact guidance as directional cues guiding cell polarization. In particular, we analyze two possible sensing strategies that a cell could apply for exploring the neighborhood around, and that determine the choice for the transition probability for the transport model. The cell can measure the guidance cues independently, and, then, choose the new orientation using the collected information, eventually weighted in different ways. Otherwise, it can measure the two directional stimuli, weighting them equally, and assuming a  conditioning of one cue on the other. Therefore, cell response is related to the choice of the sensing strategy, and the macroscopic overall effect of the two cues would also be affected. Moreover, we shall consider for the first time a non-local sensing of the fibers distribution defined at a mesoscopic level; this allows for many intermediate scenarios in the analysis about the collaborative or competitive effect of the cues. For a better understanding, we discuss how the choices made on the transition probability, together with the size of the sampling volume and the characteristics of the two cues determine the macroscopic behavior. 
Specifically,
in section 2, we shall present the mathematical framework, while in section 3 we shall introduce the two classes of models, that describe the different strategies for the sensing of a double cue, along with the corresponding macroscopic limits in various regimes, depending on the cell size and on the variability of the external cues. In section 4, some numerical simulations of the kinetic models will be presented for investigating qualitatively various scenarios in a two-dimensional setting.

\section{Mathematical framework}

\subsection{The transport model}
The cell population will be described at a mesoscopic level through the distribution density $p = p(t,\x, v,\hv)$ that, for every time $t>0$ and position $\x \in \Omega \subseteq \mathbb{R}^d$, gives the statistical distribution of the speeds $v\in [0,U]$, where $U$ is the maximal speed a cell can achieve, and of the polarization directions $\hv\in \mathbb{S}^{d-1}$, being $\mathbb{S}^{d-1}$ the unit sphere boundary in $\mathbb{R}^d$. The velocity vector, thus, will be given by $\vb=v\hv$.

Then, a macroscopic description for the cell population can be classically recovered  through the definition of moments of the distribution function $p$.  In particular, we recover the cell number density $\rho (t,\x)$
\begin{equation}\label{def_rho}
\rho(t,\x) = \int_{\mathbb{S}^{d-1}}\int_0^U p(t,\x,v,\hv) \,dv\,d\hv\,
\end{equation}
the momentum
\begin{equation}\label{def_rhoU}
\rho(t,\x)\Ub(t,\x)=\int_{\mathbb{S}^{d-1}}\int_0^U \vb\, p(t,\x,v,\hv) \,dv\,d\hv\,
\end{equation}
the cell mean velocity
\begin{equation}
\Ub(t,\x)=\dfrac{1}{\rho(t,\x)}\,\int_{\mathbb{S}^{d-1}}\int_0^U \vb\, p(t,\x,v,\hv) \,dv\,d\hv\,
\end{equation}
and the energy tensor
\begin{equation}
\mathbb{D}(t,\x)=\int_{\mathbb{S}^{d-1}}\int_0^U (\vb-\Ub)\otimes (\vb -\Ub) \,p(t,\x,v,\hv) \,dv \,d\hv.
\end{equation}
\noindent The mesoscopic model consists in the transport equation for the cell distribution
\begin{equation}\label{transport.general}
\dfrac{\partial p}{\partial t}(t,\x,v,\hv) + \vb\cdot \nabla p(t,\x,v,\hv) =  \mathcal{J} [p] (t,\x,v,\hv)
\end{equation}
where the operator $\nabla$ denotes the spatial gradient, so that the term $\vb \cdot \nabla p$ takes into account the free particle transport.
The term $\mathcal{J}[p](t,\x,v,\hv)$ is the {\it turning operator} that describes the scattering of the microscopic velocity in direction and speed. This is related to the typical microscopic dynamics of the cell, that is the \textit{run and tumble} \cite{Berg_Block_Segall,Berg}. The run and tumble prescribes an alternation of runs over straight lines and re-orientations: the choice of the new direction may be random or it may be biased by the presence of external factors, that may attract or repel the cell as well as increase the time spent in a run. The run and tumble is classically modeled by a scattering of the microscopic velocity called velocity jump process \cite{Stroock}, characterized by a \textit{turning frequency} $\mu$ and a \textit{transition probability} $T$. 
The general form of the turning operator which implements a velocity jump process at a kinetic level is given by
\begin{equation}\label{turning.operator}
\begin{split}
\mathcal{J}[p](\x,v,\hv) =& \mu(\x)\int_{\mathbb{S}^{d-1}}\int_0^U \Big[
 T(\x,v,\hv|v',\hv')p(t,\x,v',\hv')-T(\x,v',\hv'|v,\hv)p(t,\x,v,\hv)\Big]\, dv'd\hv' \,
 \end{split}
\end{equation}
where we assumed that the turning frequency does not depend on the microscopic velocity.
The transition probability $T(\x,v,\hv|v',\hv')$ is also called {\it turning kernel} and it is a conditional probability satisfying, $\forall \x \, \in \, \Omega$, 
\begin{equation}\label{normalization.T}
\int_{\mathbb{S}^{d-1}}\int_0^U T(\x,v,\hv|v',\hv') dv d\hv =1 \,,\quad \  \forall v'\in [0,U], \, \hv'\in \mathbb{S}^{d-1}.  
\end{equation}
Thanks to this property, the operator \eqref{turning.operator} reads
\[
 \mathcal{J}[p](t,\x,v,\hv) = \mu(\x) \, \left( \int_{\mathbb{S}^{d-1}}\int_0^U  T(\x,v,\hv|v',\hv')p(t,\x,v', \hv') \, dv' d\hv'- p(t,\x,v,\hv) \right) \,.
\]
For our purposes, we shall assume that the transition probability only depends on the post-tumbling velocity 
\begin{equation}\label{T_no.pre}
T(\x,v,\hv|v',\hv')=T(\x,v,\hv)
\end{equation} 
as classically done in the pioneering work concerning kinetic equations for velocity jump processes \cite{Stroock,Alt.88,Hillen.05}.
This assumption, along with the assumption on the turning frequency, is due to the fact that we shall consider directional cues which are sensed non-locally, and, therefore, the most relevant aspect will be the measured preferential direction instead than the incoming velocity.
The latter \eqref{T_no.pre} allows to write the turning operator as
\begin{equation}\label{J_r.S}
 \mathcal{J}[p](t,\x,v,\hv) = \mu(\x) \, \Big( \rho(t,\x)  T(\x,v,\hv) - p(t,\x,v,\hv) \Big) \,.
\end{equation}
The mean macroscopic velocity after a tumble is given by the average of $T$ 
\begin{equation}\label{U_T}
\begin{split}
\Ub_T(\x)&=\int_{\mathbb{S}^{d-1}}\int_0^U\vb\, T(\x,v,\hv) \,dv \,d\hv
\end{split}
\end{equation}
and the diffusion tensor by the variance-covariance matrix
\begin{equation}\label{D_T}
\begin{split}
\mathbb{D}_T(\x)&=\int_{\mathbb{S}^{d-1}}\int_0^U T(\x, v,\hv) (\vb-\Ub_T)\otimes (\vb-\Ub_T) dv \, d\hv.
\end{split}
\end{equation}

Arguing as in \cite{Petterson, Bisi.Carrillo.Lods}, we can prove
a linear version of the classical H-Theorem for the linear Boltzmann equation \eqref{transport.general}-\eqref{J_r.S} with $p^0=p(0,\x,v,\hv) \in \, L^1(\Omega \times [0,U] \times \mathbb{S}^{d-1})$. In particular the Maxwellian
\[
M(\x,v,\hv)=\rho^{\infty}(\x)T(\x,v,\hv),
\]
making the turning operator vanish, is the local asymptotic stable equilibrium of the system.
As already remarked by \cite{loy2019JMB}, this implies that $T$ is the local asymptotic equilibrium steady state of the system. Therefore $\Ub_T$ and $\mathbb{D}_T$ are the mean velocity and diffusion tensor of the cell population at equilibrium.

\subsection{Boundary conditions}
Since we are going to consider two-dimensional bounded domains without loss of cells and no cells coming in, we shall assume conservation of mass. Therefore, we will require that the chosen boundary condition is no-flux \cite{Plaza}
\begin{equation}\label{noflux}
\int_{\mathbb{S}^{d-1}}\int_0^U p(t,\x,v,\hv) \hv\cdot {\bf n}(\x)\,dv\,d\hat\vb=0, \quad \forall \x \in \partial \Omega, \quad t>0\,,
\end{equation}
being ${\bf n}(\x)$ the outward normal to the boundary $\partial \Omega$ in the point $\x$. This class of boundary conditions is part of the wider class of non-absorbing boundary conditions. 
Denoting the boundary operator as
\[
\mathcal{R}[p](t,\x, v,\hat{\vb})=p(t,\x, v',\hat{\vb}')\,,
\]
there are two important classes of kinetic boundary conditions which satisfy \eqref{noflux}: the regular
reflection boundary operators and the non-local (in velocity) boundary operators of diffusive type. We address the reader to the works \cite{Palc} and \cite{Lods} for the definition of these boundary operators. In the present work, we shall consider specular reflection boundary conditions
\begin{equation}\label{specular_reflection}
p(t,\x, v',\hat{\vb}')=
p\left(t,\x, v,\dfrac{\hat{\vb}-2(\hat{\vb}\cdot{\bf n}){\bf n}}{|\hat{\vb}-2(\hat{\vb}\cdot{\bf n}){\bf n}|}\right), \qquad {\bf n} \cdot \hv \le 0,
\end{equation}
that means that cells are reflected with an angle of $\pi/2$ when they hit the wall.  
\subsection{Macroscopic limits}
In order to investigate the overall trend of the system, the macroscopic behavior is typically analyzed. 
By integrating Eq. \eqref{transport.general} with \eqref{J_r.S} on $\mathbb{S}^{d-1}\times [0,U]$, thanks to Eq. \eqref{normalization.T}, we have that 
\[
\partial_t \rho(t,\x) +\nabla \cdot \left( \rho(t,\x)\Ub(t,\x)\right)=0\,,
\]
 $\ie$, the mass is conserved pointwise and in the entire domain, because of no-flux boundary conditions (after integration on $\Omega$). If we multiply Eq. \eqref{transport.general} with \eqref{J_r.S} by $v\hv$, and we then integrate the result on $\mathbb{S}^{d-1}\times [0,U]$, we see that the momentum is not conserved  
\[
\partial_t \rho(t,\x)\Ub(t,\x)+\nabla \cdot \left( \rho(t,\x) \mathbb{D}_T(t,\x)\right)=\mu(\x)\left(\rho(t,\x)\Ub_T(\x)-\rho(t,\x)\Ub(t,\x)\right).
\]
We can observe that, if we multiply the transport equations by increasing orders $n$ of power of $\vb$ and, then, we integrate on the velocity space, we obtain a non-closed system of macroscopic equations, since the equations describing the evolution of $n^{th}$ moment of $p$ contain the $(n+1)^{th}$ moment.
Therefore, we need some procedures to obtain a closed evolution equation (or system of equations) for the macroscopic quantities. In particular, we are interested in the evolution of $\rho(t,\x)$ in the emerging regime of the system. Therefore, we shall consider a diffusive or a hydrodynamic scaling of the transport equation \eqref{transport.general} with \eqref{J_r.S}, resulting from a proper non-dimensionalization of the system. Diffusive and hydrodynamic limits for transport equations with velocity jump processes have been widely treated in \cite{Othmer_Hillen.00, Othmer_Hillen.02, Hillen.05, loy2019JMB,bellomo2007,filbet2005}.
Formally, we introduce a small parameter $\epsilon \ll 1$ and we re-scale the spatial variable as 
\begin{equation}\label{eq:scale_space}
\boldsymbol{\xi}=\epsilon \x,
\end{equation} 
being $\boldsymbol{\xi}$ the macroscopic spatial variable.
According to the other characteristic quantities of the system of study, the macroscopic time scale $\tau$ will be 
\begin{equation}\label{eq:diff_scale}
\tau=\epsilon^2 t,
\end{equation}
that is the parabolic scaling representing a diffusion dominated phenomenon, or
\begin{equation}\label{eq:hyp_scale}
\tau=\epsilon t,
\end{equation}
that is the hyperbolic scaling that represents a drift driven phenomenon.
Up to the spatial scaling \eqref{eq:scale_space}, we have that the transition probability may be expanded as
\[
T(\boldsymbol{\xi},v,\hv)=T_0(\boldsymbol{\xi},v,\hv)+\epsilon T_1(\boldsymbol{\xi},v,\hv)+\mathcal{O}(\epsilon^2).
\]
Therefore, the corresponding means and diffusion tensors will be given by
\begin{equation}\label{UT_exp}
\Ub_T^i(\boldsymbol{\xi})=\int_{\mathbb{S}^{d-1}}\int_0^U T_i (\boldsymbol{\xi},v,\hv)\vb\, dv d\hv
\end{equation}
and 
\begin{equation}\label{DT_exp}
\mathbb{D}_T^i(\boldsymbol{\xi})=\int_{\mathbb{S}^{d-1}}\int_0^U T_i(\boldsymbol{\xi}, v,\hv) (\vb-\Ub^i_T)\otimes (\vb-\Ub^i_T) dv \, d\hv\,.
\end{equation}
Considering a Hilbert expansion of the distribution function $p$ 
\begin{equation}\label{Hilbert_exp}
p=p_0+\epsilon p_1 +\mathcal{O}(\epsilon^2)\,,
\end{equation}
if there is conservation of mass, we have that all the mass is in $p_0$ \cite{Othmer_Hillen.00}, $\ie$,
\begin{equation}\label{rho0}
\rho_0=\rho, \quad \rho_i=0 \quad \forall   i \geq 1 \, ,
\end{equation}
where $\displaystyle{\rho_i=\int_{\mathbb{S}^{d-1}}\int_0^U p_i \, dv\, d\hat{\vb}}$. 
Furthermore, for performing the diffusive limit we shall assume that $\displaystyle{\int_{\mathbb{S}^{d-1}}\int_0^U p_i \,\vb\, dv\, d\hat{\vb} =0 } \quad \forall i \geq 2$ \cite{Othmer_Hillen.00}.

The functional solvability condition that is necessary for performing a diffusive limit ($\ie$, for choosing $\tau=\epsilon^2 t$) is 
\begin{equation}\label{UT0.0}
\Ub^0_T=0,
\end{equation}
meaning that the leading order of the drift vanishes, which is coherent with the fact that the time scale $\tau=\epsilon^2 t$ is chosen because the phenomenon macroscopically is diffusion-driven.
The diffusive limit procedure prescribes to re-scale \eqref{transport.general}-\eqref{J_r.S} with \eqref{eq:scale_space}-\eqref{eq:diff_scale} and to insert \eqref{Hilbert_exp} in the re-scaled equation. By comparing equal order of $\epsilon$, we obtain the macroscopic diffusive limit, given by (dropping the dependencies)
\begin{equation}\label{eq:macro_diff}
\dfrac{\partial}{\partial {\tau}} \rho +\nabla \cdot \left( \Ub_T^1 \rho\right)=\nabla \cdot \left[ \dfrac{1}{\mu} \nabla \cdot \left(\mathbb{D}_T^0 \rho\right) \right]\,,
\end{equation}
being
\[
\mathbb{D}_T^0(\boldsymbol{\xi})=\int_{\mathbb{S}^{d-1}}\int_0^U T_0(\boldsymbol{\xi},v,\hv) \vb \otimes \vb \, dv d\hv\,
\]
the diffusion motility tensor.
Equation \eqref{eq:macro_diff} is a diffusion-advection equation, where $\Ub_T^1$ is the drift velocity of first order.
If \eqref{UT0.0} does not hold, a hyperbolic scaling is required, that gives
\begin{equation}\label{eq:macro_hyp}
\dfrac{\partial}{\partial {\tau}}\rho+\nabla \cdot \left( \rho \Ub_T^0 \right)=0\,.
\end{equation}
This is an advection equation modeling a drift driven phenomenon.
We address the reader to \cite{loy2019JMB} for further details.

Concerning the boundary conditions, at the macroscopic level \eqref{noflux} gives \cite{Plaza}
\[
\Big(  \mathbb{D}_T\nabla\rho -\rho \Ub_T^1\Big)\cdot {\bf n}=0, \quad {\rm on} \quad \partial \Omega,
\]
for the diffusive limit, whilst for the hyperbolic limit the corresponding boundary condition is 
\[
\Ub_T^0\cdot {\bf n}=0, \quad \rm{on} \quad \partial \Omega\,.
\]
\section{A mathematical model for chemotaxis on a fibers network}
In this section, we shall introduce the transition probability modeling a decision process of a cell in presence of a double directional guidance cue: a fibrous ECM and a chemoattractant. 
In particular, we shall consider amoeboid cells \cite{Wolf2003JCB} moving by contact guidance without proteolysis: cells hit the fiber and then move along the direction of the fiber itself. It has been shown experimentally, for example in the case of glioma cancer cells \cite{Chiocca}, that randomly disposed fibers imply isotropic diffusion of cells, while aligned fibers cause anisotropic diffusion of cells along the preferential direction of the fibers themselves. 
The first transport model for contact guidance was proposed by \cite{Hillen.05}, further studied and developed by \cite{Painter2008,Chauviere_Hillen_Preziosi.07, Chauviere_Hillen_Preziosi.08} and applied to the study of glioma by \cite{Hillen_Painter.13, Engwer_Stinner_Surulescu.08, Engwer_Knappitsch_Surulescu.16, Conte_Groppi_Gerardo, Engwer_Hillen_Surulescu_15}.  
The model proposed by \cite{Hillen.05} prescribes a distribution of fibers on the space of directions, given by the unit sphere in $\mathbb{R}^n$,
\begin{equation}
q= q(\x,\hv), \qquad \x \in \Omega, \quad \hv \in \mathbb{S}^{d-1}
\end{equation}
that satisfies
\begin{itemize}
\item[Q1:] $q(\x,\hv) > 0, \quad \forall \x \in\Omega,\,\, \hv \in \mathbb{S}^{d-1}$
\item[Q2:] $\displaystyle \int_{\mathbb{S}^{d-1}} q(\x,\hv) \, d\hv =1, \quad \forall \x \in \Omega$
\item[Q3:] $q(\x,\hv)=q(\x,-\hv), \quad \forall \x \in\Omega,\,\, \hv \in \mathbb{S}^{d-1}$,
\end{itemize}
where the last condition means that we are considering a non-polarized network of fibers, so that cells are able to go in both senses in every direction.
Being, then, $q(\x,\hv)$ a probability density, we can define the mean direction of the fibers
\begin{equation}\label{def:mean_q}
\E_q(\x)= \displaystyle \int_{\mathbb{S}^{d-1}} q(\x,\hv)\, \hv \, d\hv,
\end{equation}
and the diffusion tensor of the fibers, given by the variance-covariance matrix of $q$
\begin{equation}\label{def:diff_q}
\mathbb{D}_q(\x)=\displaystyle \int_{\mathbb{S}^{d-1}} q(\x,\hv)\, (\hv-\E_q)\otimes (\hv -\E_q) \, d\hv\,.
\end{equation}
As we consider a non polarized fibers network, we have that  
\begin{equation}\label{eq:mean_q.0}
\E_q(\x)=0,
\end{equation}
meaning that there is no mean direction in the dynamics.
The tensor \eqref{def:diff_q} is symmetric and positive definite, when $q$ is a regular probability distribution, and, thus, it is diagonalizable. Each eigenvalue represents the diffusivity in the direction of the corresponding eigenvector, meaning that, if the eigenvalues are equal, there is isotropic diffusion, while, if they are different, there is a preferential direction of motion, $\ie$ anisotropy. Therefore, the model introduced in \cite{Hillen.05}, as shown in \cite{Painter2008}, allows to reproduce isotropic/anisotropic diffusion on a non-polarized fibers network.

Concerning chemotaxis, we shall consider a chemoattractant in the region $\Omega$ defined by a strictly positive definite function
\begin{equation}
\cS=\cS(\x) :\Omega \longmapsto \mathbb{R}_+.
\end{equation}

We consider that the sensing performed by the cells is non-local, as they may extend their protrusions, through which they sense the environment, up to several cell diameters \cite{Berg_Purcell.77}. The maximum length $R$ of a protrusion is called {\it sensing radius} and it has been first introduced in \cite{Othmer_Hillen.02} for modeling a non-local gradient of a chemical and, then, used in a number of works (see \cite{chen2019} for a review and references therein) for describing the sensing of macroscopic quantities. In particular, in \cite{loy2019JMB} and, later, in \cite{Loy_Preziosi2} the authors propose a double bias model, in which two cues are sensed non-locally and they affect cell polarization and speed. In the present work we shall drop the sensing of a cue that affects the speed, that will be unbiased, and we will extend the model proposed in \cite{loy2019JMB} to a double sensing of cues affecting the polarization of the cell.

Therefore, in the model both $\cS$ and $q$ will be sensed non-locally by a cell that, starting from its position $\x$, extends its protrusions in every direction $\hv\in \mathbb{S}^{d-1}$ up to the distance $R$, given by the sensing radius. In particular, assuming a non-local sensing of the fibers network will allow to reproduce a wider range of migration strategies, that a cell can perform in order to 
cleverly reach the chemoattractant, with respect to a local sensing.
Therefore, we shall consider the quantities
\[
\cS(\x+\lambda \hv), \qquad q(\x+\lambda \hv, \hv), \qquad \forall \, \x\in \Omega,\quad\forall \, \hv \in \mathbb{S}^{d-1}, \quad \lambda \le R.
\]
Of course, next to the border of the domain $\Omega$, we shall always consider $\lambda$ such that $\x+\lambda \hv \in \Omega$.

In order to analyze qualitatively the impact of the non-locality at the macroscopic level, we study, as previously done in \cite{loy2019JMB, Loy_Preziosi2}, the impact of the directional cues $\cS$ and $q$ with respect to the size of the cell, that is related to its sensing radius $R$.
Thus, we introduce the characteristic length of variation of $\cS$ as
\begin{equation}\label{def:ls}
l_{\cS}:=\dfrac{1}{\max\limits_{\x \in \Omega} \frac{|\nabla \cS|}{\cS}}\,.
\end{equation}
It allows to approximate $\cS(\x+\lambda \hv)$ with a positive quantity
\begin{equation}\label{S_approx}
\cS(\x+\lambda \hv)  \sim\cS(\x)+\lambda \nabla \cS \cdot \hv \ge 0 \quad \forall \lambda \le R \quad \textit{if} \quad R<l_{\cS}
\end{equation}
where we neglected higher order terms in $\lambda$.
Beside the above defined characteristic length of variation of the chemoattractant $l_\cS$, we define an analogue quantity for the fibers distribution. We choose 
\begin{equation}\label{def:lq}
l_q:=\dfrac{1}{\max\limits_{\x \in \Omega}\, \max\limits_{\hv \in \mathbb{S}^{d-1}}\frac{|\nabla q \cdot \hv|}{q}}\,.
\end{equation} 
In this case, we can approximate $q(\x+\lambda \hv,\hv)$ with a positive quantity
\begin{equation}\label{q_approx}
q(\x+\lambda\hv, \hv) \sim q(\x,\hv)+\lambda\nabla q \cdot \hv \ge 0 \quad \forall \lambda < R \quad \textit{if} \quad R<l_q\,.
\end{equation} 
In particular, this definition of $l_q$ takes into account the variation of directionality of the fibers in space, that is what actually influences the cell orientation, more than spatial variation of the density of the extracellular matrix.
We analyze the possible scenarios depending on the relation between $R$, $l_\cS$ and $l_q$.

In analogy to \cite{loy2019JMB}, let us now introduce the parameters
\begin{equation}\label{def:etaq}
\eta_q:=\dfrac{R}{l_q}
\end{equation}
and
\begin{equation}\label{def:etas}
\eta_{\cS}:=\dfrac{R}{l_{\cS}}\,,
\end{equation}
that quantify the capability of measuring of the cell with respect to the characteristic lengths of variation of the sensed guidance cues $q$ and $\cS$. In particular, $\eta_i<1, \,\, i=q, \cS$, means that the sensing radius is smaller than the characteristic length of variation of $q$ ($\cS$, respectively) and the idea is that a single instantaneous sensing of the cell is not capable of catching the total spatial variability of $q$ ($\cS$, respectively), while if $\eta_i>1,\, \,i=q, \cS$, the sensing radius is large enough in order to capture the spatial variability of $q$ ($\cS$, respectively). If we consider the two cues separately, in the first case we expect that the sensing of $q$ ($\cS$, respectively) induces a diffusive behavior, while in the second scenario the overall behavior induced by $q$ ($\cS$, respectively) is drift-driven. 

 As we are considering the two guidance cues simultaneously affecting cell polarization, we now take into account for limit cases:
\begin{enumerate}
\item[$i)$] $\eta_q, \eta_{\cS} \gg1$;
\item[$ii)$] $\eta_q, \eta_{\cS} \ll 1$;
\item[$iii)$] $\eta_{\cS} \ll1, \eta_q\gg1$;
\item[$iv)$] $\eta_{\cS}\gg1, \eta_q\ll1$.
\end{enumerate}
In case $i)$, a Taylor expansion cannot be used, since there is no guarantee that the first order approximations are positive, as well as in case $iii)$ and $iv)$ for $q$ and $\cS$, respectively.

In order to quantify the relative contribution of chemotaxis to contact guidance, we may introduce the parameter
\begin{equation}\label{eta}
\eta =\dfrac{\eta_q}{\eta_{\cS}}
\end{equation}
that is larger than $1$ if contact guidance prevails, whilst it is smaller then $1$ if chemotaxis is stronger. Due to \eqref{def:etaq} and \eqref{def:etas}, we have that, despite its definition, $\eta$ does not depend on the size and sensing capability of the cell, as $\eta=\dfrac{\eta_q}{\eta_{\cS}}=\dfrac{l_{\cS}}{l_q}$. In particular, if $l_{\cS}$ is larger than $l_q$, $\ie$ $\eta>1$, it means that the gradient of $q$ is steeper than the one of $\cS$, thus enhancing a stronger effect of contact guidance on the dynamics. We may also observe that in case $iii)$ we have always that $\eta>1$ while in case $iv)$ we always have $\eta <1$, $\ie$ contact guidance is weaker then chemotaxis.

We shall propose two different transition probabilities describing two different sensing strategies: in the first model the sensings of $q$ and $\cS$ are independent, while in the second model a unique sensing is performed.
In the first model, we shall introduce a transition probability that is the product of two different independent sensings
\begin{equation}\label{T_indip}
T[q,\cS](\x, v,\hv)=c(\x)\int_{\mathbb{R}_+}\gamma_{\scS}(\lambda) \mathcal{S}(\x+\lambda\hv)\, d\lambda\, \int_{\mathbb{R}_+}\gamma_q(\lambda)\,q(\x+\lambda\hv,\hv)\,d\lambda\, \psi(v)\,.
\end{equation}
In this case the cell located in position $\x$ measures along the direction $\hv$ the field $\cS(\x+\lambda \hv)$ weighted by $\gamma_{\scS}$, and, independently, the quantity $q(\x+\lambda \hv,\hv)$, weighted by $\gamma_q$. The \textit{sensing functions} $\gamma_{\scS}$ and $\gamma_q$ have compact support in $[0,R]$ and they may be Dirac deltas centered in $R$, if the cell only measures the guidance cues on its membrane (only on $\x+R\hv$ for every $\hv$), or Heaviside functions if the cell measures and gives the same weight to $q$ and $\cS$ from $\x$ to $\x+R\hv$ in every direction.  
Formally the transition probability might be seen as the product of the independent probabilities of $q$ and $\cS$, $\ie$ $T[q,\cS]=\hat{T}[q]\,\hat{T}[\cS]$. 

The second model prescribes a simultaneous averaging of the guidance cues $\cS$ and $q$, $\ie$,
\begin{equation}\label{T_dep}
T[q,\cS](\x, v,\hv)=c(\x)\int_{\mathbb{R}_+}\gamma(\lambda) \mathcal{S}(\x+\lambda\hv)\,q(\x+\lambda\hv,\hv)d\lambda\, \psi(v)\,.
\end{equation}
This transition probability describes a cells in position $\x$ that measures in the direction $\hv$ the two quantities $\cS(\x+\lambda \hv)$ and $q(\x+\lambda \hv)$, weighting both with $\gamma$, that is a sensing function. Formally, as the two sensing are not independent and, therefore, factorized, we have a conditioning of $\cS$ given $q$ and viceversa, $\ie$, $T[q,\cS]=\tilde{T}[\cS|q]\,\tilde{T}[q]=\tilde{T}[q|\cS]\,\tilde{T}[\cS]$.

\noindent In \eqref{T_indip} and \eqref{T_dep}, $c(\x)$ is a normalization coefficient. Moreover the probability density $\psi$ is the distribution of the speeds on the interval $[0,U]$ and satisfies 
\[
\displaystyle \int_0^U \psi(v) dv =1\,.
\]
We introduce its mean speed
\begin{equation}
\bar{U}=\displaystyle \int_0^U v \,\psi(v) \,dv
\end{equation}
and the second moment
\begin{equation}
D=\displaystyle \int_0^U v^2\, \psi(v)\, dv\,,
\end{equation}
such that the variance of $\psi$ is given by $\sigma^2_{\psi}=\dfrac{1}{2}(D-\bar{U}^2)$. 

We shall refer to the transport model \eqref{transport.general}-\eqref{J_r.S} with \eqref{T_indip} as \textit{non-local independent sensing model}, in which the cell averages the two cues independently according to two different sensing functions $\gamma_q$, $\gamma_{\scS}$. On the other hand, the transport model \eqref{transport.general}-\eqref{J_r.S} with \eqref{T_dep} is defined as \textit{non-local dependent sensing model}, describing cells that sense the two cues at the same time and average them with a unique sensing kernel $\gamma$. 
In the next sections we shall analyze the macroscopic limits for the two models in the scenarios $i)-iv)$ and we shall compare the two models.

\subsection{Amoeboid motion and chemotaxis: non-local independent sensing}
We first consider the non-local independent sensing case  \eqref{transport.general}-\eqref{J_r.S} with \eqref{T_indip}. We recall the expression of the transition probability
\[
T[q,\cS](\x, v,\hv)=c(\x)\int_{\mathbb{R}_+}\gamma_{\scS}(\lambda) \mathcal{S}(\x+\lambda\hv)\, d\lambda\, \int_{\mathbb{R}_+}\gamma_q(\lambda)\,q(\x+\lambda\hv,\hv)\,d\lambda\, \psi(v)\,.
\]

\noindent The average of $T$, that will be the equilibrium velocity of the cell population, is given by
\begin{equation}\label{U_T_indip}
\Ub_T(\x)=c(\x)\,\bar{U} \int_{\mathbb{S}^{d-1}}\hv \left(\int_{\mathbb{R}_+}\gamma_{\scS}(\lambda) \mathcal{S}(\x+\lambda\hv) \,d\lambda\, \int_{\mathbb{R}_+}\gamma_q(\lambda)\,q(\x+\lambda\hv,\hv)\,d\lambda \right)d\hv\,.
\end{equation}

\paragraph{Case $i)$} In this case, we shall choose 
\[
\epsilon=\min \left\lbrace \frac{1}{\eta_q}, \frac{1}{\eta_\cS} \right\rbrace\,.
\]
As a consequence of the fact that $T$ cannot be expanded in powers of $\epsilon$ after re-scaling with \eqref{eq:scale_space}, we have that $\Ub_T^0=\Ub_T$ given by \eqref{U_T_indip}. Therefore, we have to perform a hyperbolic scaling that leads to the following macroscopic equation for the cells macroscopic density:
\begin{equation}\label{eq:macro_ind_i}
\dfrac{\partial}{\partial \tau} \rho(\tau,\boldsymbol{\xi})+ \nabla \cdot (\rho(\tau,\boldsymbol{\xi}) \Ub_T(\boldsymbol{\xi}))=0\,,
\end{equation}
with $\Ub_T(\boldsymbol{\xi})$ given by the re-scaling of \eqref{U_T_indip} with \eqref{eq:scale_space}.

\paragraph{Case $ii)$} In this case, we can expand both $\cS(\x+\lambda\hv)$ and $q(\x+\lambda\hv,\hv)$ and consider the approximations \eqref{S_approx} and  \eqref{q_approx} for $\lambda< \min \lbrace l_q,l_\cS \rbrace$.
Therefore, we approximate the transition probability by substituting \eqref{S_approx} and \eqref{q_approx} in \eqref{T_indip}, and, thus, we obtain the following approximation for the turning kernel $T[q,\cS]$, that reads
\begin{equation}\label{T_indip_ii}
\begin{split}
T[q,\cS](\x, v,\hv)=&c(\x)\Big[\Gamma_0^\cS\,\Gamma_0^q\,\cS(\x)\,q(\x,\hv)+\Gamma_0^\cS\,\Gamma_1^q\,\cS(\x)\,\nabla q \cdot\hv +\Gamma_1^\cS\,\Gamma_0^q\,q(\x,\hv)\,\nabla\cS\cdot \hv \Big]\psi(v)\,
\end{split}
\end{equation}
where we neglected higher orders terms in $\lambda$.
In the latter
\[
c(\x)=\dfrac{1}{\cS(\x)\,\Gamma_0^\cS\,\Gamma_0^q}
\]
and 
\[
\Gamma_i^\cS:=\int_{\mathbb{R}_+}\lambda^i\gamma_{\scS}(\lambda)\,d\lambda\quad\quad i=0,1
\]
\[
\Gamma_i^q:=\int_{\mathbb{R}_+}\lambda^i\gamma_q(\lambda)\,d\lambda\quad\quad i=0,1\,.
\]
The quantities $\Gamma_0^q, \Gamma_0^\cS$ are the weighted (by $\gamma_q, \gamma_{\scS}$) measures of the sensed linear tracts in every direction, whilst $\Gamma_1^q, \Gamma_1^\cS$ are the averages of $\gamma_q, \gamma_{\scS}$ on $[0,R]$.

\noindent We can, then, introduce the small parameter
\begin{equation}
\epsilon =\min \lbrace \eta_q, \eta_\cS \rbrace
\end{equation}
and re-scale the space variable as $\boldsymbol{\xi}=\epsilon \x$, getting
\begin{equation}\label{T0_ii}
T_0[q,\cS](\boldsymbol{\xi},v,\hv)= q(\boldsymbol{\xi},\hv) \psi(v)\,,
\end{equation}
meaning that the equilibrium is determined by the fibers distribution, and
\[
T_1[q,\cS](\boldsymbol{\xi},v,\hv)=\left[\Gamma^q\,\nabla q \cdot\hv +\Gamma^\cS\,q(\boldsymbol{\xi},\hv)\,\dfrac{\nabla\cS}{\cS(\boldsymbol{\xi})}\cdot \hv \right]\psi(v)
\]
where
\[
\Gamma^\cS:=\dfrac{\Gamma_1^\cS}{\Gamma_0^\cS}\text{,}\qquad\Gamma^q:=\dfrac{\Gamma_1^q}{\Gamma_0^q}\,.
\]
Because of \eqref{eq:mean_q.0} and \eqref{T0_ii}, we have that $\Ub_0^T(\boldsymbol{\xi})=0$, meaning that we are in a diffusive regime, and the diffusive limits leads to the advection-diffusion equation \eqref{eq:macro_diff}. The explicit form for the zero-order macroscopic diffusion tensor is 
\begin{equation}\label{DT_0_ind.ii}
\mathbb{D}_T^0(\boldsymbol{\xi})=D\int_{\mathbb{S}^{d-1}}q(\boldsymbol{\xi},\hv)\hv\otimes\hv\,d\hv=D\,\mathbb{D}_q(\boldsymbol{\xi})\,,
\end{equation}
and for the macroscopic first-order velocity is
\begin{equation}\label{UT_ind.ii}
\begin{split}
\Ub_T^1(\boldsymbol{\xi})&=\bar{U}\int_{\mathbb{S}^{d-1}}\left(\Gamma^q\,\nabla q\cdot \hv + \Gamma^\cS\,\dfrac{\nabla \cS}{\cS(\boldsymbol{\xi})}\cdot \hv\,q(\boldsymbol{\xi},\hv)\right)\hv d\hv\\[0.3cm]
&=\bar{U}\,\Gamma^q\int_{\mathbb{S}^{d-1}}\left(\nabla q\cdot \hv\right)\, \hv d\hv+\bar{U}\,\Gamma^\cS\,\dfrac{\nabla \cS}{\cS} \int_{\mathbb{S}^{d-1}}\hv\otimes \hv\,q(\boldsymbol{\xi},\hv) d\hv\\
&=\bar{U}\,\left[\Gamma^q\,\nabla \cdot \mathbb{D}_q+\,\Gamma^\cS\,\mathbb{D}_q\,\dfrac{\nabla \cS}{\cS}\right].
\end{split}
\end{equation}
Therefore, the diffusion-advection equation \eqref{eq:macro_diff} reads (dropping the dependencies)
\begin{equation}\label{macro_indip.ii}
\dfrac{\partial}{\partial \tau}\rho+ \nabla\cdot\left[\left(\chi^\cS\,\mathbb{D}_q\nabla \cS+\chi^q\nabla \cdot \mathbb{D}_q\right)\rho\right]=\nabla\cdot\left[\dfrac{1}{\mu}\,\nabla\cdot \big(D\,\mathbb{D}_q\,\rho\big)\right]\,,
\end{equation}
where
\begin{equation}\label{chis_ind_ii}
\chi^\cS(\boldsymbol{\xi}):=\dfrac{\bar{U}\,\Gamma^\cS\,}{\cS(\boldsymbol{\xi})}\quad\text{,}\quad\chi^q:=\bar{U}\,\Gamma^q 
\end{equation}
are the sensitivities. The diffusion represented by the motility tensor of the cells \eqref{DT_0_ind.ii} only depends on the fibers distribution, while the advective term has two contributions differently weighted by the sensitivities \eqref{chis_ind_ii}. We remark that, in this regime, we obtain the same macroscopic behavior postulated by Keller and Segel \cite{Keller_Segel}, with the logarithmic chemotactic sensitivity $\chi_{\cS}$ given in \eqref{chis_ind_ii}.
 The term $\mathbb{D}_q\nabla \cS$ depends on both the fibers distribution and the chemotactic field; it never vanishes if $\nabla \cS$ is not the null vector, since it may be proved that $\mathbb{D}_q$ is invertible. In the case of randomly disposed fibers, corresponding to the isotropic case, $\ie$, when $\mathbb{D}_q$ is proportional to the identity matrix, then $\mathbb{D}_q\nabla \cS$ is parallel to $\nabla \cS$, that, thus, represents the anisotropy direction. On the other hand, when $\mathbb{D}_q$ is anisotropic, if $\nabla \cS$ is not parallel to the eigenvector corresponding to the highest eigenvalue of $\mathbb{D}_q$, then the migration does not follow the dominant direction of the fibers, but rather its projection on $\nabla \cS$. Moreover, the second contribution in the drift term, $\ie$, $\nabla \cdot \mathbb{D}_q$, is a measure of the velocity field induced by the spatial variation of the distribution of the fiber directions, that determines the microscopic velocities of the cells. This term vanishes if the fibers distribution is homogeneous in space. Therefore, if $q$ is homogeneous in space, even in case of competing cues, $\ie$, $\textbf{E}_{q}\perp \nabla \cS$, in general the advective term $\Ub_T^1$ does not vanish, while in case of cooperating cues, $\ie, \, \nabla \cS$ is an eigenvector of $\mathbb{D}_q$ with eigenvalue $D_{\nabla \cS}$, migration is in direction $\nabla \cS$ with a kinetic factor $\chi_{\scS} D_{\nabla \cS}$. In intermediate scenarios, migration happens in the projection $\mathbb{D}_q\nabla \cS$, but, if $q$ is not homogeneous, the dynamics is more complex and, even in case of cooperation, we cannot conclude anything about additivity effects.

\paragraph{Case $iii)$}
In this case, we can only expand with Taylor series the chemoattractant, as in \eqref{S_approx}, and the turning kernel \eqref{T_indip} may be approximated as
\begin{equation}\label{T_indip.iii}
\begin{split}
T[q,\cS](\x, v,\hv)=&c(\x)\Big[\cS(\x)\,\Gamma_0^\cS\,\int_{\mathbb{R}_+}\gamma_q(\lambda)q(\x+\lambda\hv,\hv)\,d\lambda+\Gamma_1^\cS\,(\nabla\cS\cdot \hv)\int_{\mathbb{R}_+}\gamma_q(\lambda)q(\x+\lambda\hv,\hv)\,d\lambda\Big]\psi(v)
\end{split}
\end{equation}
where we neglected higher order terms in $\lambda$. Here, the normalization coefficient reduces to
\[
c(\x)=\dfrac{1}{\Gamma_0^\cS\,\Gamma_0^q\,\cS(\x)}\,.
\]
In this case we may choose
\[
\epsilon=\min \left \lbrace \dfrac{1}{\eta_q}, \eta_{\cS}\right \rbrace,
\]
and, re-scaling the space variable as \eqref{eq:scale_space}, we get
\begin{equation}\label{ii_indip_T0}
T_0[q,\cS](\boldsymbol{\xi}, v,\hv)=\dfrac{1}{\Gamma_0^q}\int_{\mathbb{R}_+}\gamma_q(\lambda)q(\boldsymbol{\xi}+\lambda\hv,\hv)\,d\lambda \, \psi(v)
\end{equation}
and 
\[
T_1[q,\cS](\boldsymbol{\xi}, v,\hv)=\dfrac{\Gamma^\cS}{\Gamma_0^q}\,\left(\dfrac{\nabla\cS}{\cS}\cdot \hv\right)\int_{\mathbb{R}_+}\gamma_q(\lambda)q(\boldsymbol{\xi}+\lambda\hv,\hv)\,d\lambda\, \psi(v).
\]
Equation \eqref{ii_indip_T0} indicates that the equilibrium distribution is a non-local average of the fibers distribution according to the sensing kernel $\gamma_q$ and normalized by the measure of the sensed linear tract $\Gamma_0^q$ over the direction $\hv$. Its average is
\[
\Ub_T^0(\boldsymbol{\xi})=\dfrac{\bar{U}}{\Gamma_0^q}\int_{\mathbb{R}_+}\gamma_q(\lambda) \E_q(\boldsymbol{\xi}+\lambda \hv) \, d\lambda
\]
that vanishes as $\boldsymbol{\xi}+\lambda\hv \in \Omega$ and \eqref{eq:mean_q.0} holds true.
Therefore, we perform the diffusive limit that leads to \eqref{eq:macro_diff} with
\begin{equation*}
\begin{split}
\mathbb{D}_T^0(\boldsymbol{\xi})&=D\int_{\mathbb{S}^{d-1}}\dfrac{1}{\Gamma_0^q}\int_{\mathbb{R}_+}\gamma_q(\lambda)\,q(\boldsymbol{\xi}+\lambda\hv,\hv)\,d\lambda \,\hv\otimes\hv\,d\hv\,.
\end{split}
\end{equation*}
Let us now define
\begin{equation}\label{def_Dl}
\mathbb{D}_q^{\lambda}(\boldsymbol{\xi})=\int_{\mathbb{S}^{d-1}}q(\boldsymbol{\xi}+\lambda\hv,\hv)\,\hv\otimes\hv\,d\hv\,,
\end{equation}
that, for each point $\boldsymbol{\xi}$, is the diffusion tensor of the fibers on a circle of radius $\lambda$,
and 
\begin{equation}
\bar{\mathbb{D}}_q^0=\dfrac{1}{\Gamma_0^q}\int_{\mathbb{R}_+}\gamma_q(\lambda)\mathbb{D}_q^{\lambda} \, d\lambda\,,
\end{equation}
that is a weighted diffusion tensor of the fibers in the whole neighborhood sensed by the cells,
so that 
\begin{equation}\label{DT_0_indip.iii}
\mathbb{D}_T^0(\boldsymbol{\xi})=D\bar{\mathbb{D}}_q^0(\boldsymbol{\xi})
\end{equation}
and
\begin{equation}\label{UT_indip.iii}
\begin{split}
\Ub_T^1(\boldsymbol{\xi})&=\bar{U}\,c(\boldsymbol{\xi})\,\int_{\mathbb{S}^{d-1}}\left(\Gamma_1^\cS\,(\nabla \cS\cdot \hv)\int_{\mathbb{R}_+}\gamma_q(\lambda)\,q(\boldsymbol{\xi}+\lambda\hv,\hv)\,d\lambda\right)\,\hv\, d\hv\\[0.3cm]
&=\bar{U}\,c(\boldsymbol{\xi})\,\Gamma_1^\cS \nabla \cS \int_{\mathbb{R}_+}\gamma_q(\lambda)\,\int_{\mathbb{S}^{d-1}}\hv\otimes \hv\,q(\boldsymbol{\xi}+\lambda\hv,\hv) \, d\hv\,d\lambda=\\[0.3cm] 
&=\bar{U}\,\Gamma^\cS\,\bar{\mathbb{D}}_q^0(\boldsymbol{\xi})\dfrac{\nabla \cS}{\cS(\boldsymbol{\xi})}=\chi^\cS(\boldsymbol{\xi})\bar{\mathbb{D}}_q^0(\boldsymbol{\xi})\nabla \cS\,.
\end{split}
\end{equation}
We have defined the chemotactic sensitivity as
\[
\chi^\cS(\boldsymbol{\xi}):=\dfrac{\bar{U}\,\Gamma^\cS\,}{\cS(\boldsymbol{\xi})}\,,
\]
that is a function of the chemical alone, as it is the cue inducing a diffusive behavior.
Here, the advection velocity is related to a non-local average of the diffusion tensor of the fibers $\bar{\mathbb{D}}_q^0$ projected on $\nabla \cS$, and it cannot be decomposed into two contributions because of the large size of the cell with respect to the spatial variability of the fibers distribution. Therefore, in this case the additivity effect of the two cues is not evident and the possible scenarios are many more.\\

\noindent \textbf{Remark}
If we consider $\gamma_q=\delta(\lambda-0)$  we obtain a local sensing of fibers. Without chemotaxis we would have the classical model for contact guidance \cite{Hillen.05}, that gives rise, at the macroscopic level, to a fully anisotropic diffusive equation. The presence of a non-local chemoattractant, even when $R<l_\cS$, gives rise to a drift correction term proportional to $\mathbb{D}_q\nabla \cS$.

\paragraph{Case $iv)$}
The last case allows only for the Taylor expansion of the distribution function $q$, as in \eqref{q_approx}. Therefore, the turning kernel may be approximated as
\begin{equation}\label{T_indip.iv}
\begin{split}
T[q,\cS](\x, v,\hv)=&\Big[c_0(\x)\,\Gamma_0^q\,q(\x,\hv)\,\int_{\mathbb{R}_+}\gamma_{\scS}(\lambda)\cS(\x+\lambda\hv)\,d\lambda+c_1(\x)\Gamma_1^q\,(\nabla q\cdot \hv)\int_{\mathbb{R}_+}\gamma_{\scS}(\lambda)\cS(\x+\lambda\hv)\,d\lambda\Big]\psi(v)
\end{split}
\end{equation}
where
\[
c_0(\x)^{-1}:=2\int_{\mathbb{S}^{d-1}}\Gamma_0^q\,q(\x,\hv)\int_{\mathbb{R}_+}\gamma_{\scS}(\lambda)\cS(\x+\lambda\hv)\,d\lambda\,d\hv\]
and
\[
c_1(\x)^{-1}:=2\int_{\mathbb{S}^{d-1}}\Gamma_1^q\,(\nabla q \cdot \hv)\,\int_{\mathbb{R}_+}\gamma_{\scS}(\lambda)\cS(\x+\lambda\hv)\,d\lambda\,d\hv\,,
\]
both different from zero.
In this case we may choose
\[
\epsilon=\min \left \lbrace \dfrac{1}{\eta_{\cS}}, \eta_{q}\right \rbrace
\]
and, by re-scaling \eqref{T_indip.iv} with \eqref{eq:scale_space}, we get $T[q,\cS]=T_0[q,\cS]$.
Hence $\Ub_T^0(\boldsymbol{\xi})$ does not vanish in $\Omega$, as it is given by 
\begin{equation}\label{UT_0:indip.iv}
\begin{split}
\Ub_T^0(\boldsymbol{\xi})=\dfrac{\bar{U}\,\Gamma_0^q}{c_0(\boldsymbol{\xi})}\,\int_{\mathbb{S}^{d-1}}\hv\,q(\boldsymbol{\xi},\hv)\,\,\int_{\mathbb{R}_+}\gamma_{\scS}(\lambda)\,\cS(\boldsymbol{\xi}+\lambda\hv)\,d\lambda\,d\hv\\[8pt]
+\dfrac{\bar{U}\,\Gamma_1^q}{c_1(\boldsymbol{\xi})}\int_{\mathbb{S}^{d-1}}\hv \otimes \hv \,\nabla q \int_{\mathbb{R}_+} \gamma_{\scS}(\lambda)\,\cS(\boldsymbol{\xi}+\lambda \hv) \, d\lambda \, d\hv\,,
\end{split}
\end{equation}
and the macroscopic equation is given by \eqref{eq:macro_hyp}.
The mean velocity \eqref{UT_0:indip.iv} is a linear combination of a non-local measure of the chemoattractant $\cS$ over the fibers network and a non-local measure of $\cS$ weighted by the directional average of the spatial variability of the fiber direction.\\

\noindent \textbf{Remark}
If we consider a local sensing for the chemoattractant, $\ie$ $\gamma_{\scS}=\delta(\lambda-0)$, we obtain a macroscopic advection-diffusion equation, where the macroscopic velocity is induced by the spatial variation of the distribution of fiber directions $\nabla \cdot \mathbb{D}_q$, and the measure of $\cS$ does not affect the choice of the direction. In this case, if $\nabla q$ vanishes, the model reduces to a fully anisotropic diffusive equation \cite{Hillen.05}.

\subsection{Amoeboid motion and chemotaxis: non-local dependent sensing}
Concerning the non-local dependent sensing case \eqref{transport.general}-\eqref{J_r.S} with \eqref{T_dep}, we recall the expression of the transition probability
\[
T[q,\cS](\x, v,\hv)=c(\x)\int_{\mathbb{R}_+}\gamma(\lambda) \mathcal{S}(\x+\lambda\hv)\,q(\x+\lambda\hv,\hv)d\lambda\, \psi(v)\,,
\]
with
\[
c(\x):=\int_{\mathbb{S}^{d-1}}\int_{\mathbb{R}_+}\gamma(\lambda) \mathcal{S}(\x+\lambda\hv)\,q(\x+\lambda\hv,\hv)d\lambda\,.
\] 
The macroscopic velocity is here given by
\begin{equation}\label{U_T:dep}
\Ub_T(\x)=c(\x)\,\bar{U}\,\int_{\mathbb{S}^{d-1}}\hv\,\int_{\mathbb{R}_+}\gamma(\lambda) \mathcal{S}(\x+\lambda\hv) \,q(\x+\lambda\hv,\hv)d\lambda\,d\hv\,.
\end{equation}
The macroscopic limits can be performed as in the previous section and the choice of the parameter $\epsilon$ will be the same for the cases $i)-iv)$, since it does not depend on the kind of model (independent or dependent sensing), but only on $\eta_{\cS}$ and $\eta_{q}$.
\paragraph{Case $i)$}
In this case we cannot consider the expansions \eqref{q_approx} and \eqref{S_approx}, and, thus, we cannot expand the turning kernel, whose non vanishing average is given by \eqref{U_T:dep}. Therefore, we  perform a hyperbolic limit leading to \eqref{eq:macro_hyp} with macroscopic velocity \eqref{U_T:dep}.

\paragraph{Case $ii)$}
When, instead, the maximum sensing radius $R$ is smaller than both the characteristic lengths, we may consider the positive expansions \eqref{q_approx} and \eqref{S_approx} and substitute them in \eqref{T_dep}. Neglecting the higher order terms in $\lambda$, we get the approximation
\begin{equation}\label{T_dep.ii}
T[q,\cS](\x, v,\hv)=c(\x)\Big[\cS(\x)\,\Gamma_0\,q(\x,\hv)+\cS(\x)\,\Gamma_1\,\nabla q \cdot\hv +\Gamma_1\,q(\x,\hv)\,\nabla\cS\cdot \hv\Big]\,\psi(v)
\end{equation}
with
\[
c(\x)=\dfrac{1}{\cS(\x)\,\Gamma_0}
\]
and 
\[
\Gamma_i:=\int_0^R\lambda^i\gamma(\lambda)\,d\lambda\,,\qquad i=0,1\,.
\]
Re-scaling the space variable as in \eqref{eq:scale_space}, we find 
\[
T_0[q,\cS](\boldsymbol{\xi},v,\hv)=q(\boldsymbol{\xi},\hv)\psi(v)
\]
and
\[
T_1[q,\cS](\boldsymbol{\xi},v,\hv)=\Gamma\Big[\nabla q \cdot\hv +q(\boldsymbol{\xi},\hv)\,\dfrac{\nabla\cS}{\cS}\cdot \hv\Big]\,\psi(v)\,
\]
with
\[
\Gamma:=\dfrac{\Gamma_1}{\Gamma_0}.
\]
Therefore, 
$\Ub_0^T(\boldsymbol{\xi})=0$, because of \eqref{eq:mean_q.0}, and we can perform a diffusive scaling that leads to the zero-order macroscopic diffusion tensor
\begin{equation}\label{DT_0_dep.ii}
\mathbb{D}_T^0(\boldsymbol{\xi})=D\,\mathbb{D}_q(\boldsymbol{\xi})\,,
\end{equation}
and to the macroscopic first-order velocity 
\begin{equation}\label{UT1_dep.ii}
\Ub_T^1(\boldsymbol{\xi})=\bar{U}\,\Gamma\,\nabla\cdot \mathbb{D}_q(\boldsymbol{\xi})+\bar{U}\,\Gamma\,\mathbb{D}_q(\boldsymbol{\xi})\,\,\dfrac{\nabla \cS}{\cS}\,.
\end{equation}
The macroscopic advection-diffusion equation \eqref{eq:macro_diff} now reads (dropping the dependencies)
\begin{equation}\label{macro_dep.ii}
\dfrac{\partial}{\partial \tau}\rho+\nabla\cdot\left[\chi\,\left(\nabla\cdot\mathbb{D}_q\,+\mathbb{D}_q\dfrac{\nabla \cS}{\cS}\right)\,\rho\right]=\nabla\cdot\left[\dfrac{1}{\mu}\,\nabla\cdot \big(D\,\mathbb{D}_q\,\rho\big)\right]
\end{equation}
where
\[
\chi:=\bar{U}\Gamma\,.
\]
Similar considerations to the case $ii)$ of the non-local independent sensing model may be done, except that there is a unique sensitivity $\chi$ that weights equally the two contributions to the advection term \eqref{UT1_dep.ii}.

\paragraph{Case $iii)$}
In this case, we expand only the chemoattractant $\cS(\x+\lambda\hv)$, as in \eqref{S_approx}, and the turning kernel \eqref{T_dep} can be approximated as
\begin{equation}\label{T_dep.iii}
\begin{split}
T[q,\cS](\x, v,\hv)=&c(\x)\Big[\cS(\x)\,\int_{\mathbb{R}_+}\gamma(\lambda)q(\x+\lambda\hv,\hv)\,d\lambda+(\nabla\cS\cdot \hv)\int_{\mathbb{R}_+}\lambda\,\gamma(\lambda)q(\x+\lambda\hv,\hv)\,d\lambda\Big]\psi(v)
\end{split}
\end{equation}
with 
\[
c(\x):=\dfrac{1}{\Gamma_0\,\cS(\x)}\,.
\]
Re-scaling the space variable as in \eqref{eq:scale_space}, we find
\[
T_0[q,\cS](\boldsymbol{\xi}, v,\hv)=\dfrac{1}{\Gamma_0}\,\int_{\mathbb{R}_+}\gamma(\lambda)q(\boldsymbol{\xi}+\lambda\hv,\hv)\,d\lambda \, \psi(v),
\]
and 
\[
T_1[q,\cS](\boldsymbol{\xi}, v,\hv)=\dfrac{1}{\Gamma_0}\,\left(\dfrac{\nabla\cS}{\cS}\cdot \hv\right)\,\int_{\mathbb{R}_+}\lambda\,\gamma(\lambda)q(\boldsymbol{\xi}+\lambda\hv,\hv)\,d\lambda\,\psi(v)\,.
\]
The macroscopic velocity of zero order is then
\begin{equation}\label{UT_dep.iii.0}
\Ub_T^0(\boldsymbol{\xi})=\dfrac{\bar{U}}{\Gamma_0}\int_{\mathbb{S}^{d-1}}\int_{\mathbb{R}_+}\gamma(\lambda)\,q(\boldsymbol{\xi}+\lambda\hv,\hv)\,d\lambda\,\hv\,d\hv\,,
\end{equation}
and, again, it vanishes  because of $\boldsymbol{\xi}+\lambda\hv \in \Omega$ and \eqref{eq:mean_q.0}.
Therefore, the macroscopic diffusion-advection equation is given by
\eqref{eq:macro_diff} with
\begin{equation}\label{DT_0_2.3}
\mathbb{D}_T^0(\boldsymbol{\xi})=\dfrac{D}{\Gamma_0}\,\int_{\mathbb{R}_+}\mathbb{D}_q^{\lambda}(\boldsymbol{\xi})\,\gamma(\lambda)d\lambda=D\bar{\mathbb{D}}_q^0
\end{equation}
and
\begin{equation}\label{UT_dep.iii.1}
\Ub_T^1(\boldsymbol{\xi})=\dfrac{\bar{U}}{\Gamma_0}\,\int_{\mathbb{R}_+}\lambda\,\mathbb{D}_q^{\lambda}(\boldsymbol{\xi})\,\gamma(\lambda)\,d\lambda\,\dfrac{\nabla \cS}{\cS(\boldsymbol{\xi})}=\bar{U}\bar{\mathbb{D}}_q^1(\boldsymbol{\xi})\dfrac{\nabla \cS}{\cS(\boldsymbol{\xi})},
\end{equation}
where we defined
\begin{equation}
\bar{\mathbb{D}}_q^1(\boldsymbol{\xi})=\dfrac{1}{\Gamma_0}\,\int_{\mathbb{R}_+}\lambda\,\mathbb{D}_q^{\lambda}(\boldsymbol{\xi})\,\gamma(\lambda)d\lambda
\end{equation}
as an average of the weighted diffusion tensor of the fibers in the whole neighborhood sensed by the cells, differently form the case $iii)$ of the non-local independent model.

\paragraph{Case $iv)$}
In this case, again, we can only consider the positive approximation \eqref{q_approx}, and the transition probability rewrites as
\begin{equation}\label{T_dep.iv}
\begin{split}
T[q,\cS](\x, v,\hv)=&\Big[c_0(\x)q(\x,\hv)\,\int_{\mathbb{R}_+} \gamma(\lambda)\,\cS(\x+\lambda\hv)\,d\lambda+c_1(\x)\nabla q\cdot \hv \,\int_{\mathbb{R}_+}\lambda\,\gamma(\lambda)\,\cS(\x+\lambda\hv)\,d\lambda\Big]\psi(v)
\end{split}
\end{equation}
where 
\[
c_0(\x)^{-1}:=2\int_{\mathbb{S}^{d-1}}q(\x,\hv)\int_{\mathbb{R}_+}\gamma(\lambda)\cS(\x+\lambda\hv)\,d\lambda\,d\hv
\]
and 
\[
c_1(\x)^{-1}:= 2\int_{\mathbb{S}^{d-1}}(\nabla q \cdot \hv)\,\int_{\mathbb{R}_+}\lambda\,\gamma(\lambda)\cS(\x+\lambda\hv)\,d\lambda\,d\hv\,,
\]
both different from zero. As before, by re-scaling \eqref{T_dep.iv} with \eqref{eq:scale_space}, we get $T[q,\cS]=T_0[q,\cS]$ and we have that the average velocity $\Ub_T^0=\Ub_T \ne0$. In particular, it is given by 
\begin{equation}\label{UT_0:dep.iv}
\begin{split}
\Ub_T(\boldsymbol{\xi}):&=\dfrac{\bar{U}}{c_0(\boldsymbol{\xi})}\int_{\mathbb{S}^{d-1}}\hv \,q(\boldsymbol{\xi},\hv)\int_{\mathbb{R}_+}\gamma(\lambda)\,\cS(\boldsymbol{\xi}+\lambda\hv)\,d\lambda\,d\hv\,\\
&+\dfrac{\bar{U}}{c_1(\boldsymbol{\xi})}\int_{\mathbb{S}^{d-1}} \hv \otimes \hv\,\nabla q(\boldsymbol{\xi},\hv) \int_{\mathbb{R}_+}\lambda\,\gamma(\lambda)\cS(\boldsymbol{\xi}+\lambda \hv)  \, d\lambda \, d\hv
\end{split}
\end{equation}
and, thus, we perform a hyperbolic limit leading to \eqref{eq:macro_hyp}.
The mean velocity \eqref{UT_0:dep.iv} is a linear combination of a non-local measure of the chemoattractant $\cS$ over the fibers network and a non-local average of $\cS$ weighted by the directional average of the spatial variability of the fiber direction.\\
\begin{table}[!htbp]
\begin{scriptsize}
\hspace{0cm}\begin{tabular}{|c|c|c|}
\hline
\rule{0pt}{3ex}Case & non-local independent sensing \eqref{transport.general}-\eqref{J_r.S}-\eqref{T_indip}  & non-local dependent sensing \eqref{transport.general}-\eqref{J_r.S}-\eqref{T_dep} \\[1.5ex]
 \hline\hline
\rule{0pt}{3ex}$i)$ & drift dominated & drift dominated \\[2ex]
 &\!$\Ub_T= c\bar{U}\!\!\displaystyle\int_{\mathbb{S}^{d-1}}\!\!\!\hv\!\!\displaystyle\int_{ 0}^{ R}\!\!\!\!\gamma_{\cS}(\lambda) \mathcal{S}(\boldsymbol{\xi}+\lambda\hv)d\lambda\!\!\int_{ 0}^{ R}\!\!\!\!\gamma_q(\lambda)\,q(\boldsymbol{\xi}+\lambda\hv,\hv)d\lambda d\hv$\!&\!\!$\Ub_T=c\bar{U}\!\!\displaystyle\int_{\mathbb{S}^{ d-1}}\!\!\hv\!\!\int_{0}^R\!\!\!\!\gamma(\lambda) \mathcal{S}(\boldsymbol{\xi}+\lambda\hv) q(\boldsymbol{\xi}+\lambda\hv,\hv)d\lambda\,d\hv$\!\\[3ex]

\hline 
\rule{0pt}{3ex} $ii)$& drift-diffusion & drift-diffusion \\[2ex]
 & $\mathbb{D}_T^0=D\,\mathbb{D}_q$& $\mathbb{D}_T^0=D\,\mathbb{D}_q$\\[2ex]
 & $\Ub_T^1=\bar{U}\,\left[\Gamma^q\,\nabla \cdot \mathbb{D}_q+\,\Gamma^{\tiny\cS}\,\mathbb{D}_q\,\dfrac{\nabla \cS}{\cS}\right]  $& $\Ub_T^1=\bar{U}\Gamma\,\left[\nabla\cdot \mathbb{D}_q+\,\mathbb{D}_q\,\,\dfrac{\nabla \cS}{\cS}\right]$\\[3ex]

\hline
\rule{0pt}{3ex} $iii)$& drift-diffusion & drift-diffusion  \\[2ex]
&$\mathbb{D}_T^0= D\bar{\mathbb{D}}_q^0$ &$\mathbb{D}_T^0=D\bar{\mathbb{D}}^0_q$\\[3ex]
&$\Ub_T^1=\bar{U}\,\Gamma^{\tiny\cS}\,\bar{\mathbb{D}}^0_q\dfrac{\nabla \cS}{\cS}$ &$\Ub_T^1=\bar{U}\,\bar{\mathbb{D}}_q^1\,\dfrac{\nabla \cS}{\cS}$\\[3ex]

\hline
\rule{0pt}{3ex} $iv)$&  drift dominated & drift dominated  \\[2ex]
&$\Ub_T=\dfrac{\bar{U}\,\Gamma_0^q}{c_0}\,\displaystyle\int_{\mathbb{S}^{d-1}}\!\!\!\hv\,q\,\,\int_{0}^R\!\!\!\!\gamma_{\tiny \cS}(\lambda)\,\cS(\boldsymbol{\xi}+\lambda\hv)\,d\lambda\,d\hv$
&$\Ub_T:=\dfrac{\bar{U}}{c_0}\displaystyle\int_{\mathbb{S}^{d-1}}\!\!\!\hv \,q\int_{0}^R\!\!\!\!\gamma(\lambda)\,\cS(\boldsymbol{\xi}+\lambda\hv)\,d\lambda\,d\hv\,$  \\[3ex]
& $+\dfrac{\bar{U}\,\Gamma_1^q}{c_1}\displaystyle\int_{\mathbb{S}^{d-1}}\!\!\!\hv \otimes \hv \,\nabla q \int_{0}^R\!\!\!\!\gamma_{\scS}(\lambda)\,\cS(\boldsymbol{\xi}+\lambda \hv) \, d\lambda \, d\hv$ &$+\dfrac{\bar{U}}{c_1}\displaystyle\int_{\mathbb{S}^{d-1}} \!\!\!\hv \otimes \hv\,\nabla q \int_{0}^R\!\!\!\!\lambda\,\gamma(\lambda)\cS(\boldsymbol{\xi}+\lambda \hv)  \, d\lambda \, d\hv$ \\[3ex]
\hline
\end{tabular}
\caption{Summary of the models (dropping the local dependencies in $\boldsymbol{\xi}$). }
\end{scriptsize}
\label{Model_summary}
\end{table}

\subsubsection{Comments}

We can observe that, if $\gamma_q=\gamma_{\scS}=\gamma=\delta(\lambda-R)$, the two non-local transport models for independent and dependent sensing are the same, while, if the sensing kernels are not dirac deltas (even if $\gamma_q=\gamma_{\scS}=\gamma$), the transport models are always different.
Instead, at the macroscopic level, with any choice of the sensing functions the models coincide only in case $ii)$. 
In this case, in fact, the macroscopic limits are different only if $\gamma_q \neq \gamma_{\scS}$, while in the cases $iii)$ and $iv)$ they are different if the sensing kernel are not dirac deltas (even if $\gamma_{\scS}=\gamma_q=\gamma$). The relevant difference concerns the macroscopic transport velocities (see \eqref{UT_indip.iii} and \eqref{UT_dep.iii.1} for the case $iii$, and \eqref{UT_0:indip.iv} and \eqref{UT_0:dep.iv} for the case $iv$). In fact, in the cases $iii)$ and $iv)$, for the non-local dependent sensing model, as only one cue is considered non-locally and both cues are averaged with the same sensing function $\gamma$, we have a weighted average on $\lambda$ of the non-local quantities, that results in the weighted averages \eqref{UT_dep.iii.1} and the second term of \eqref{UT_0:dep.iv}. These remarks are summarized in Table \ref{Model_equal}. 

\begin{table}[!htbp]
\footnotesize
\begin{center}
\begin{tabular}{c|c|c|c}
 & $\gamma_q=\gamma_{\scS}=\gamma=\delta$ & $\gamma_q=\gamma_{\scS}=\gamma\ne\delta$ &$\gamma_q\ne\gamma_{\scS}$\\[1.5ex]
\hline
Meso models \eqref{transport.general}-\eqref{J_r.S}-\eqref{T_indip} and \eqref{transport.general}-\eqref{J_r.S}-\eqref{T_dep}& = &$\ne$  &$\ne$\\[1.5ex]
\hline
Macro models case $i)$& =& $\ne$ &$\ne$\\[1.5ex]
\hline
Macro models case $ii)$&  = & = &$\ne$\\[1.5ex]
\hline
Macro models case $iii)$&  = &$\ne$  &$\ne$\\[1.5ex]
\hline
Macro models case $iv)$&  = &$\ne$  &$\ne$\\[1.5ex]
\end{tabular}
\caption{ Summary of the comparison of the models for different choices of the sensing functions. $=$ indicates the cases in which the models coincide, while $\ne$ the ones in which the models are different.}
\label{Model_equal}
\end{center}
\end{table}
 
\section{Numerical simulations}
We shall now propose two-dimensional numerical simulations in order to illustrate the behavior of the kinetic transport models for non-local independent sensing and non-local dependent sensing.
In particular, we shall integrate numerically the transport equation as in \cite{loy2019JMB} and, then, we shall compute the macroscopic density \eqref{def_rho}.
Concerning the fibers network, a classical used distribution is the Von Mises distribution \cite{mardia2009}
\[
\tilde{q}(\x,\hv)=\dfrac{1}{2\pi I_0(k(\x))}e^{k(\x)\,\ub(\x)\cdot \hv}
\]
where $I_\nu(k)$ is the modified Bessel function of first kind of order $\nu$ and 
\[
\ub(\x)=(\cos(\theta_q(\x)),\sin(\theta_q(\x))).
\]
It can be proved that ${\bf E}_{\tilde{q}}(\x)=\ub(\x)$ \cite{HMPS2017MBE}, and, therefore, $\theta_q(\x)$ is the mean direction in the space $[0,2\pi)$ of the fibers located at point $\x$. As we are dealing with cell migrating on a non-polarized network of fibers, we shall consider the symmetric version, namely the Bimodal Von Mises distribution
\[
q\left(\x,\hv\right)=\dfrac{1}{4\pi I_0(k(\x))}\left(e^{k(\x)\,\ub(\x)\cdot \hv}+e^{-k(\x)\,\ub(\x)\cdot \hv}\right)\,,
\]
that also satisfies Q3; its variance is \cite{HMPS2017MBE}
\[
\mathbb{D}_q(x)=\dfrac{1}{2}\left(1-\dfrac{I_2(k)}{I_0(k)} \right)\mathbb{I}_2+\dfrac{I_2(k)}{I_0(k)}\ub \otimes \ub,
\]
where $\mathbb{I}_2$ is the identity tensor in $\mathbb{R}^{2\times 2}$, while $k$ and $\ub$ are functions of $\x$.
Moreover, the variance in the space $[0,2\pi)$ is the scalar
\[
D_q(\x)=\dfrac{1}{2}\displaystyle \int_{0}^{2\pi} q (\theta-\theta_q)^2  \, d\theta=\left(1-\dfrac{I_1(k)}{I_0(k)}\right)
\]
that represents the degree of alignment of the fibers at point $\x$.
\subsection{Test 1: local ECM sensing and non-local chemotaxis}
\begin{figure}[!ht] 
\begin{subfigure}{0.32\textwidth}
        \centering
        \includegraphics[width=\textwidth]{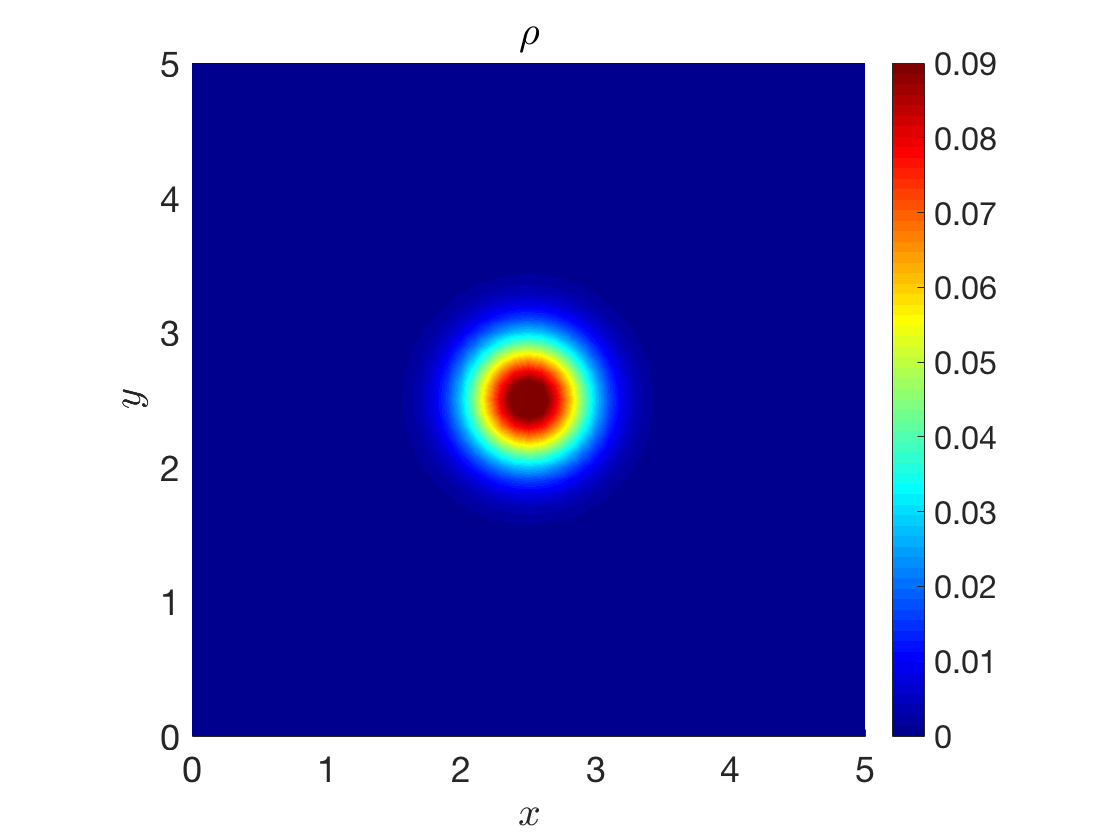}
        \caption{\scriptsize Initial cell distribution}
        \label{TM_rho_t0}
    \end{subfigure}
   \begin{subfigure}{0.32\textwidth}
        \centering
        \includegraphics[width=\textwidth]{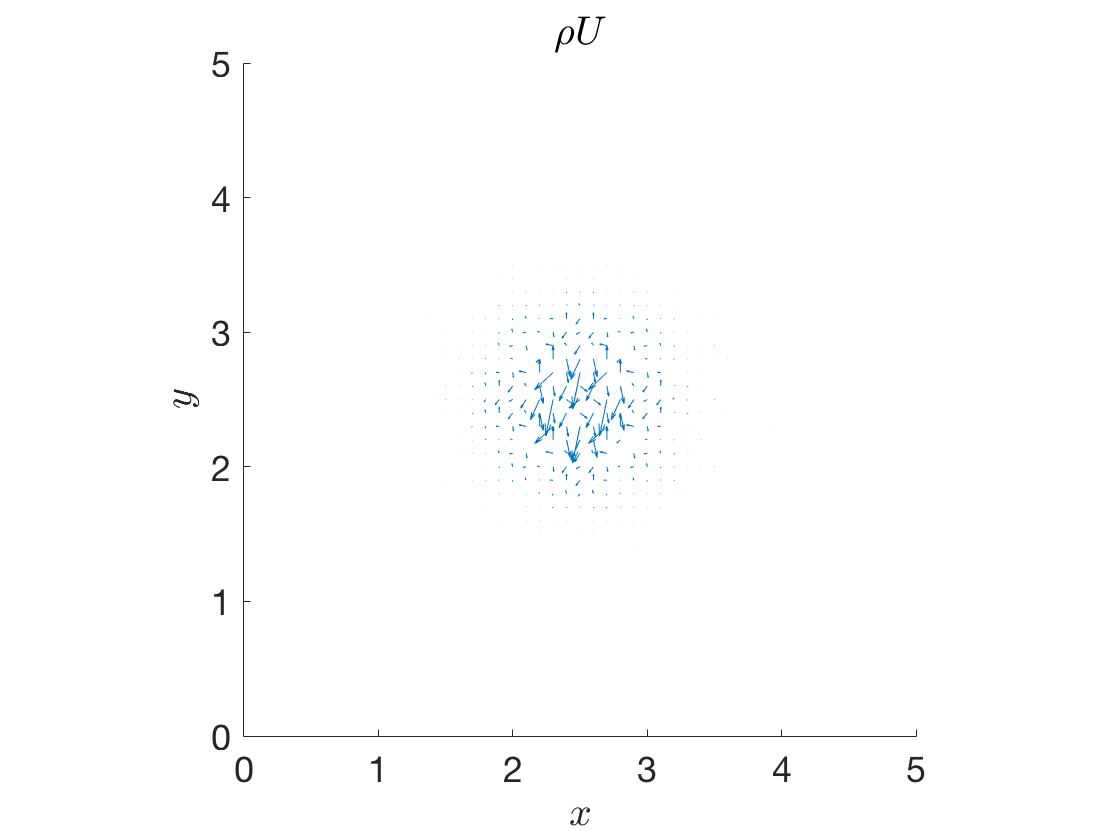}
        \caption{\scriptsize Initial average polarization}
        \label{TM_rhoU_t100}   
     \end{subfigure} 
     \begin{subfigure}{0.32\textwidth}
        \centering
        \includegraphics[width=0.8\textwidth]{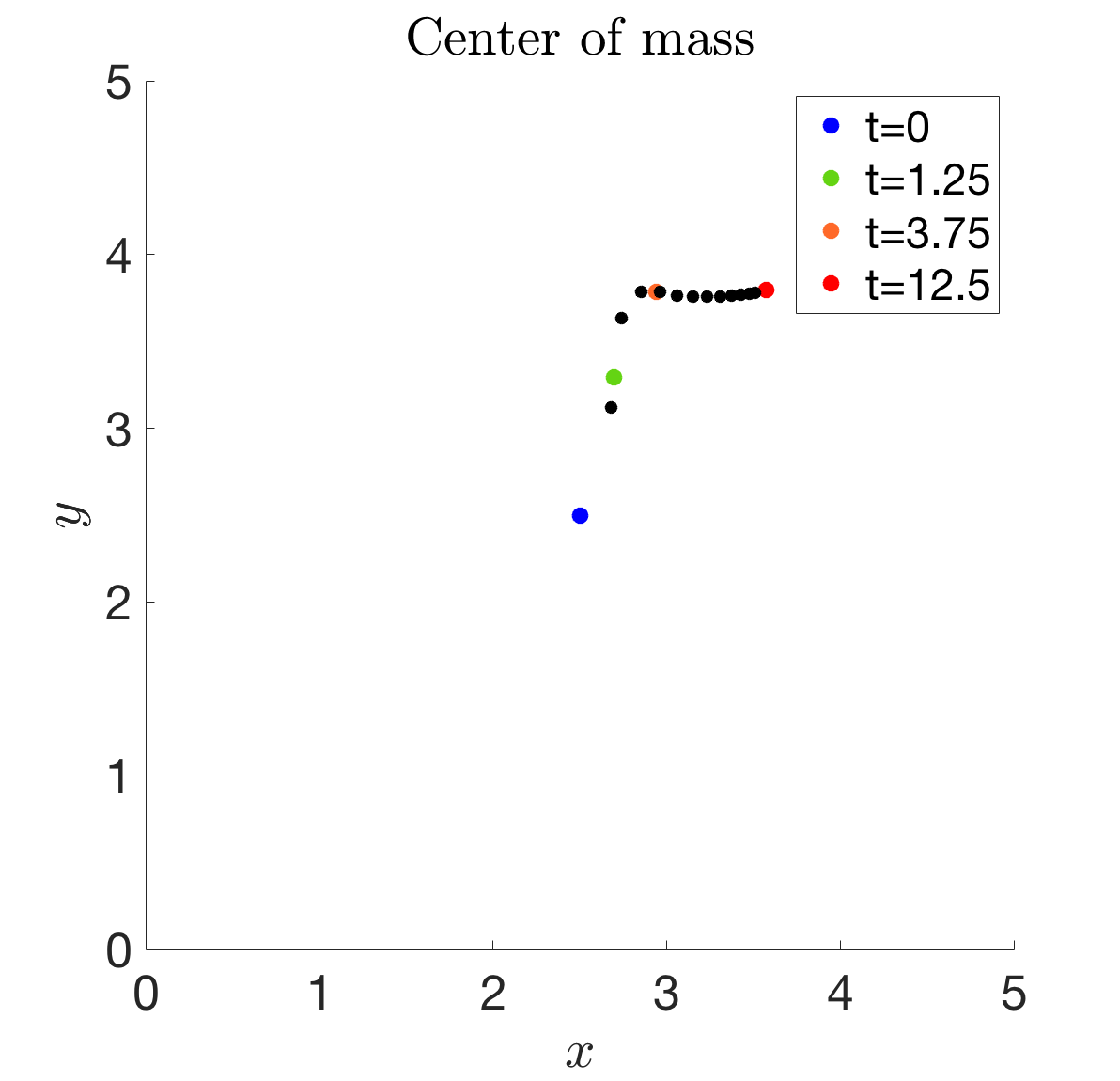}   
        \caption{\scriptsize Center of mass: trajectory.}
        \label{TM_CM_T}
     \end{subfigure}
    \begin{subfigure}{0.32\textwidth}
        \centering
        \includegraphics[width=\textwidth]{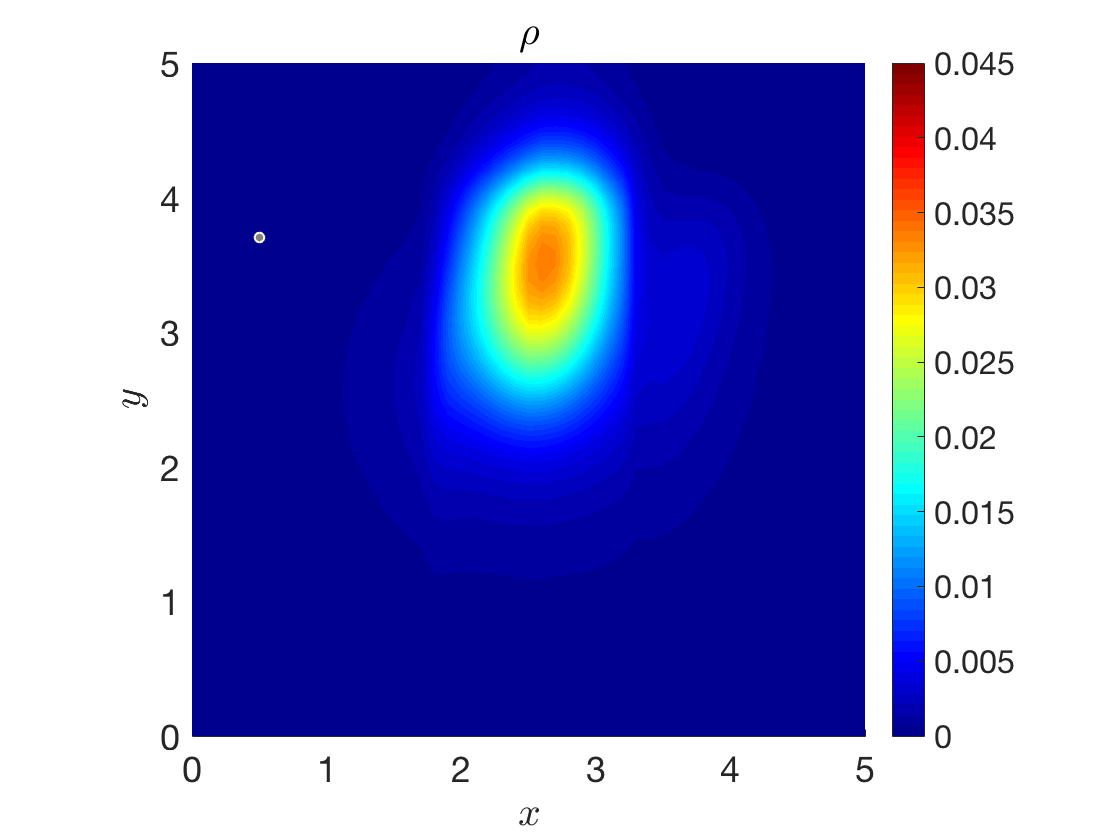}
        \caption{\scriptsize t=1.25}
        \label{TM_t50}
    \end{subfigure}
   \begin{subfigure}{0.32\textwidth}
        \centering
        \includegraphics[width=\textwidth]{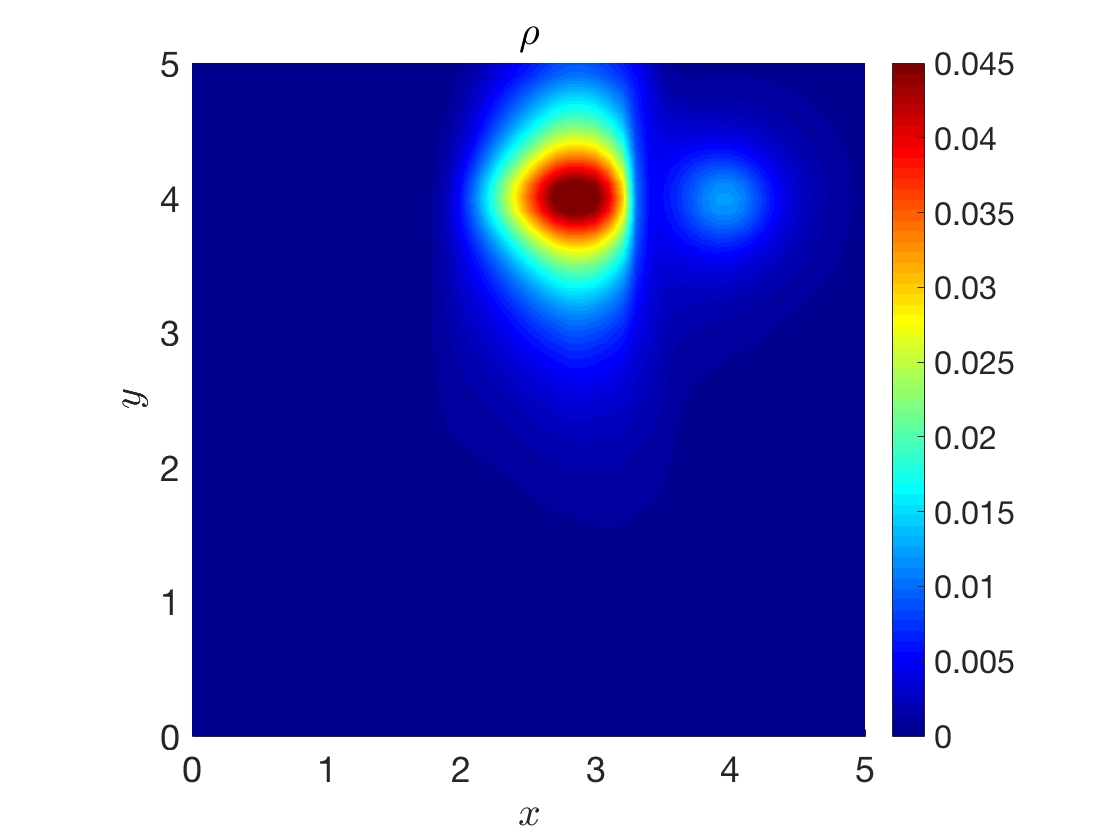}
        \caption{\scriptsize t=3.75}
        \label{TM_t150}   
     \end{subfigure} 
     \begin{subfigure}{0.32\textwidth}
        \centering
        \includegraphics[width=\textwidth]{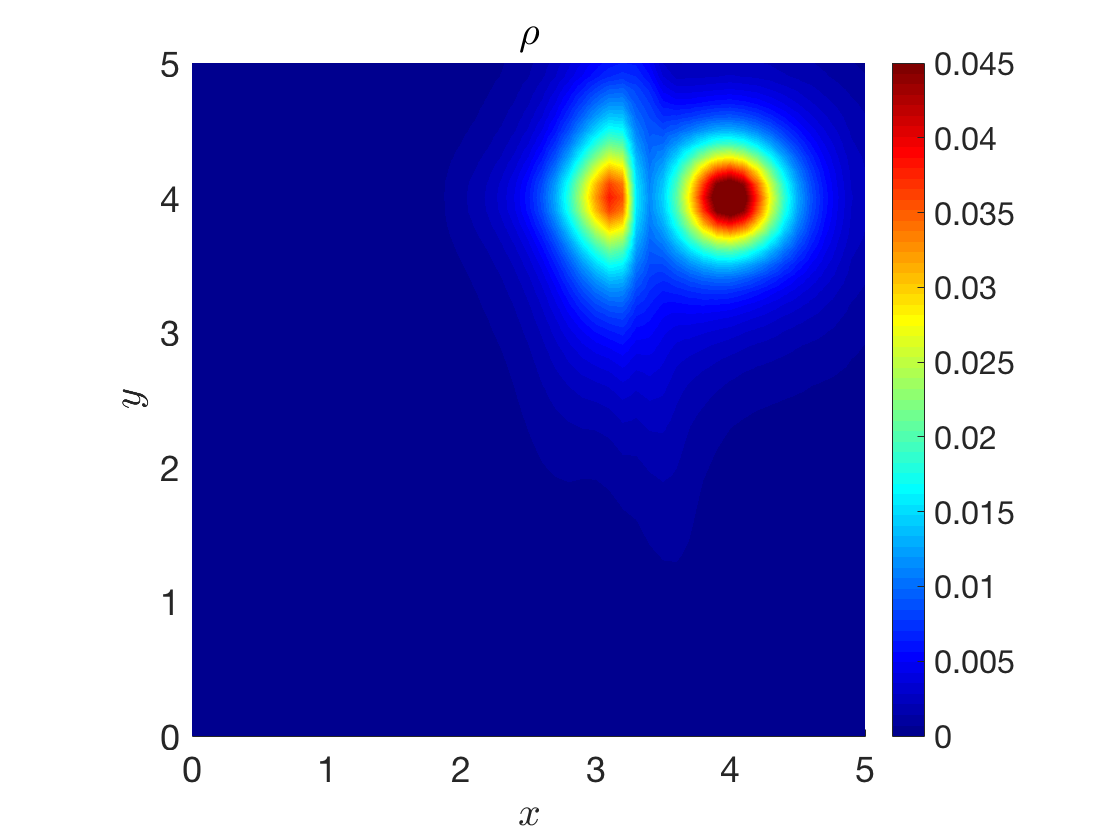}   
        \caption{\scriptsize t=12.5}
        \label{TM_t500}
     \end{subfigure}
      \begin{subfigure}{0.32\textwidth}
        \centering
        \includegraphics[width=\textwidth]{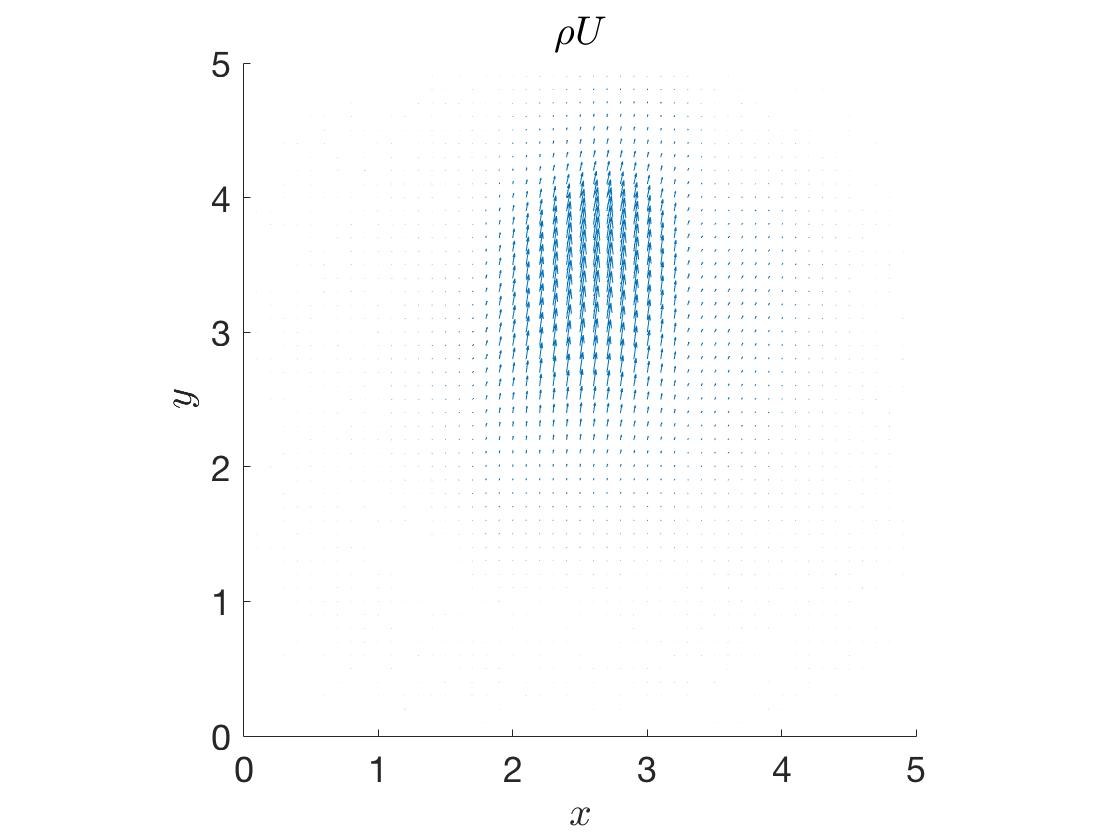}
        \caption{\scriptsize t=1.25}
        \label{TM_theta50}
    \end{subfigure}
   \begin{subfigure}{0.32\textwidth}
        \centering
        \includegraphics[width=\textwidth]{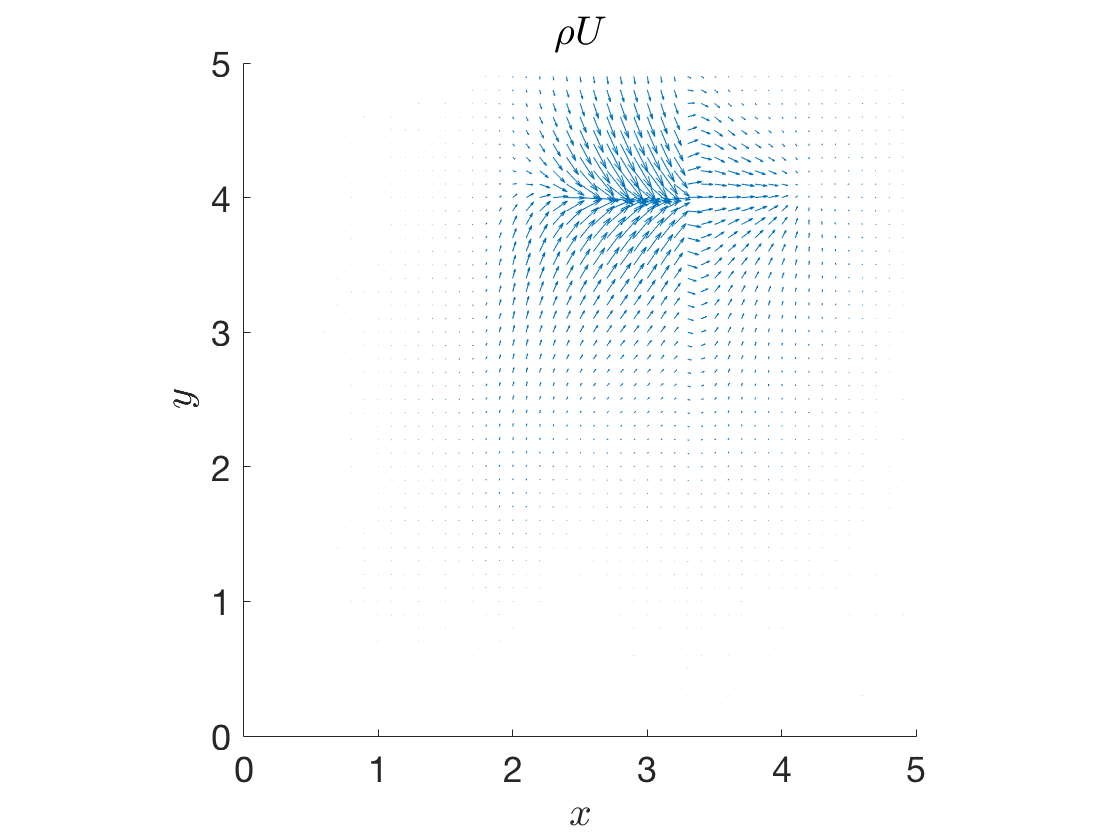}
        \caption{\scriptsize t=3.75}
        \label{TM_theta150}   
     \end{subfigure} 
     \begin{subfigure}{0.32\textwidth}
        \centering
        \includegraphics[width=\textwidth]{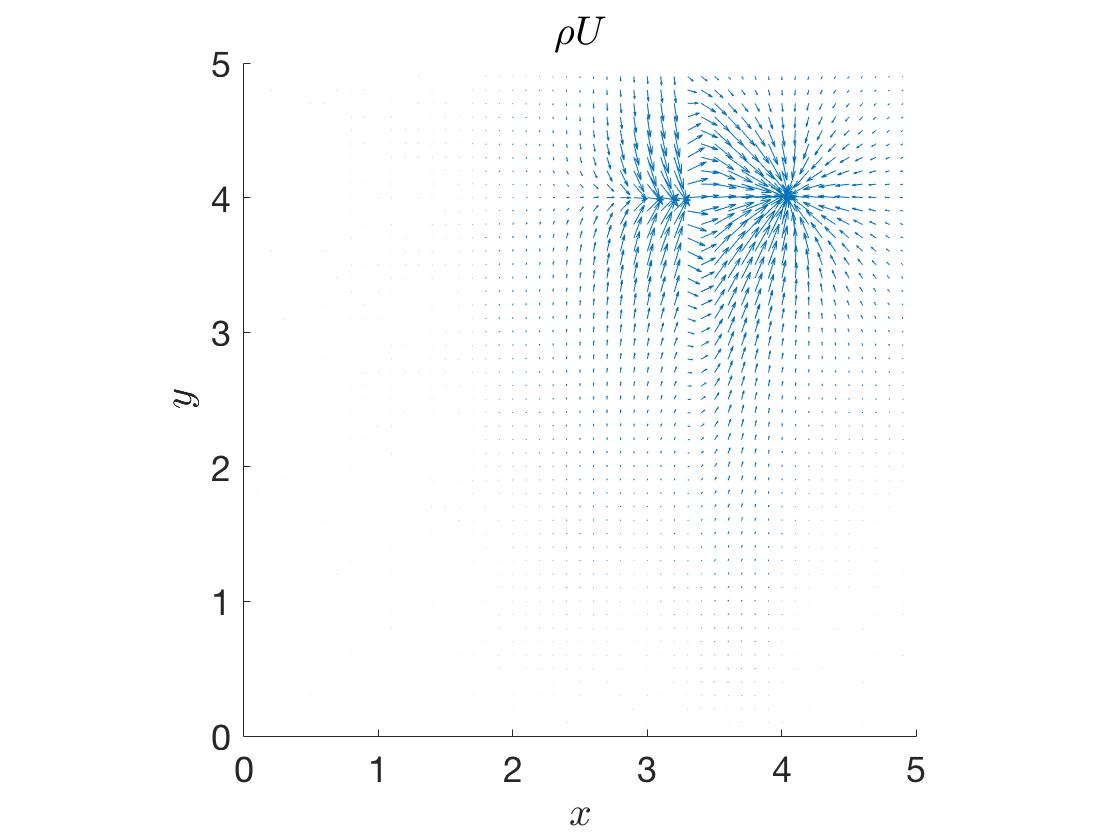}   
        \caption{\scriptsize t=12.5}
        \label{TM_theta500}
     \end{subfigure}
   \caption{{\bf Test 1} Evolution of the initial distribution given in (a) for the case of local $q$ and non-local chemoattractant $\cS$ with sensing function $\gamma_{\scS}=\delta(\lambda-R)$. In (b), $\cS$ is a Gaussian centred in $(4,4)$ and with $m_{\cS}=10$ and $\sigma_{\cS}^2=0.1$. The sensing radius of the cells is set to $R=0.5$. (c): trajectory of the center of mass of the cell population, where each black dot is plotted every $\Delta t=1$. Figs. (d)-(f): evolution of the macroscopic density.  Figs. (g)-(i): polarizations of the cells.}
       \label{TM}
\end{figure}

 As a first example, we shall present the particular case in which the sensing of $q$ is local. This illustrates the effect of a second directional cue when dealing with a cell population migrating by contact guidance and evaluating the local alignment of the fibers over a non-polarized network. Formally, we are dealing with \eqref{T_indip} in which $\gamma_q=\delta(\lambda-0)$. In particular, we shall consider a region  
\begin{equation}\label{Omega_q}
\Omega_q=\left\lbrace \x=(x,y)\in \Omega \quad s.t. \quad x_1 \le x \le x_2 \right\rbrace
\end{equation}
with $x_1=1.8$ and $x_2=3.2$ in which the fibers are strongly aligned along the direction identified by $\theta_q=\pi/2$. In particular, for $(x,y) \in \Omega_q$, $k(x,y)=700$, such that $D_q=5\cdot 10^{-3}$ . In the rest of the domain $\Omega-\Omega_q$ fibers are uniformly distributed.
The chemoattractant has a Gaussian profile 
\begin{equation}\label{S.gauss}
\cS(x,y)=\dfrac{m_{\cS}}{\sqrt{2\pi \sigma_{\cS}^2}}e^{-\dfrac{\left((x,y)-(x_{\cS},y_{\cS})\right)^2}{2\sigma_{\cS}^2}}.
\end{equation}
In particular, in Test 1 (see Fig. \ref{TM}) we choose $(x_{\cS},y_{\cS})=(4,4), \, m_{\cS}=10, \sigma_{\cS}^2=0.1$. The initial condition for the cell population is a Gaussian 
\begin{equation}\label{ci_cells_gaussian}
\rho_0(x,y)=r_0 e^{-\dfrac{\left((x,y)-(x_0,y_0) \right)^2}{2\sigma_0^2}}
\end{equation}
with $r_0=0.1$ and $\sigma_0^2=0.1$. In this first test, the initial condition for the cell population is centered in $(x_0,y_0)=(2.5,2.5)$, $\ie$, the center of the region $\Omega_q$ (see Fig. \ref{TM_rho_t0}). Without chemoattractant, because of the presence of highly aligned fibers, we would expect that cells diffuse anisotropically in the preferential direction of the fibers $\pm\pi/2$, forming the well known ellipsis \cite{Painter2008}, that represents cells moving with the same probability along direction $\pi/2$ and $-\pi/2$. In the present case, due to the presence of a chemoattractant, the symmetry is broken, and, even if $q$ describes a non-polarized fibers network, there is a preferential sense of motion (see Fig. \ref{TM_t50}-\ref{TM_t500}). In particular, cells migrate along the fibers in the direction identified by $\theta_q=\pi/2$, corresponding to the preferential sense imposed by the presence of the chemoattractant in the upper-right corner of the domain $\Omega$. Given this directional setting, the cell population dynamics is also greatly affected by the strength of the chemoattractant, that depends on $m_{\cS}$ and $\sigma_{\cS}^2$, the degree of the alignment $D_q$, that depends on $k(x,y)$, and by the sensing radius $R$. Another important aspect is the sensing function $\gamma_{\scS}$, that influences the transient dynamics and, especially, the relaxation time. This appears to be double in the case of a Heaviside function, since the kernel $\gamma_{\scS}$ doubles when computed with a Heaviside function instead of a Dirac delta (see also \cite{loy2019JMB}). 

We also analyzed the average polarization of the cells at every position $\x$, that is given by the momentum \eqref{def_rhoU}. The microscopic directions of cells are initially randomly distributed and they start from a vanishing initial speed (see Fig. \ref{TM_rhoU_t100}). Then, they start to align along the fibers and to migrate upward  in the direction individuated by the angle $\pi/2$, since cells sense the chemoattractant (see Figs. \ref{TM_theta50}-\ref{TM_theta150}). Eventually when cells reach the level $y=4$, the microscopic directions polarize towards the chemoattractant (see Fig. \ref{TM_theta500}). 
The center of mass plotted in Fig. \ref{TM_CM_T} stays in the region $\Omega_q$ during the migration of cells along the fibers bundle in $\Omega_q$, and it moves out of $\Omega_q$ only when it reaches $y=4$. The black dots are plotted every $\Delta t=1$ and it is clear that the highest acceleration happens when cells are on the bundle of fibers, while they are slowed down when they start to move out of the fibers stripe $\Omega_q$.

\subsection{Test 2: non-local ECM sensing and chemotaxis}

\begin{figure}[!htbp] 
\begin{subfigure}{0.49\textwidth}
        \centering
        \includegraphics[width=\textwidth]{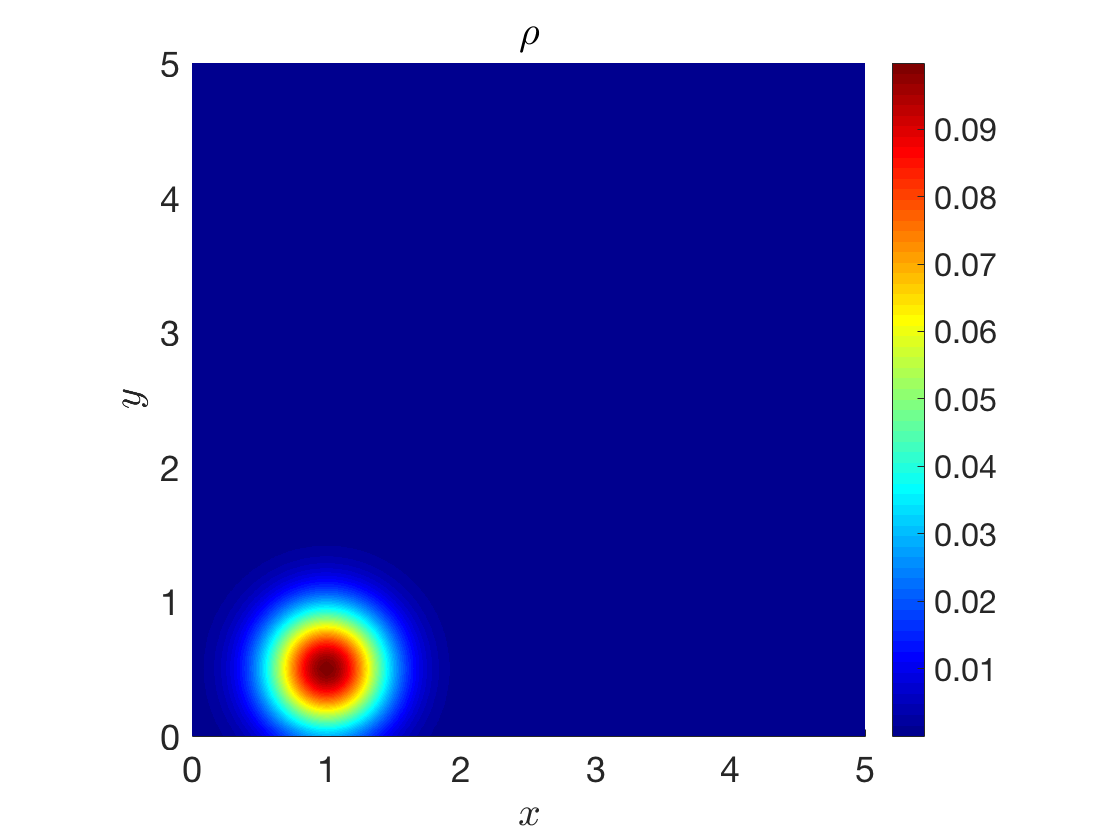}
        \caption{\scriptsize Initial condition for cells}
        \label{InCon_cells}
    \end{subfigure}
   \begin{subfigure}{0.49\textwidth}
        \centering
        \includegraphics[width=0.75\textwidth]{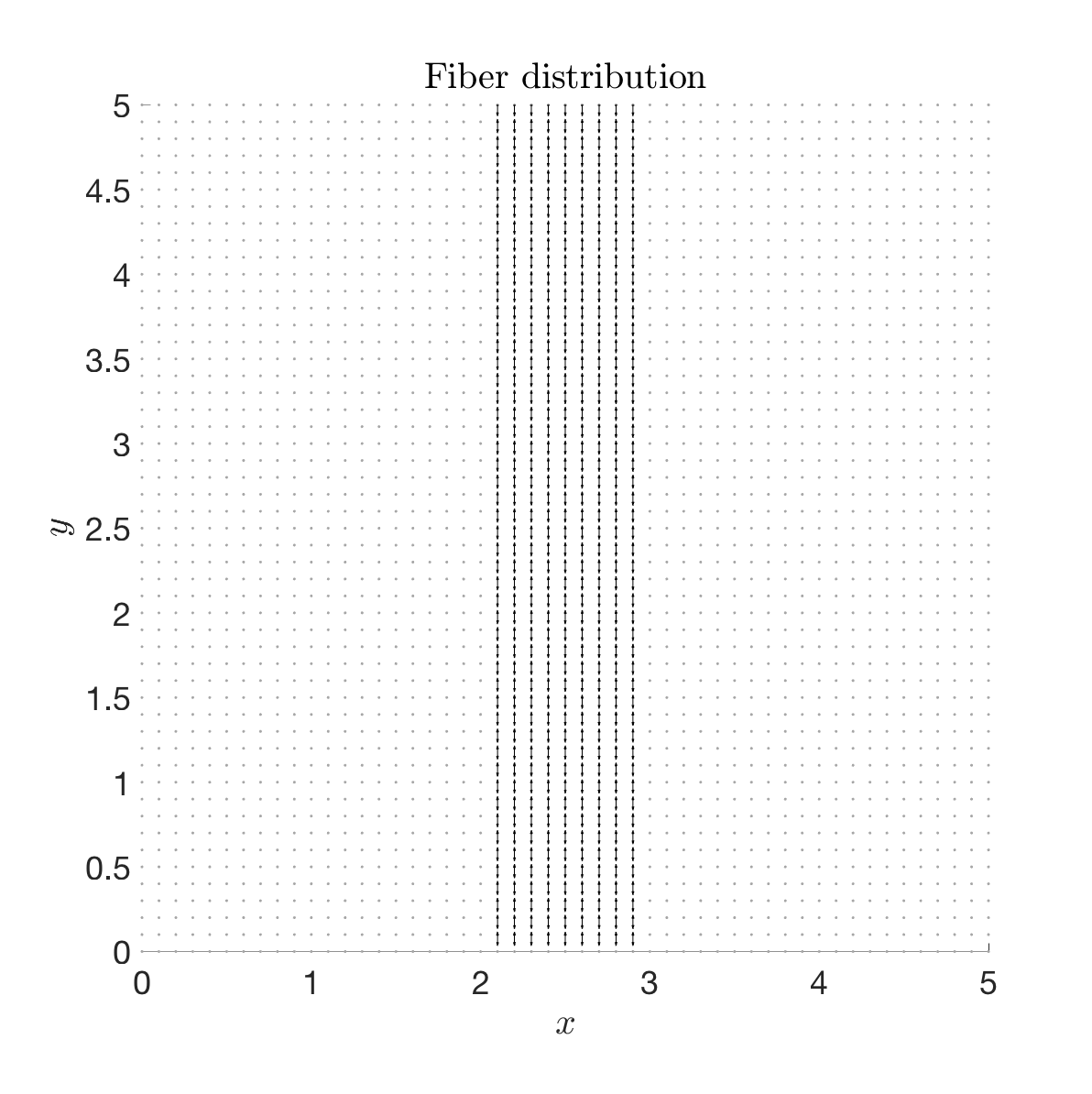}
        \caption{\scriptsize Initial fiber distribution}
        \label{InCon_fiber}   
     \end{subfigure} 
\begin{subfigure}{0.24\textwidth}
        \centering
        \includegraphics[width=\textwidth]{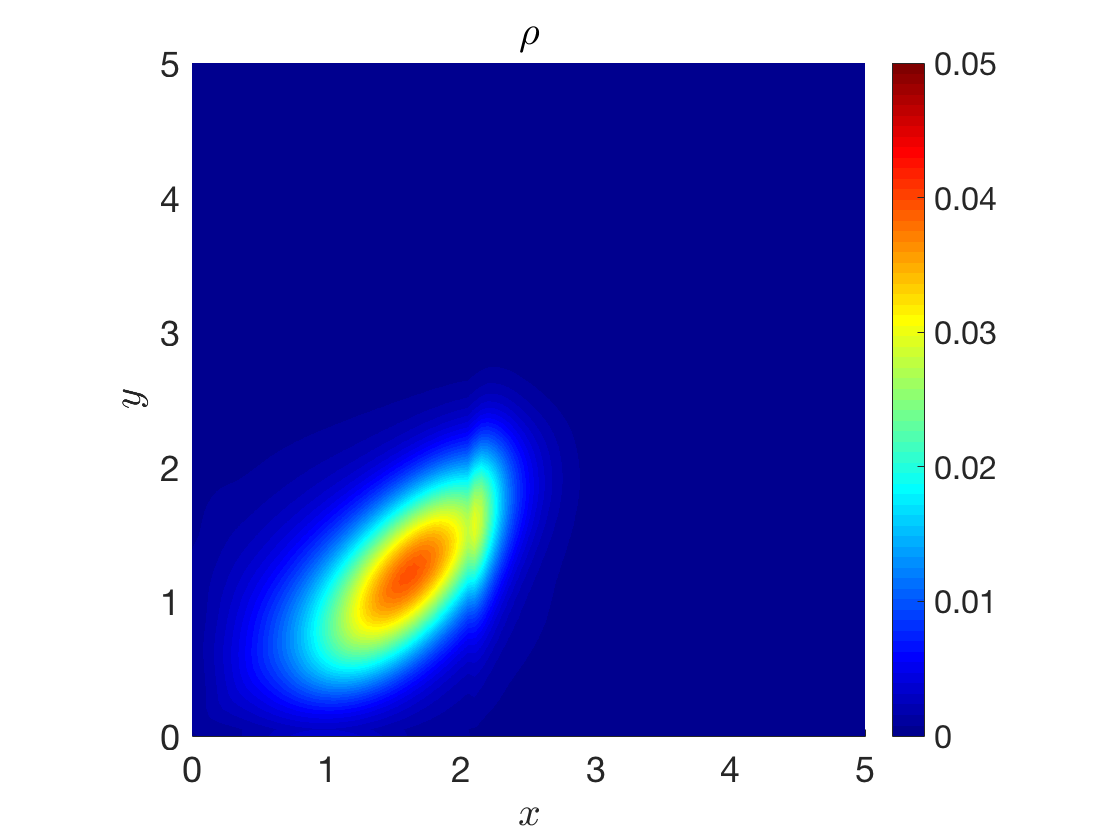}
        \caption{\scriptsize t=1.25}
        \label{TM_cellout_t100}
    \end{subfigure}
   \begin{subfigure}{0.24\textwidth}
        \centering
        \includegraphics[width=\textwidth]{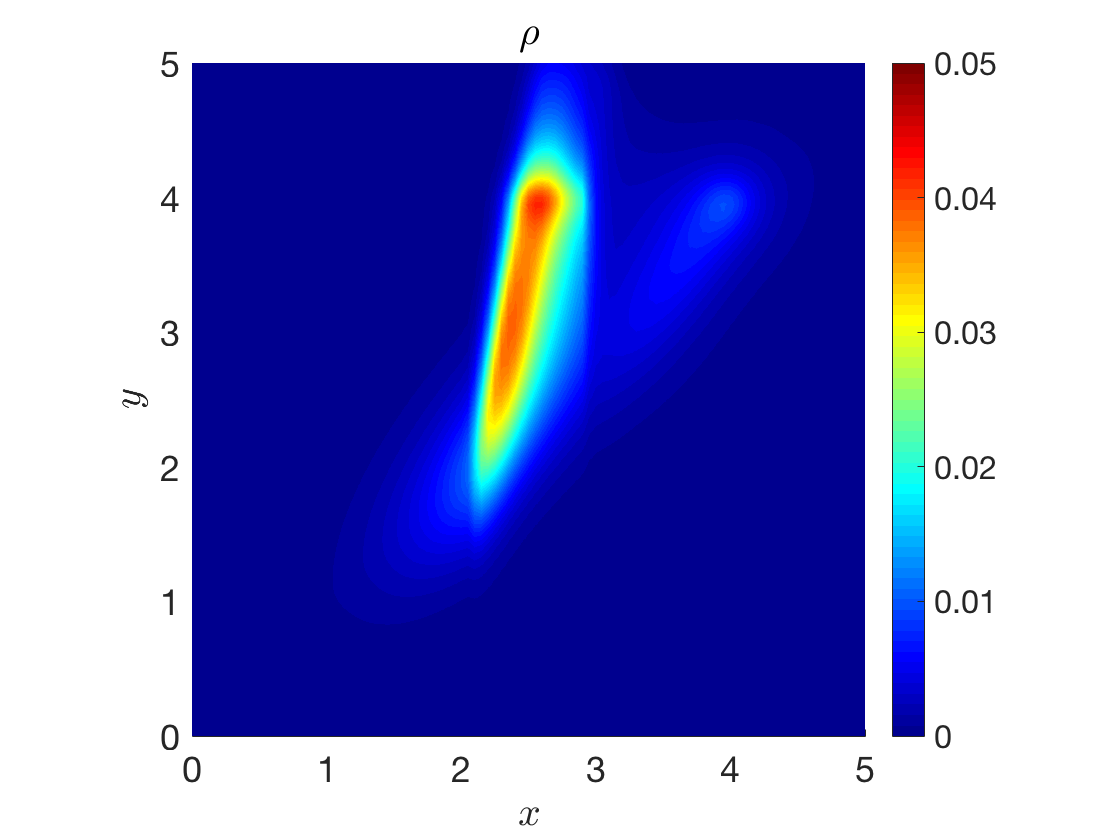}
        \caption{\scriptsize t=3.75}
        \label{TM_cellout_t300}   
     \end{subfigure} 
     \begin{subfigure}{0.24\textwidth}
        \centering
        \includegraphics[width=\textwidth]{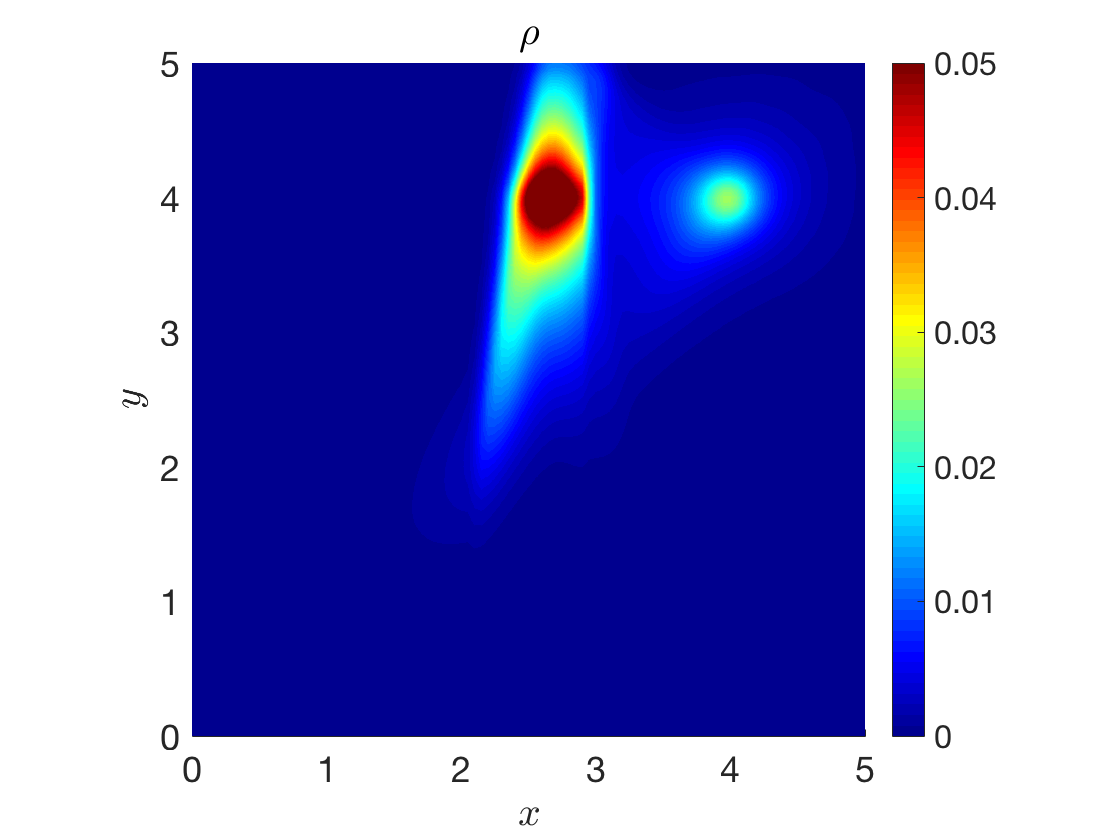}   
        \caption{\scriptsize t=5}
        \label{TM_cellout_t400}
     \end{subfigure}
       \begin{subfigure}{0.24\textwidth}
        \centering
        \includegraphics[width=\textwidth]{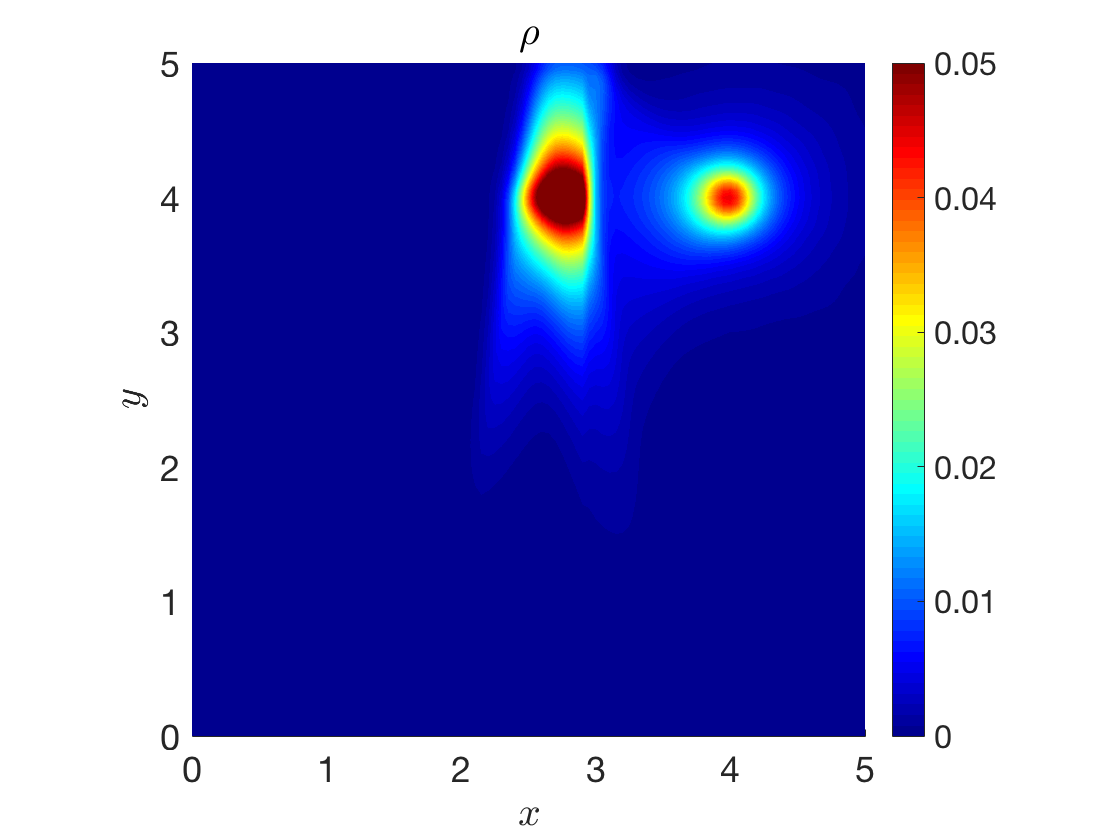}   
        \caption{\scriptsize t=6.25}
        \label{TM_cellout_t500}
     \end{subfigure}
\begin{subfigure}{0.24\textwidth}
        \centering
        \includegraphics[width=\textwidth]{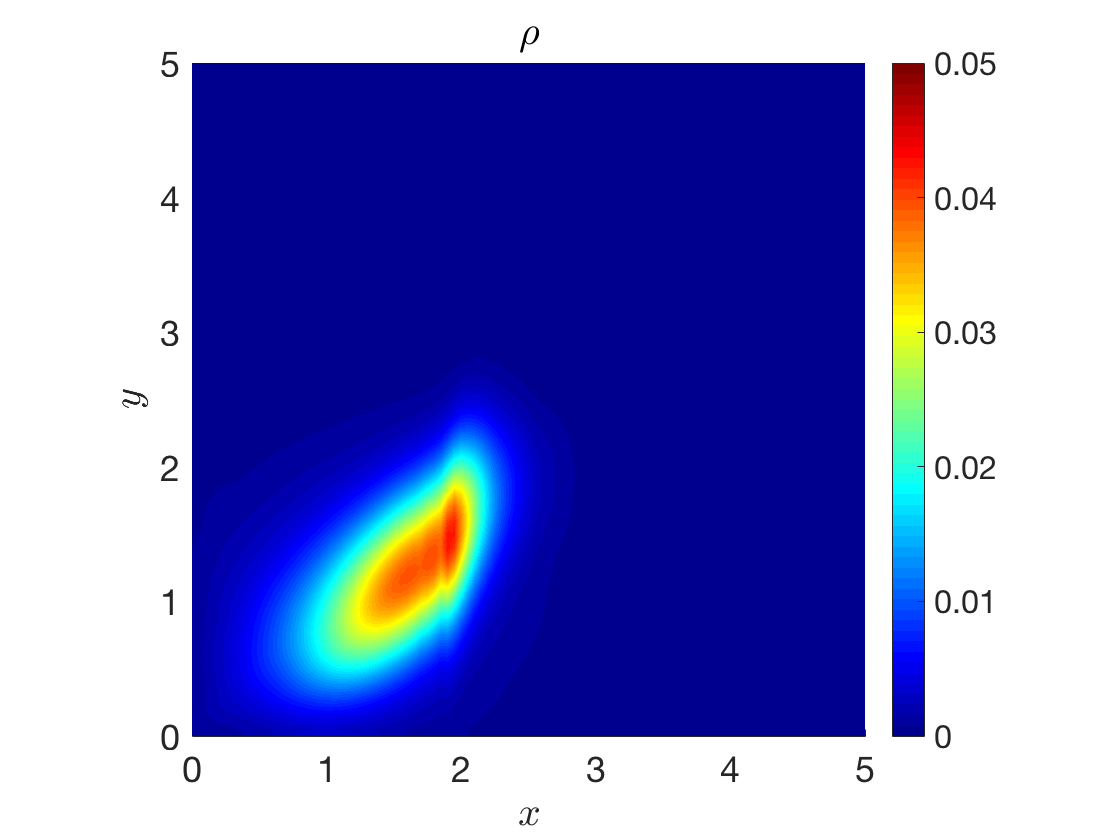}
        \caption{\scriptsize t=1.25}
        \label{M23_cellout_t100}
    \end{subfigure}
   \begin{subfigure}{0.24\textwidth}
        \centering
        \includegraphics[width=\textwidth]{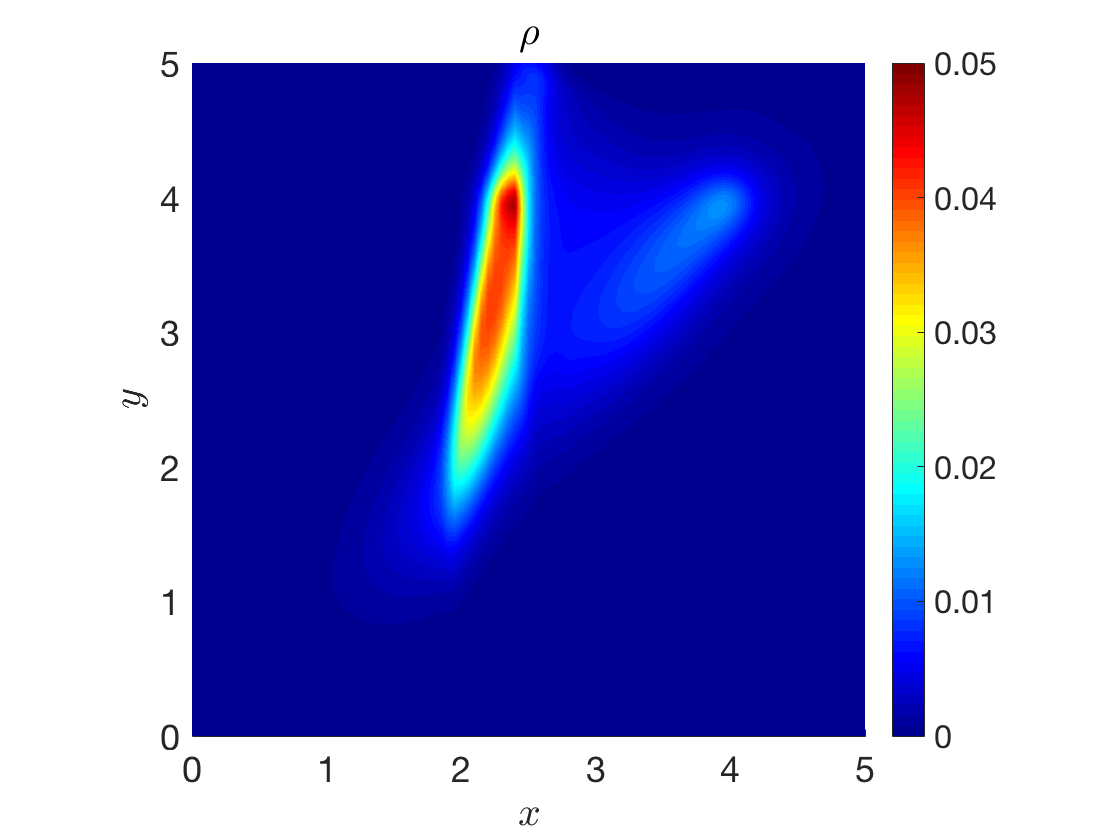}
        \caption{\scriptsize t=3.75}
        \label{M23_cellout_t300}   
     \end{subfigure} 
     \begin{subfigure}{0.24\textwidth}
        \centering
        \includegraphics[width=\textwidth]{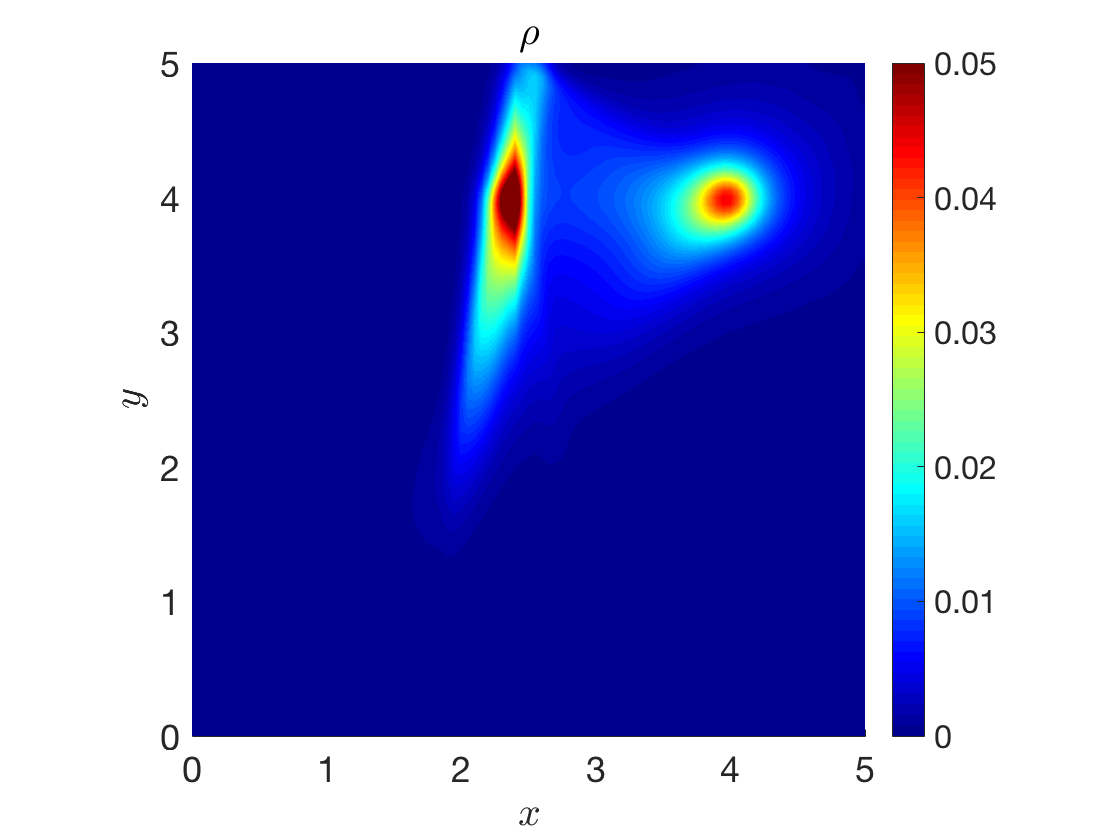}   
        \caption{\scriptsize t=5}
        \label{M23_cellout_t400}
     \end{subfigure}
       \begin{subfigure}{0.24\textwidth}
        \centering
        \includegraphics[width=\textwidth]{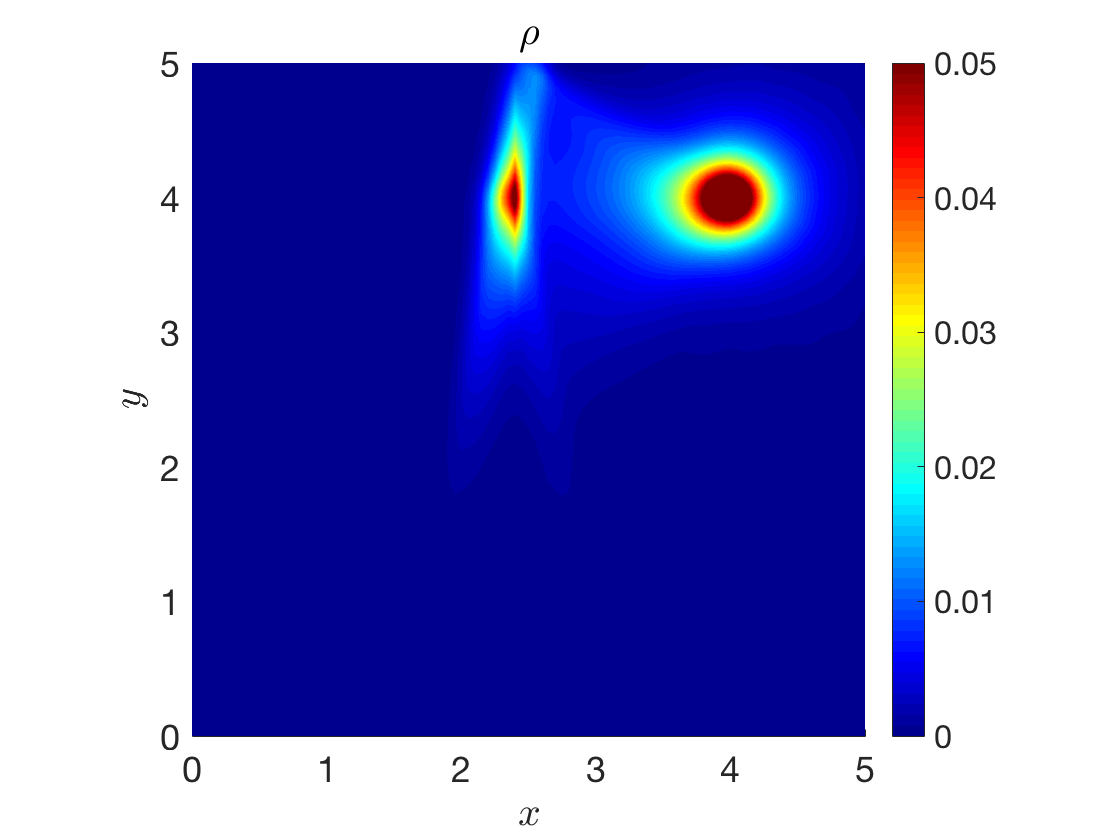}   
        \caption{\scriptsize t=6.25}
        \label{M23_cellout_t500}
     \end{subfigure}
\begin{subfigure}{0.24\textwidth}
        \centering
        \includegraphics[width=\textwidth]{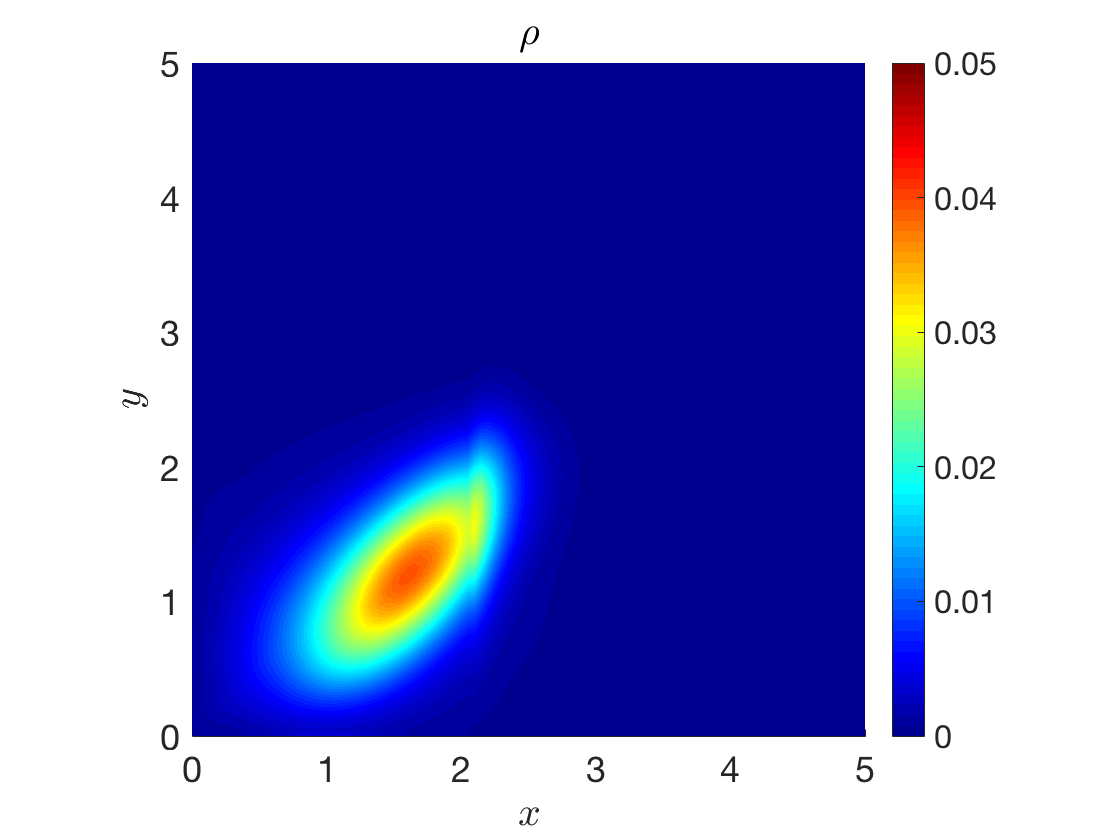}
        \caption{\scriptsize t=1.25}
        \label{M2_cellout_t100}
    \end{subfigure}
   \begin{subfigure}{0.24\textwidth}
        \centering
        \includegraphics[width=\textwidth]{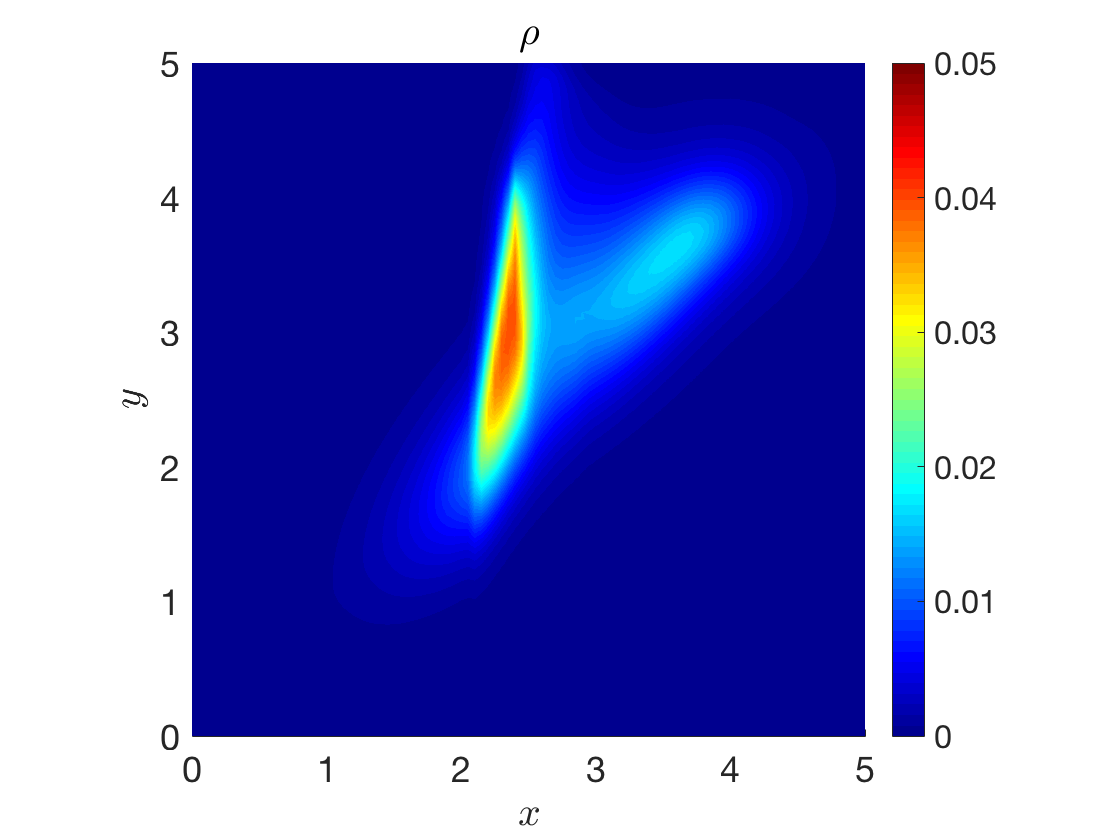}
        \caption{\scriptsize t=3.75}
        \label{M2_cellout_t300}   
     \end{subfigure} 
     \begin{subfigure}{0.24\textwidth}
        \centering
        \includegraphics[width=\textwidth]{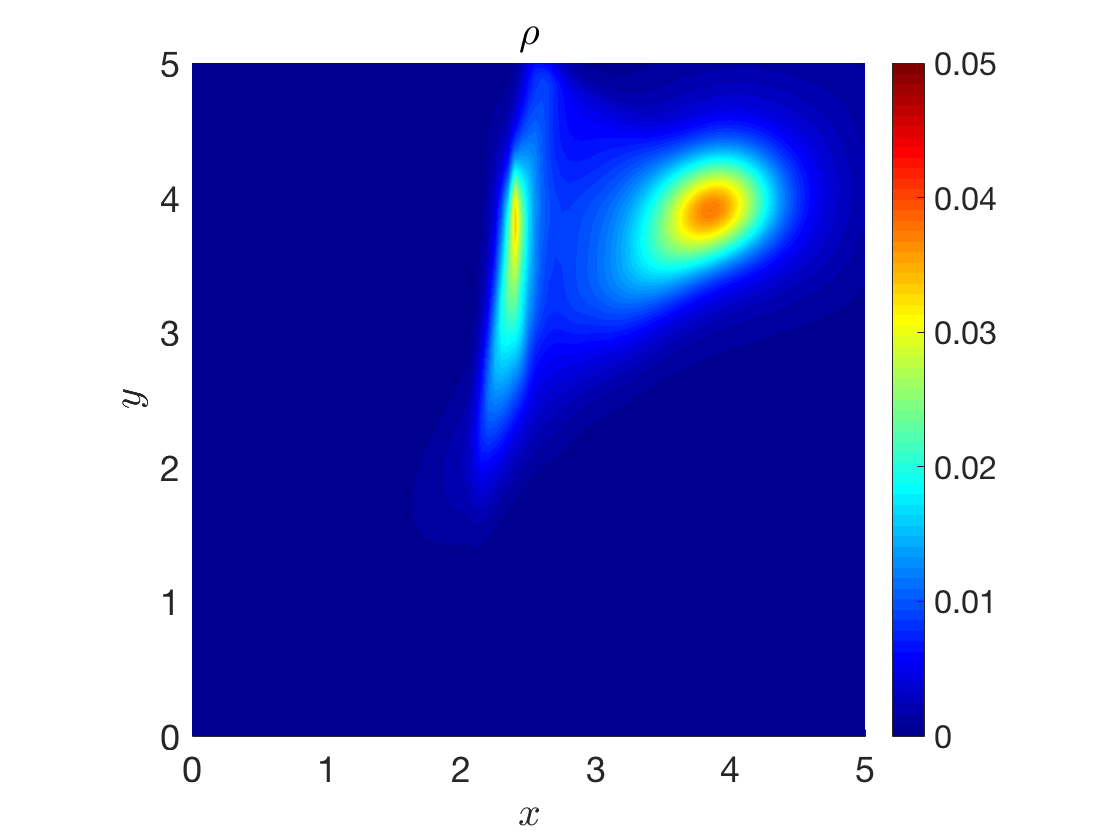}   
        \caption{\scriptsize t=5}
        \label{M2_cellout_t400}
     \end{subfigure}
       \begin{subfigure}{0.24\textwidth}
        \centering
        \includegraphics[width=\textwidth]{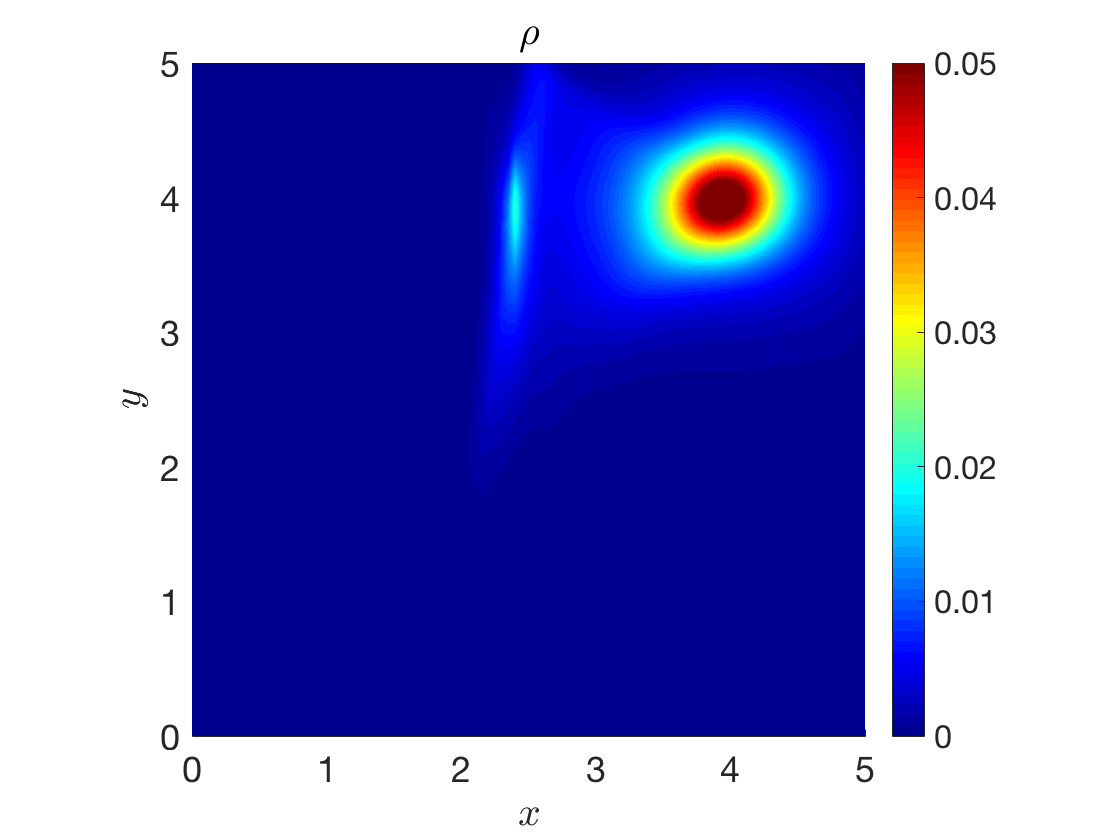}   
        \caption{\scriptsize t=6.25}
        \label{M2_cellout_t500}
     \end{subfigure}
    \begin{subfigure}{0.24\textwidth}
        \centering
        \includegraphics[width=\textwidth]{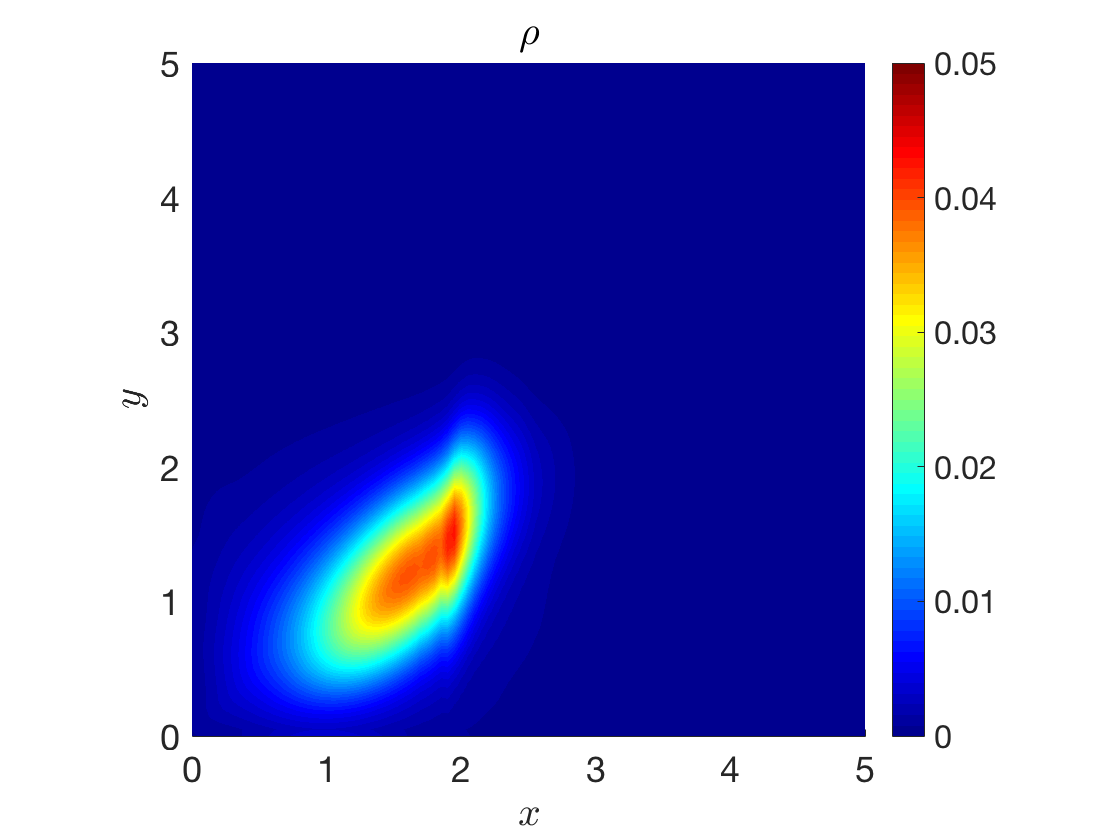}
        \caption{\scriptsize t=1.25}
        \label{M3_cellout_t100}
    \end{subfigure}
   \begin{subfigure}{0.24\textwidth}
        \centering
        \includegraphics[width=\textwidth]{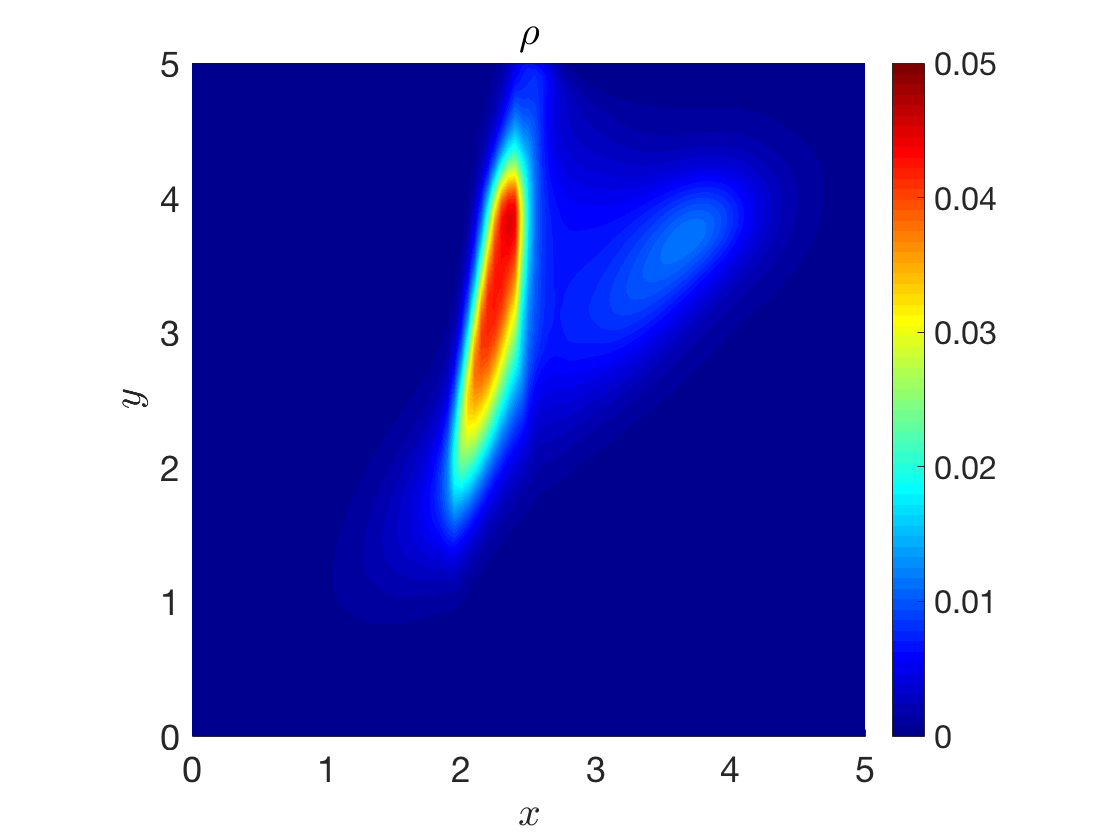}
        \caption{\scriptsize t=3.75}
        \label{M3_cellout_t300}   
     \end{subfigure} 
     \begin{subfigure}{0.24\textwidth}
        \centering
        \includegraphics[width=\textwidth]{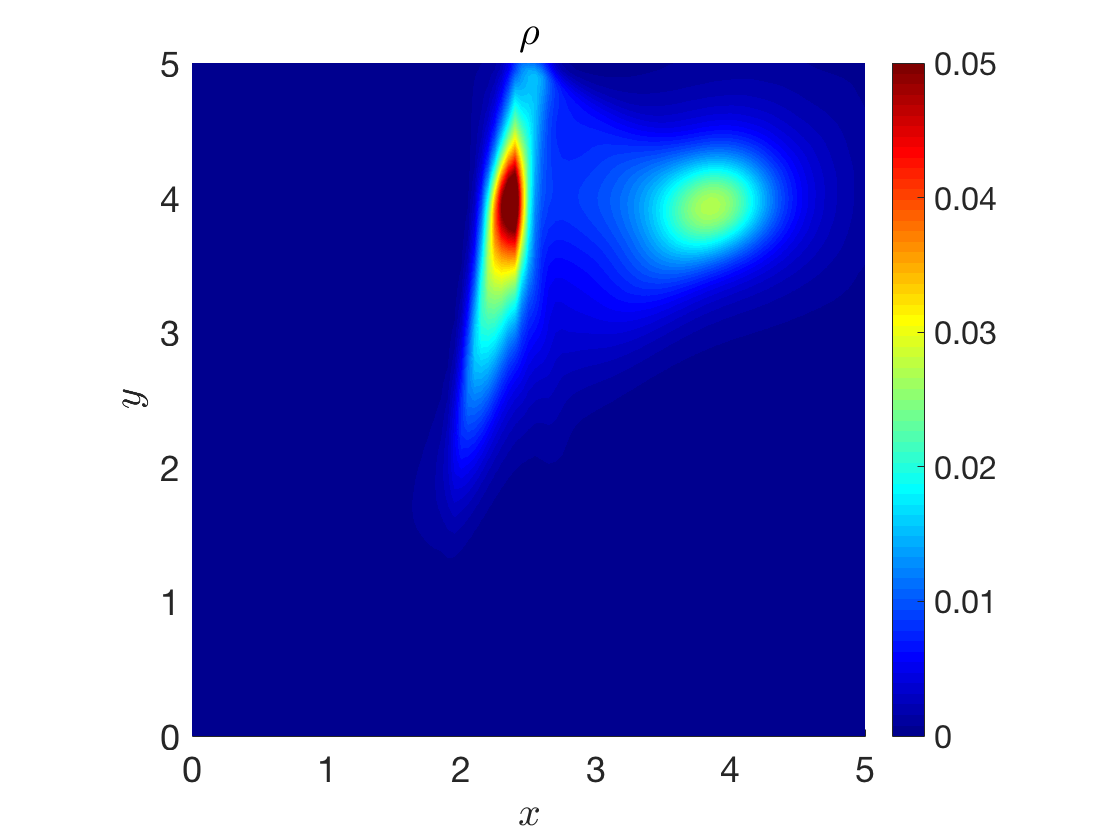}   
        \caption{\scriptsize t=5}
        \label{M3_cellout_t400}
     \end{subfigure}
       \begin{subfigure}{0.24\textwidth}
        \centering
        \includegraphics[width=\textwidth]{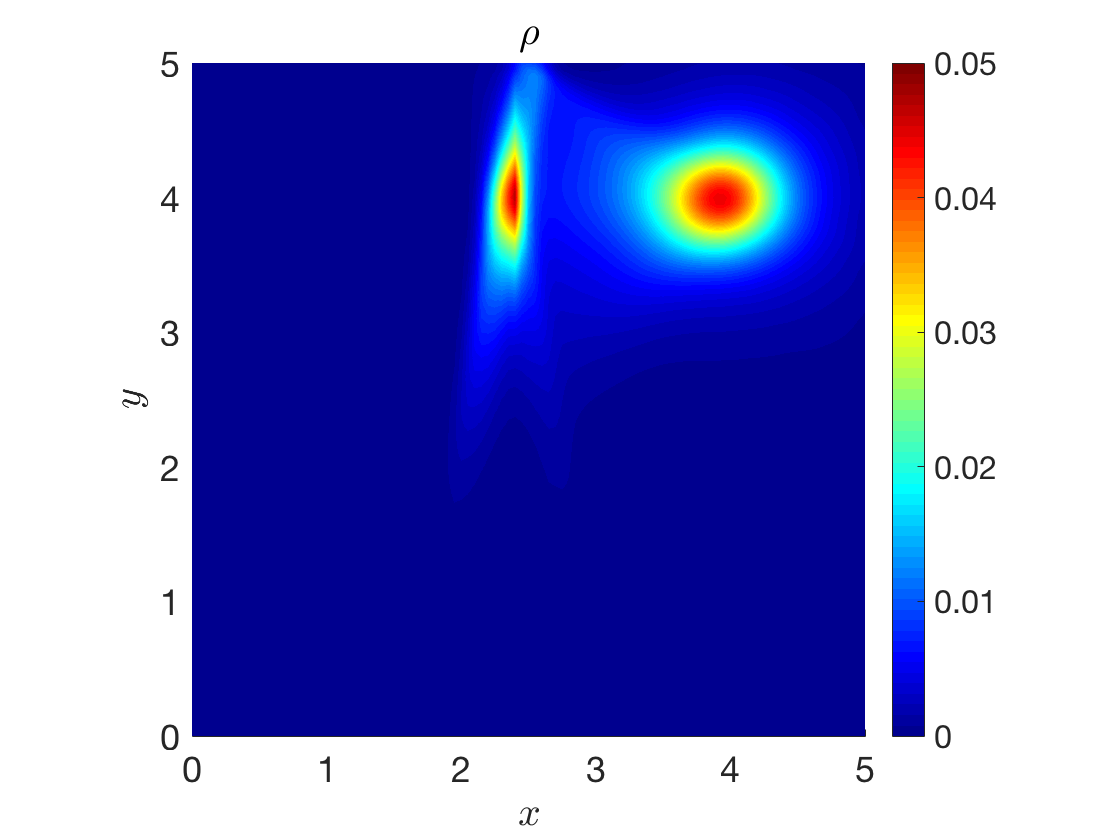}   
        \caption{\scriptsize t=6.25}
        \label{M3_cellout_t500}
     \end{subfigure}
    \caption{ \textbf{Test 2} Time evolution of the initial distribution given in Fig. \ref{InCon_cells} in the four settings 1-4. The sensing radius of the cells is $R=0.5$ and the chemoattractant is \eqref{S.gauss} with $m_{\cS}=10, \sigma_{\cS}^2=0.05$ and $(x_\cS, y_\cS)=(4,4)$. Setting 1 is represented in Figs. (c)-(f): local $q$ and non-local chemoattractant, $\gamma_{\scS}=\delta(\lambda-R)$. Setting 2 is represented in Figs. (g)-(j): non-local $q$ and  $\cS$ with sensing functions $\gamma_q=\gamma_{\scS}=\delta(\lambda-R)$. Setting 3 is represented in Figs. (k)-(n): non-local $q$ and $\cS$, independent sensing with $\gamma_q=\gamma_{\scS}=H(R-\lambda)$. Setting 4 is represented in Figs. (o)-(r): non-local $q$ and $\cS$, dependent sensing with $\gamma=H(R-\lambda)$.}
\label{Comparison_cellout}
\end{figure}

As a second test, we present both the non-local independent sensing model and the non-local dependent sensing model. We shall now consider a non-local sensing of the distribution of fibers. In particular, we assume fibers distributed similarly to the previous test, $\ie$, fibers shall be highly aligned in $\Omega_q$ given, this time, by $x_1=2.1$ and $x_2=2.9$ (see Fig. \ref{InCon_fiber}). Here, for $(x,y) \in \Omega_q$, $k(x,y)=100$, that corresponds to $D_q=0.0025$, and $\theta_q(x,y)=\pi/2$. In the region $\Omega-\Omega_q$ fibers are uniformly distributed. The initial condition of the cell population is \eqref{ci_cells_gaussian} with in $(x_0,y_0)=(1,0.5)$ (see Fig. \ref{InCon_cells}) while the chemoattractant is located as in Test 1, with $m_{\cS}=10$ and $\sigma_{\cS}^2=0.05$. We shall compare the dynamics of the cells in four settings: 
\begin{enumerate}
\item  local fiber distribution and non-local chemoattractant, as in Test 1, $\ie$, \eqref{T_indip} with $\gamma_q=\delta(\lambda-0)$ and  $\gamma_{\scS}=\delta(\lambda-R)$;
\item non-local sensing with a Dirac Delta for both $q$ and $\cS$; this corresponds to both \eqref{T_indip} and \eqref{T_dep} with $\gamma_q=\gamma_{\scS}=\gamma=\delta(\lambda-R)$; 
\item non-local independent sensing with Heaviside sensing functions for both $\cS$ and $q$, $\ie$, \eqref{T_indip} with $\gamma_q=\gamma_{\scS}=H(R-\lambda)$; 
\item non-local dependent sensing for $q$ and $\cS$, dealing with \eqref{T_dep} and $\gamma=H(R-\lambda)$.
\end{enumerate}
Results of these simulations are shown in Fig. \ref{Comparison_cellout}. We can observe that, in the 1-4 settings, cells start from $(1,0.5)$, they are attracted by the chemoattractant and, on their way towards $\cS$, they cross the aligned fibers region $\Omega_q$ and climb up this region in the direction $\pi/2$. Eventually, in all the cases, cells reach the chemoattractant, but the dynamics, as well as the transient time, is influenced by the different sensing kernels, even though the differences are not extremely appreciable, and by the local or non-local sensing strategy. Although settings 3 and 4 in Fig. \ref{Comparison_cellout}, that are related to the case of independent and dependent cues, respectively, do not show very strong differences, in case 3 (see Figs. \ref{M2_cellout_t100}-\ref{M2_cellout_t500}) the tendency of going in both the direction $\pi/2$, determined by the fibers, and $\pi/4$, determined by the chemoattractant, appears more marked because of the independent sensing. In contrast, this behavior results the least evident in the case in which cells deal with a local sensing of the fibers (setting 1), resulting also in a general slow down of the dynamics.

\subsection{Test 3. non-local independent sensing model: comparison of the cases $i)-iv)$}
We now present a comparison of the macroscopic behaviors of the cells, depending on the relation between $R$, $l_\cS$ and $l_q$, $\ie$, we compare the cases $i),ii),iii)$ and $iv)$. In particular, we shall do this for the non-local independent sensing model 
with $\gamma_q=\gamma_{\scS}=H(R-\lambda)$, as this is the case in which the transport model is different from the dependent sensing model.
Additionally, the independence of the two sensings allows to visualize more efficiently the two distinct directional effects (contact guidance and chemotaxis). 

\begin{figure}[!h] 
\centering
	\begin{subfigure}{0.45\textwidth}
        \centering
        \includegraphics[width=0.8\textwidth]{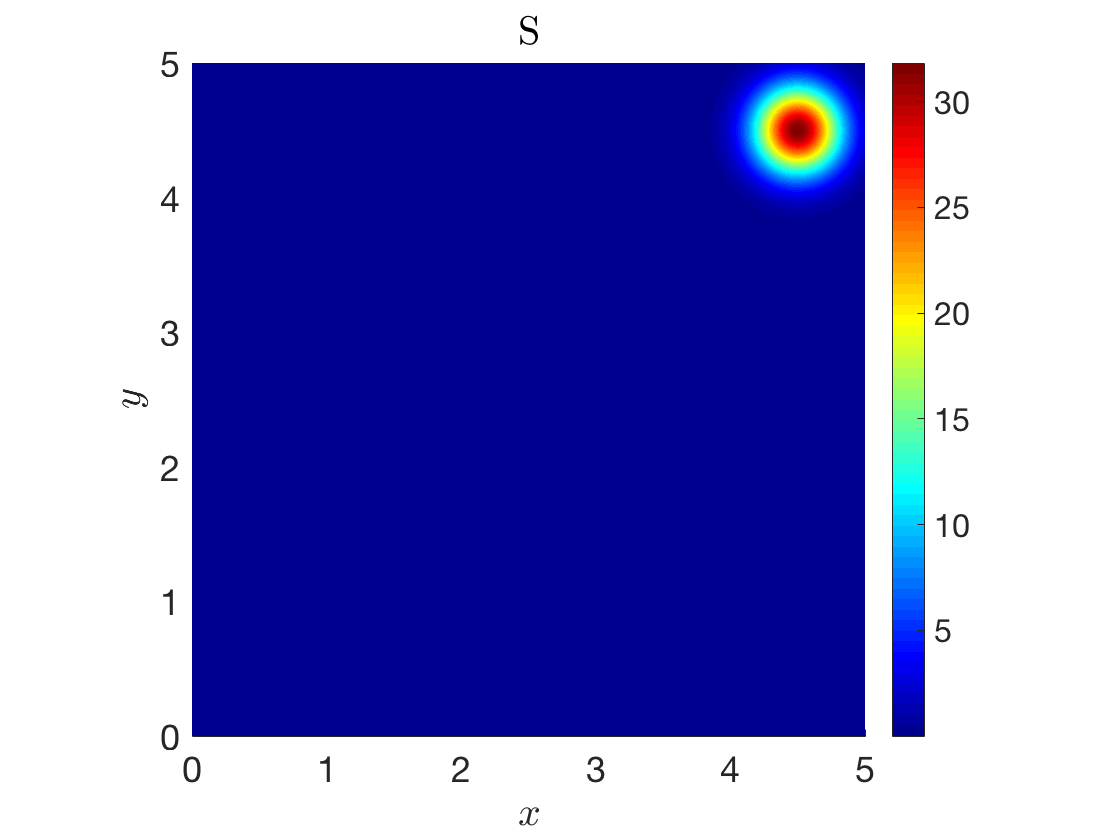}
        \caption{\scriptsize Chemoattractant $\cS$ with $\sigma_{\cS}^2=0.05$.}
        \label{S005}
    \end{subfigure}
   	\begin{subfigure}{0.45\textwidth}
        \centering
        \includegraphics[width=0.8\textwidth]{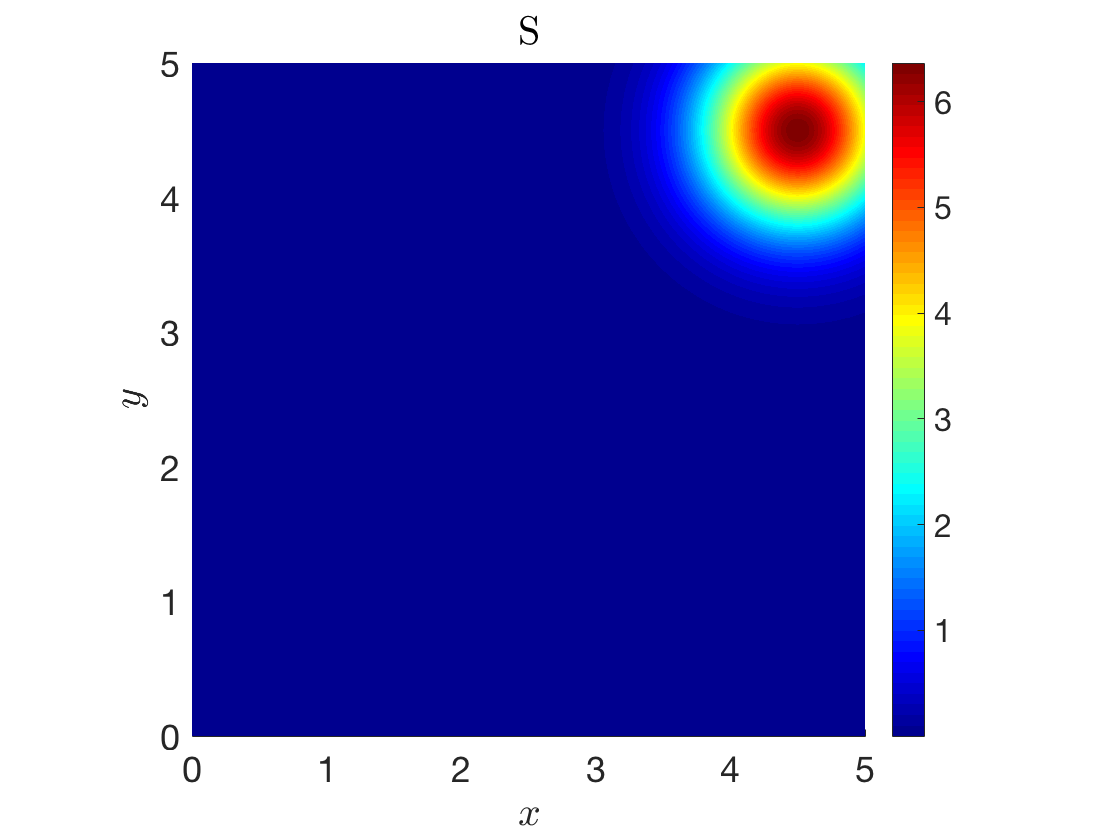}
        \caption{\scriptsize Chemoattractant $\cS$ with $\sigma_{\cS}^2=0.25$.}
        \label{S025}   
     \end{subfigure} 
     \begin{subfigure}{0.45\textwidth}
        \centering
        \includegraphics[width=0.8\textwidth]{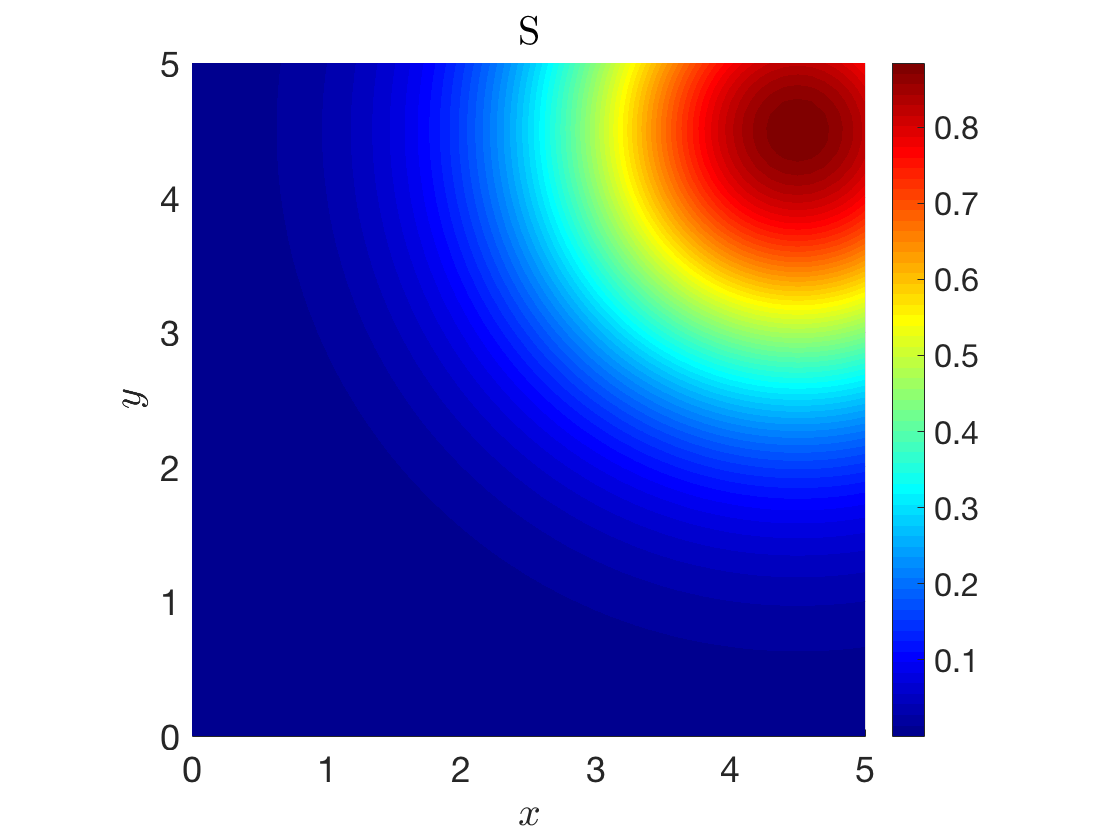}
        \caption{\scriptsize Chemoattractant $\cS$ with $\sigma_{\cS}^2=1.8$.}
        \label{S18}
     \end{subfigure}
     \begin{subfigure}{0.45\textwidth}
        \centering
        \includegraphics[width=0.7\textwidth]{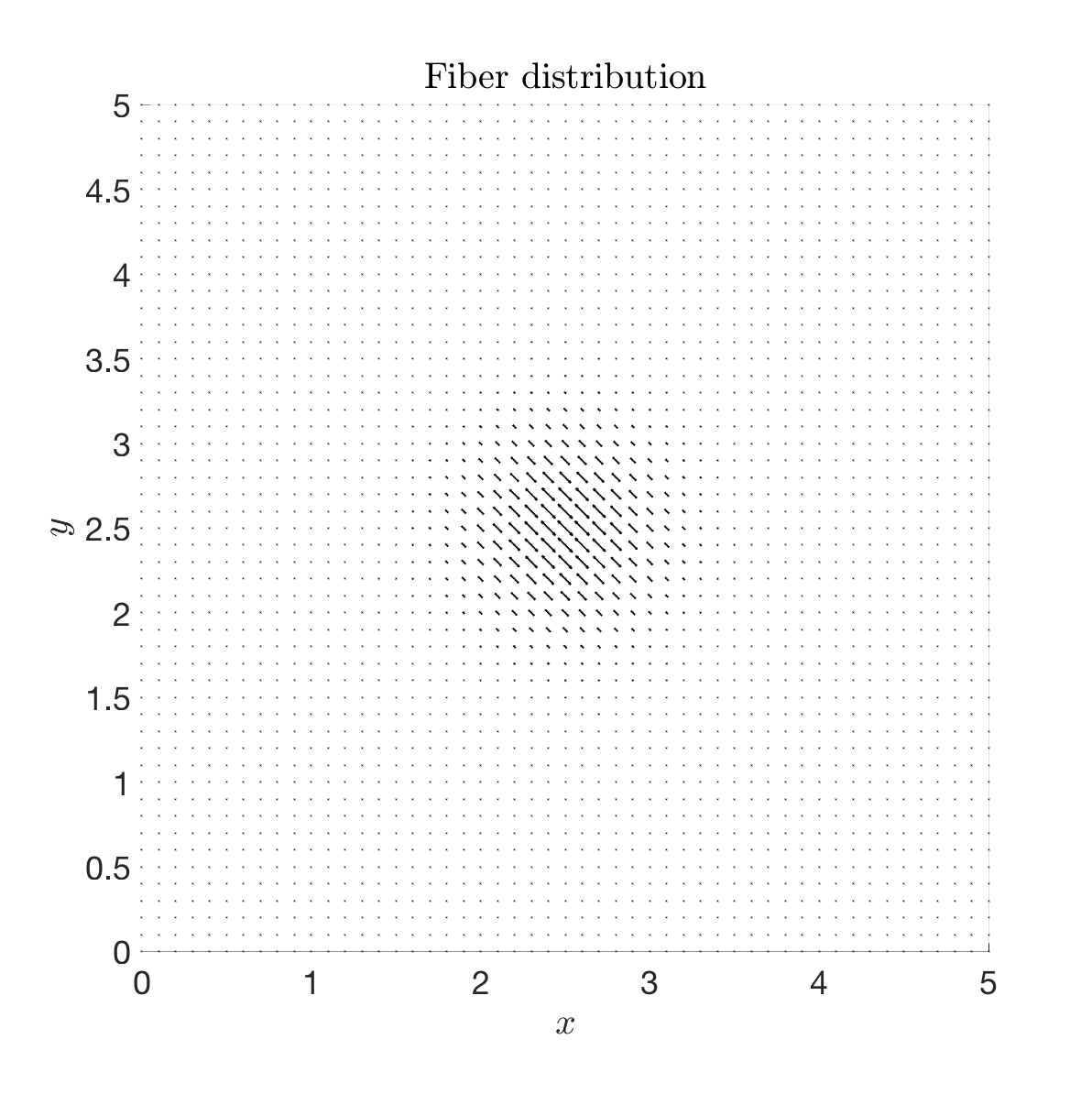}
        \caption{\scriptsize Fibers distribution}
        \label{k_gauss}   
     \end{subfigure} 
     \caption{ \textbf{Test 3} Three different chemoattractants used for comparing models $i)-iv)$. The chemoattractant profile is given by \eqref{S.gauss} with $m_{\cS}=10$ and (a) $\sigma_{\cS}^2=0.05$, corresponding to $l_{\cS}=0.002$, (b) $\sigma_{\cS}^2=0.25$, corresponding to $l_{\cS}=0.055$, and (c) $\sigma_{\cS}^2=1.8$, corresponding to $l_{\cS}=0.25$. The fibers distribution in sketched in (d). }
\end{figure}

\begin{figure}[!htbp]
\begin{subfigure}{0.32\textwidth}
        \centering
        \includegraphics[width=\textwidth]{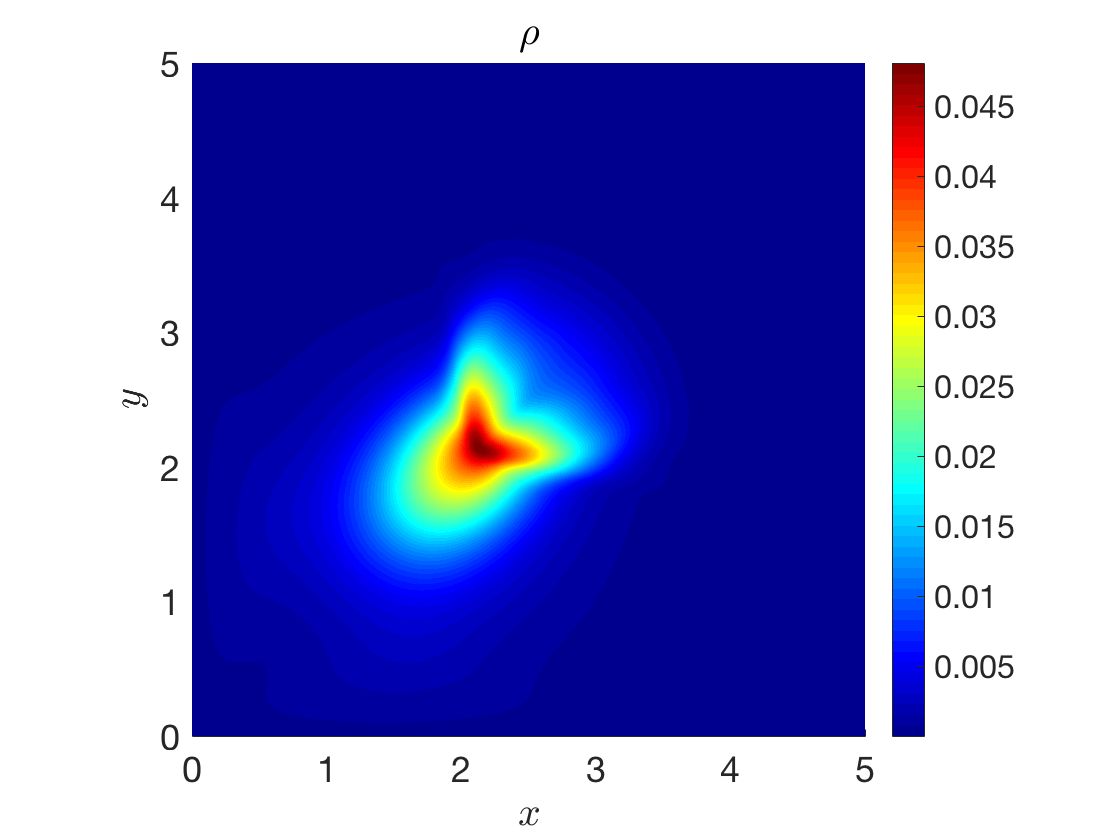}
        \caption{\scriptsize t=1.25}
        \label{i_t50}
    \end{subfigure}
     \begin{subfigure}{0.32\textwidth}
        \centering
        \includegraphics[width=1\textwidth]{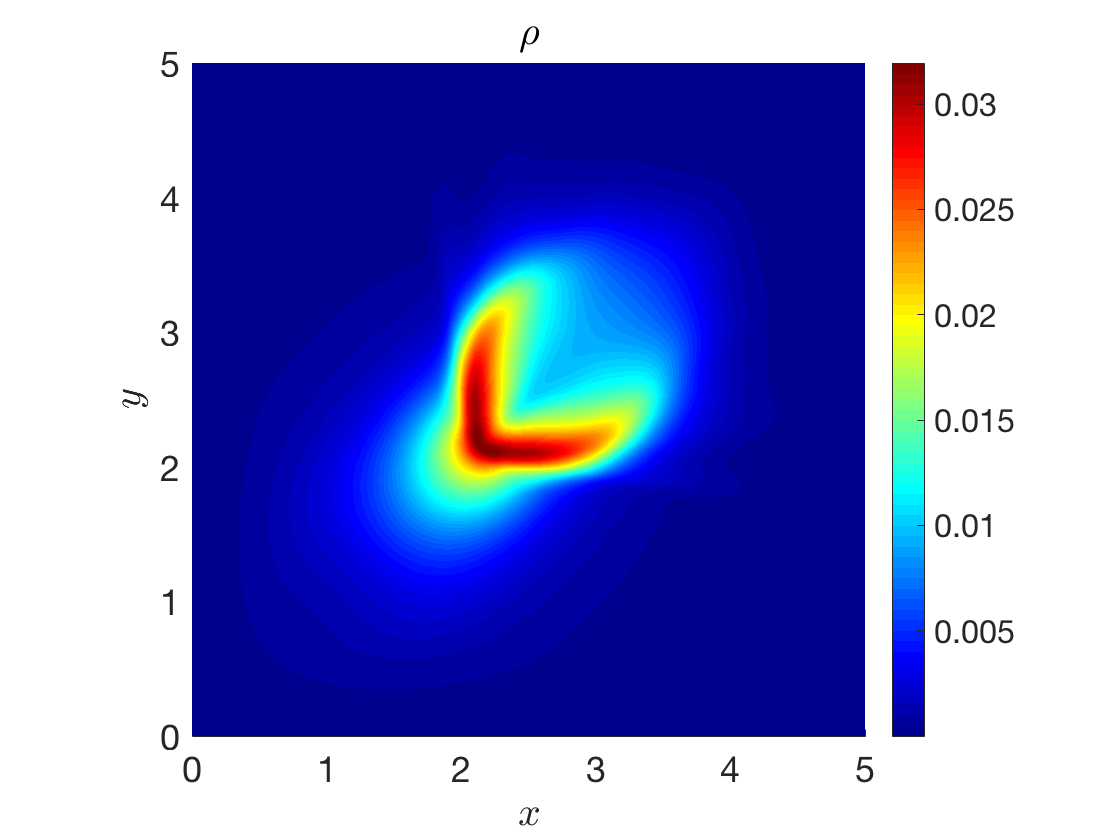}
        \caption{\scriptsize t=1.875}
        \label{i_t75}
    \end{subfigure}
    \begin{subfigure}{0.32\textwidth}
        \centering
        \includegraphics[width=1\textwidth]{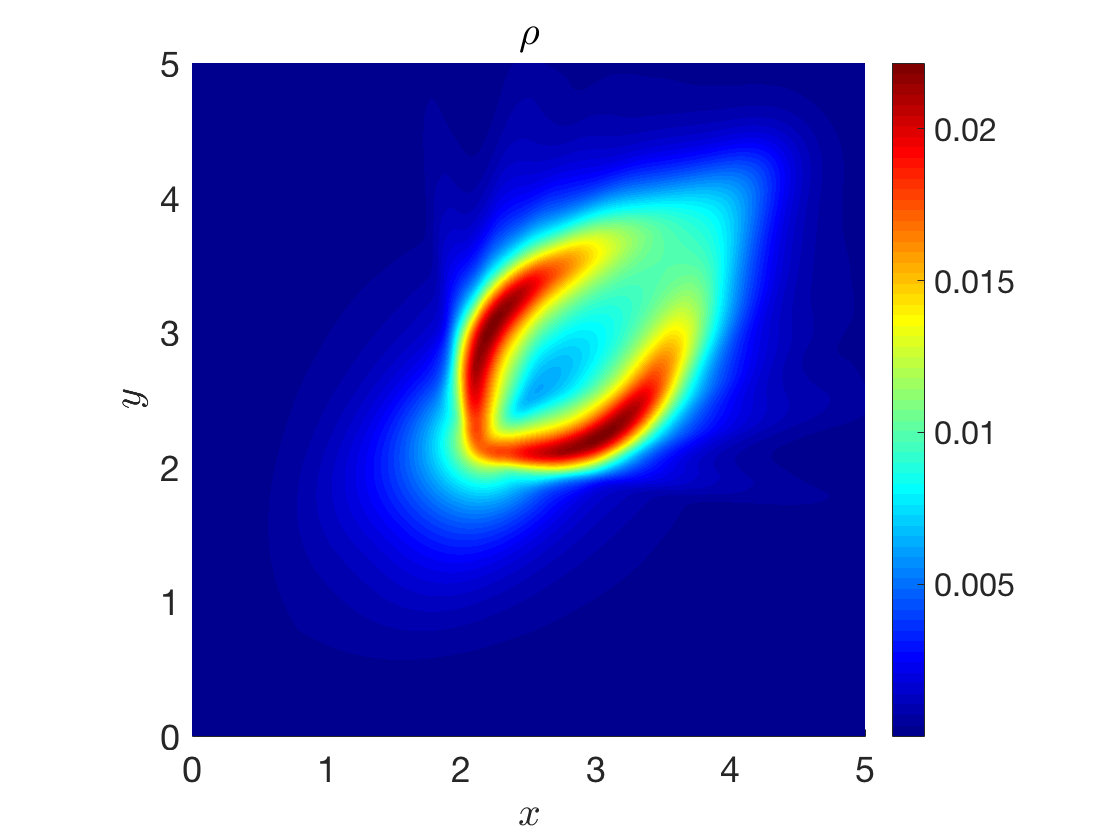}
        \caption{\scriptsize t=2.5}
        \label{i_t100}
    \end{subfigure}
    
   \begin{subfigure}{0.32\textwidth}
        \centering
        \includegraphics[width=\textwidth]{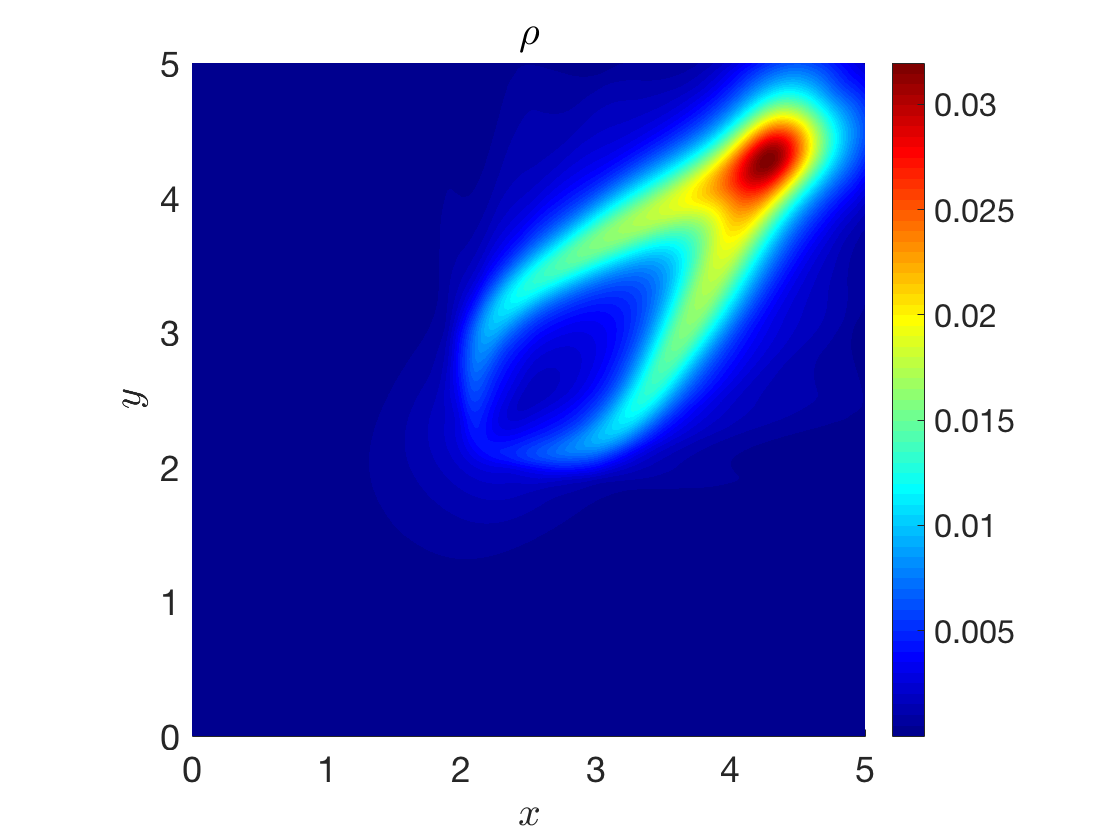}
        \caption{\scriptsize t=3.75}
        \label{i_t150}   
     \end{subfigure} 
     \begin{subfigure}{0.32\textwidth}
        \centering
        \includegraphics[width=\textwidth]{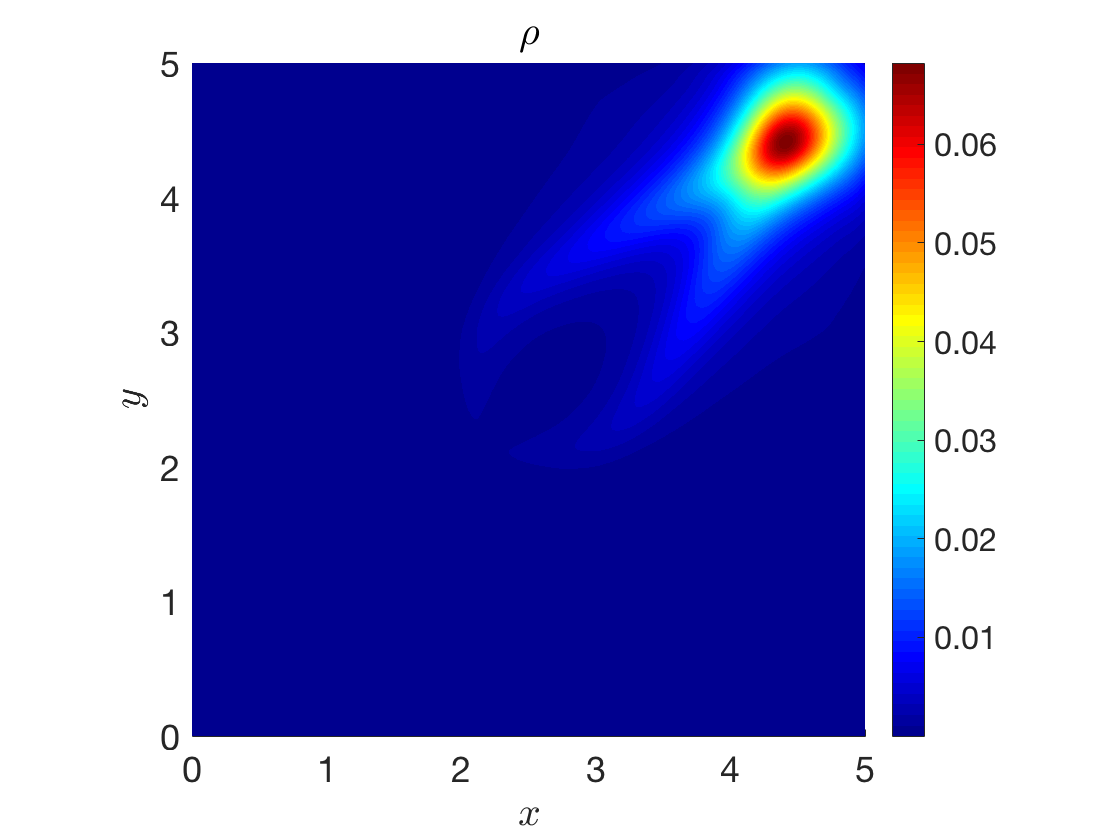}   
        \caption{\scriptsize t=5}
        \label{i_t200}
     \end{subfigure}
       \begin{subfigure}{0.32\textwidth}
        \centering
        \includegraphics[width=\textwidth]{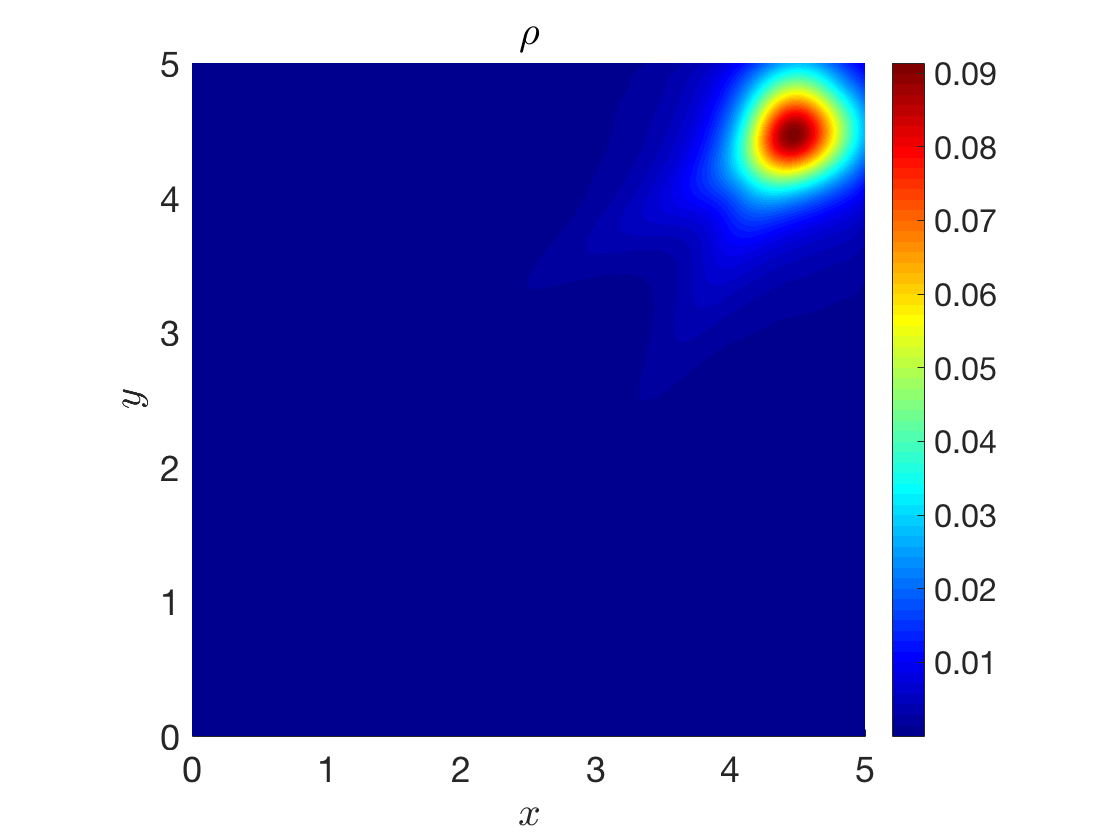}   
        \caption{\scriptsize t=6.25}
        \label{i_t250}
     \end{subfigure}
    \caption{ \textbf{Test 3} Case $i)$ with non-local $q$ and $\cS$, sensed with an independent sensing through the kernels $\gamma_q=\gamma_{\scS}=H(R-\lambda)$. $\cS$ is given in Fig. \ref{S005} with $m_{\cS}=10$ and $\sigma_{\cS}^2=0.05$, so that $l_\cS=0.002$. The fibers distribution $q$ has a space dependent parameter $k$ given by \eqref{k_gaussian} with $m_{k}=100$, so that $l_q\approx 0.0031$. The sensing radius of the cells is $R=0.7$.}
       \label{Sim_i_varS005}
\end{figure}

\begin{figure}[!h] 
\begin{subfigure}{0.32\textwidth}
        \centering
        \includegraphics[width=\textwidth]{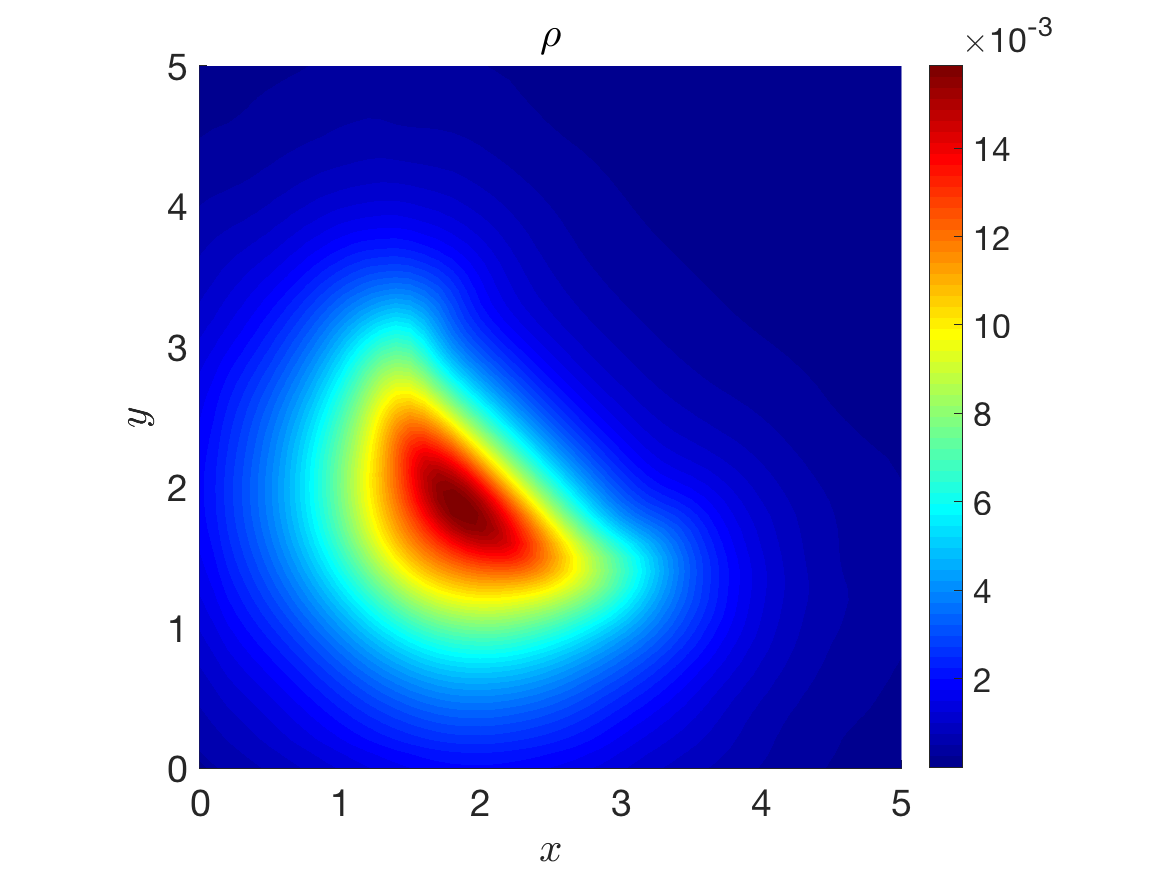}
        \caption{\scriptsize t=2.5}
        \label{i_t100_S18}
    \end{subfigure}
\begin{subfigure}{0.32\textwidth}
        \centering
        \includegraphics[width=\textwidth]{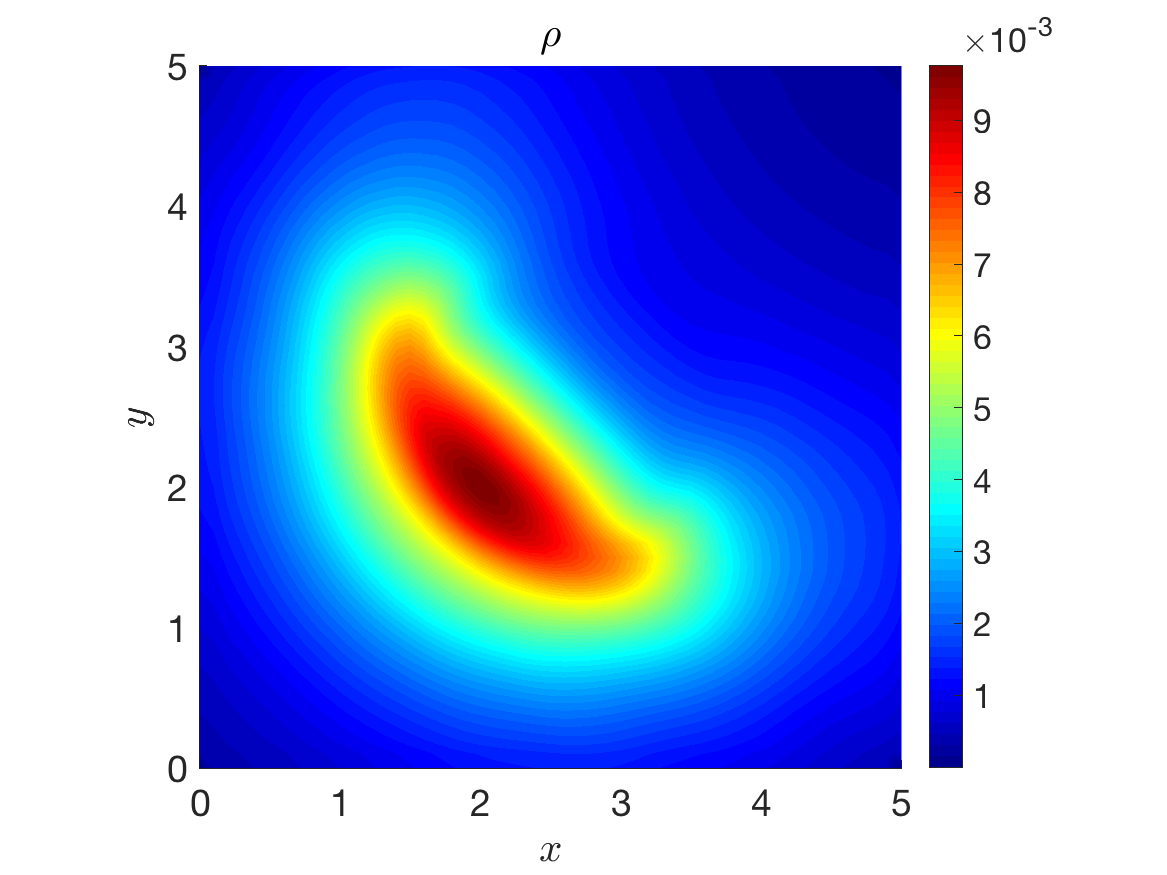}
        \caption{\scriptsize t=5}
        \label{i_t200_S18}
    \end{subfigure}
    \begin{subfigure}{0.32\textwidth}
        \centering
        \includegraphics[width=\textwidth]{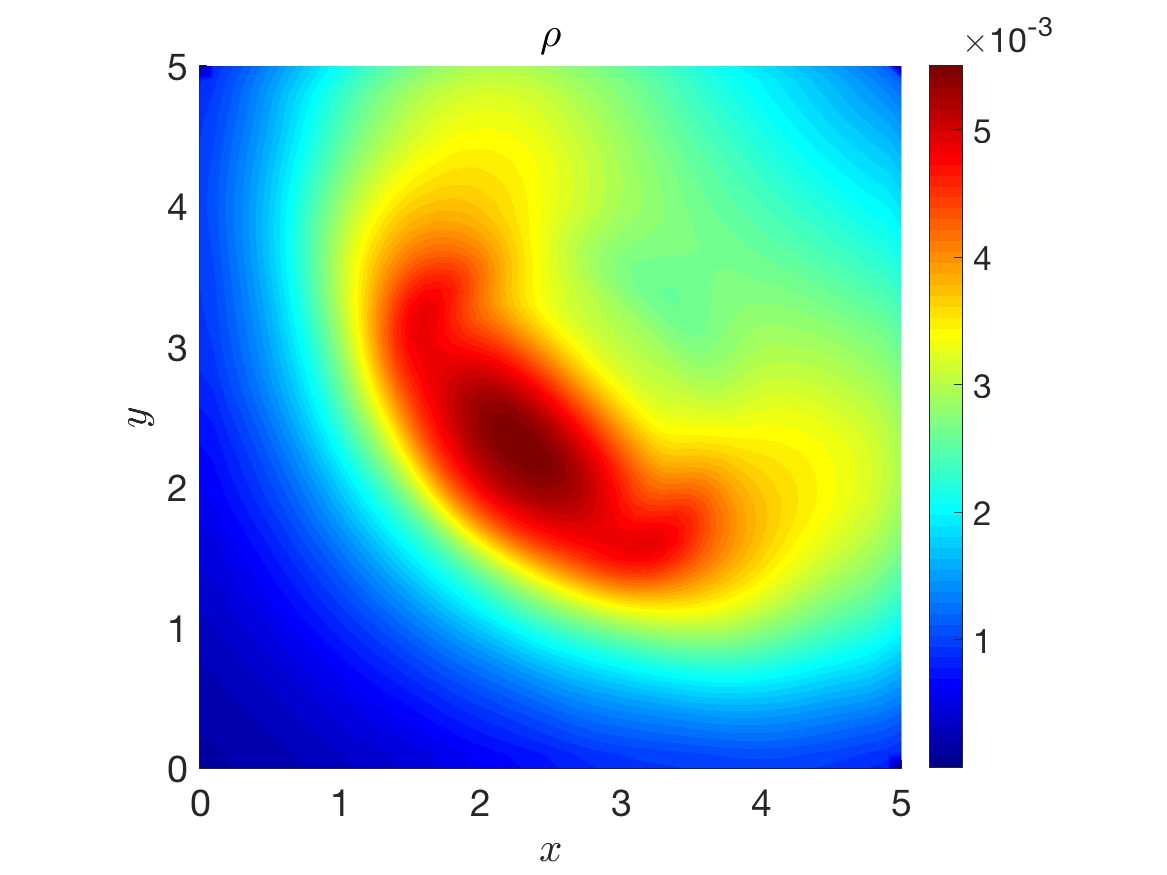}
        \caption{\scriptsize t=10}
        \label{i_t400_S18}
    \end{subfigure}
    
   \begin{subfigure}{0.32\textwidth}
        \centering
        \includegraphics[width=\textwidth]{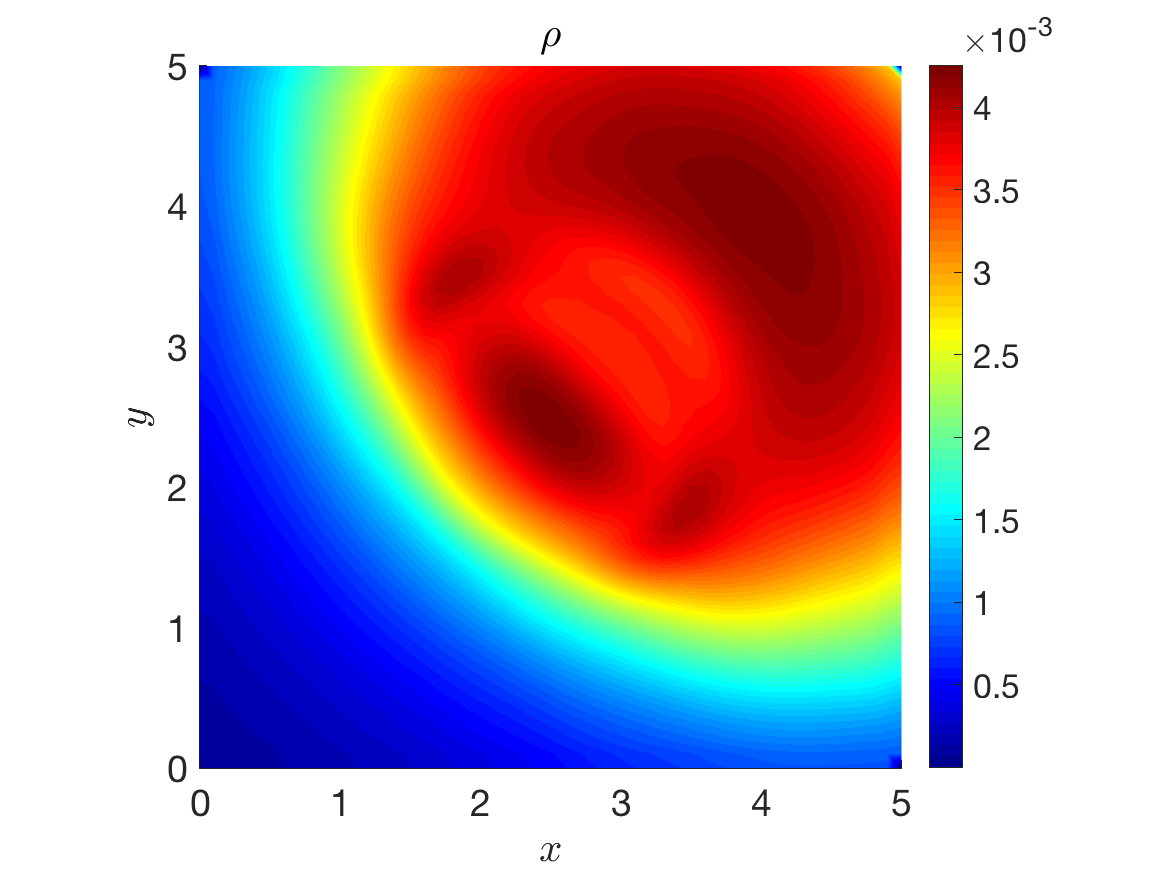}
        \caption{\scriptsize t=15}
        \label{i_t600_S18}   
     \end{subfigure} 
     \begin{subfigure}{0.32\textwidth}
        \centering
        \includegraphics[width=\textwidth]{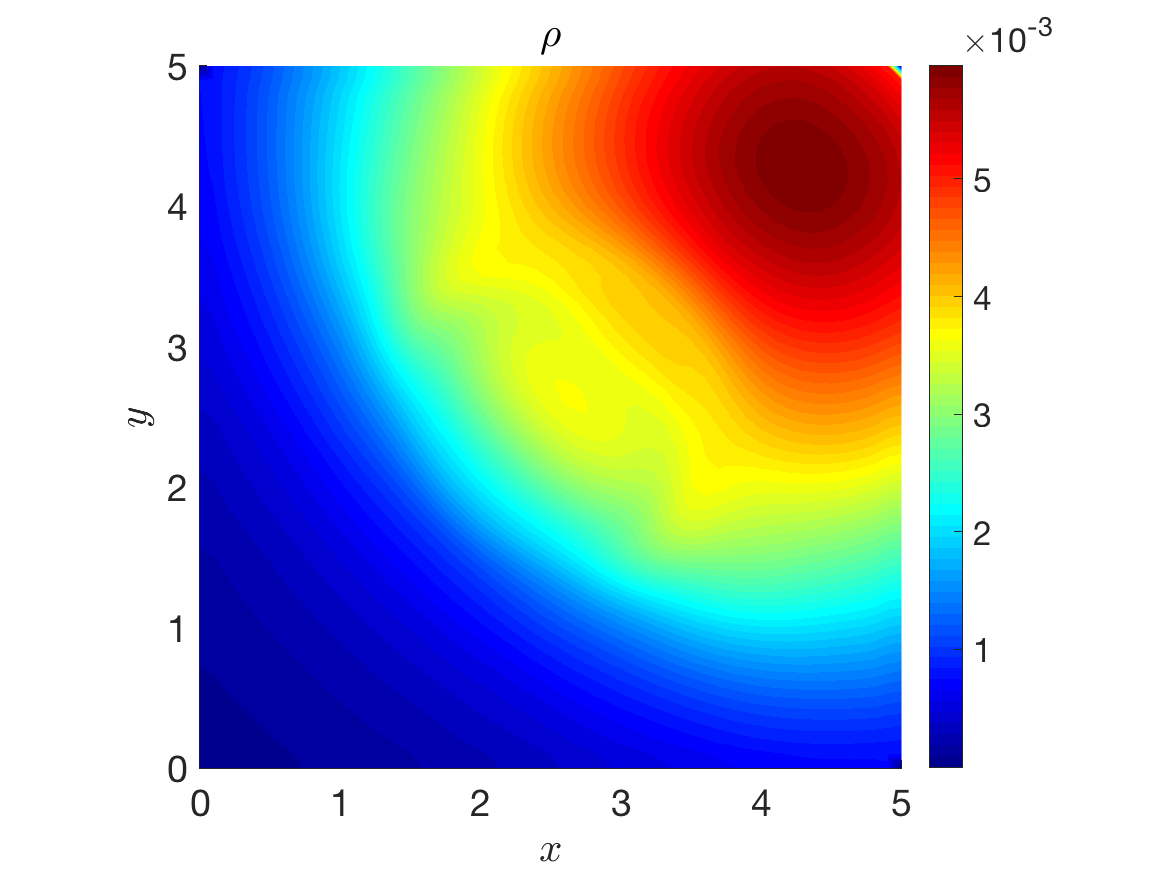}   
        \caption{\scriptsize t=22.5}
        \label{i_t900_S18}
     \end{subfigure}
       \begin{subfigure}{0.32\textwidth}
        \centering
        \includegraphics[width=\textwidth]{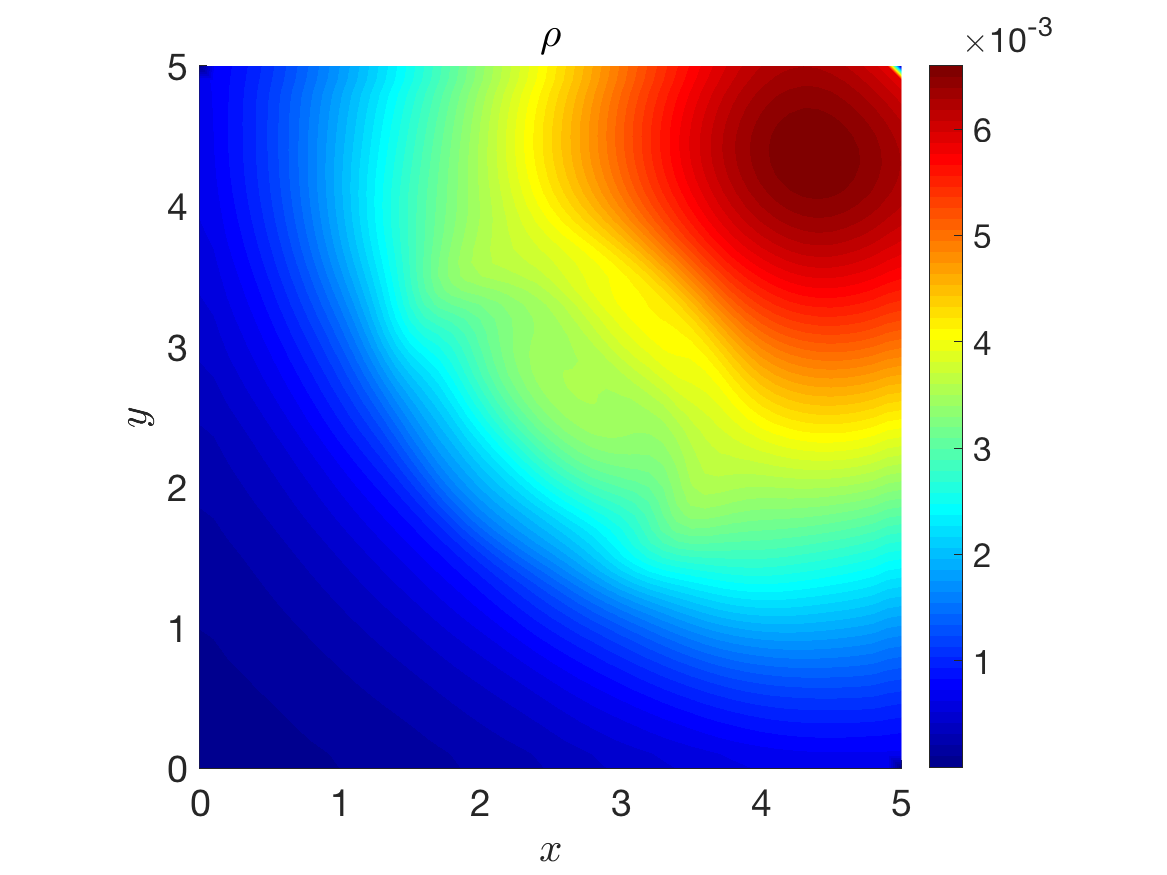}   
        \caption{\scriptsize t=27.5}
        \label{i_t1100_S18}
     \end{subfigure}
    \caption{ \textbf{Test 3} Case $i)$ with non-local $q$ and $\cS$, independent and sensing with $\gamma_q=\gamma_{\scS}=H(R-\lambda)$. $\cS$ is given in Fig. \ref{S18}, that corresponds to $l_\cS=0.25$, while for the fiber distribution $m_k=100$, so that $l_q\approx 0.0031$. The sensing radius of the cells is $R=0.7$. }
       \label{Sim_i_varS18}
\end{figure}

\begin{figure}[!h] 
\begin{subfigure}{0.32\textwidth}
        \centering
        \includegraphics[width=\textwidth]{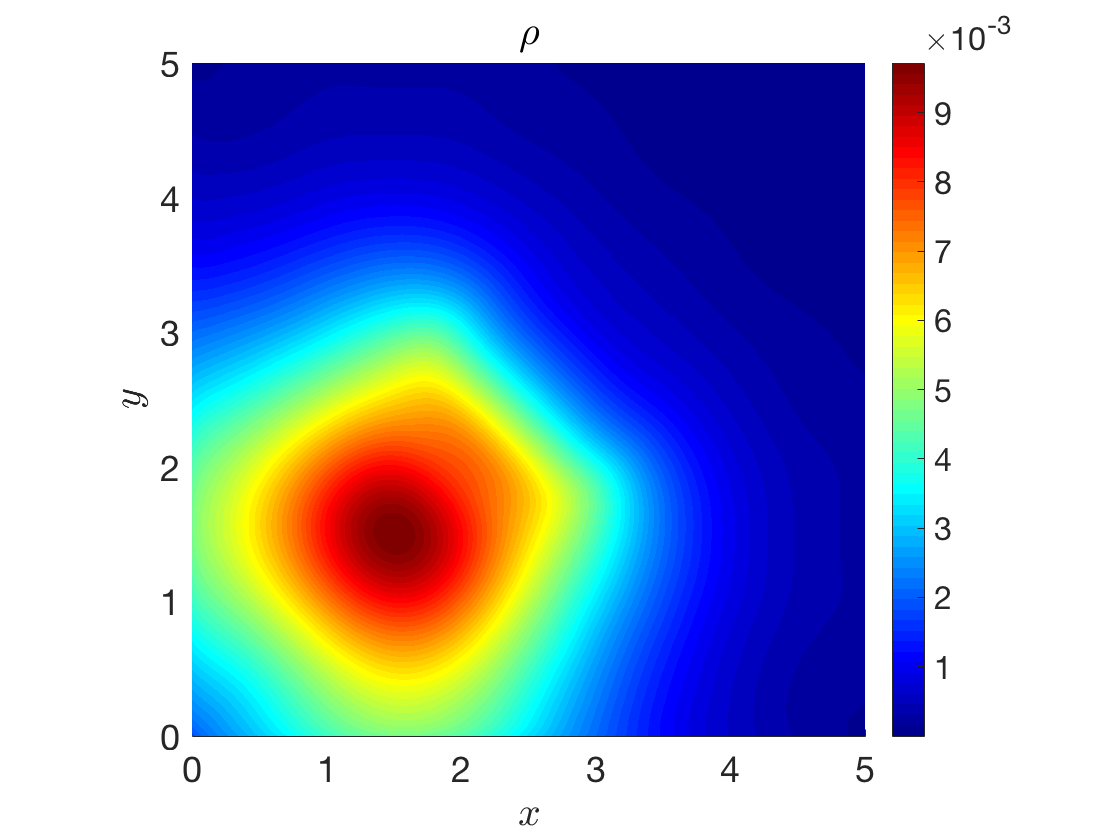}
        \caption{\scriptsize t=2.5}
        \label{ii_t1000}
    \end{subfigure}
    \begin{subfigure}{0.32\textwidth}
        \centering
        \includegraphics[width=\textwidth]{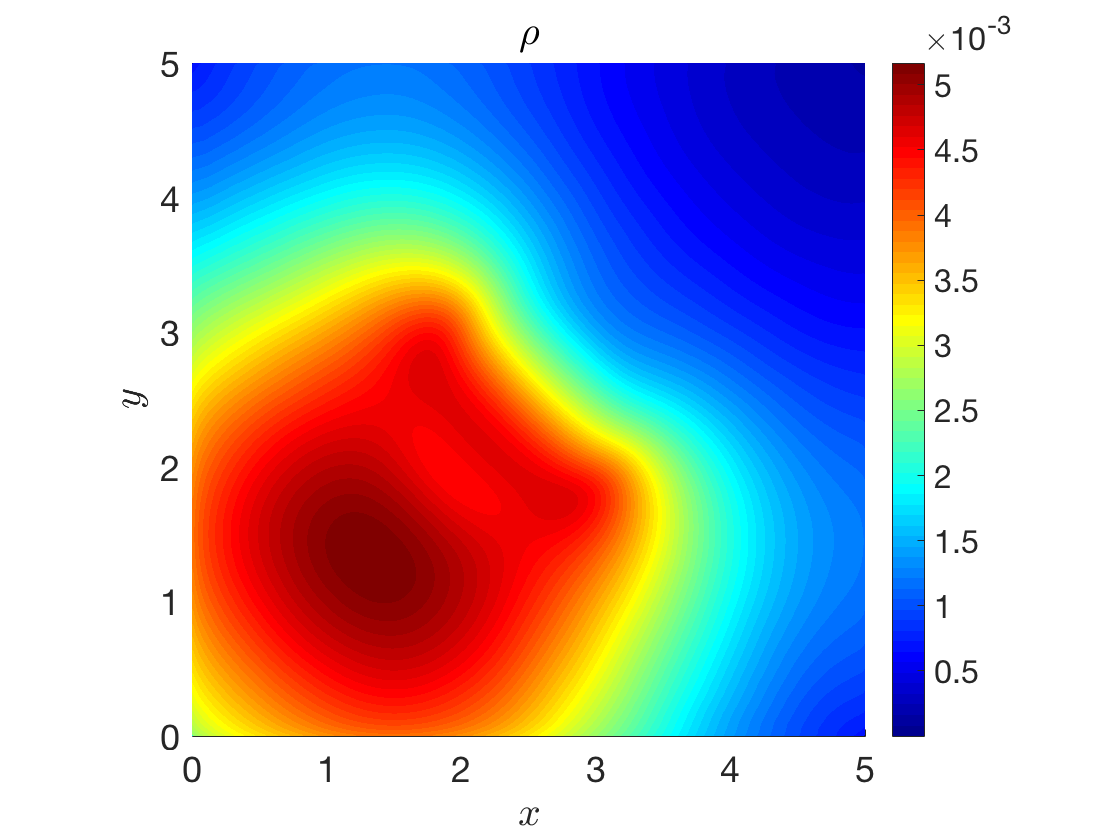}
        \caption{\scriptsize t=5}
        \label{ii_t2000}
    \end{subfigure}
   \begin{subfigure}{0.32\textwidth}
        \centering
        \includegraphics[width=\textwidth]{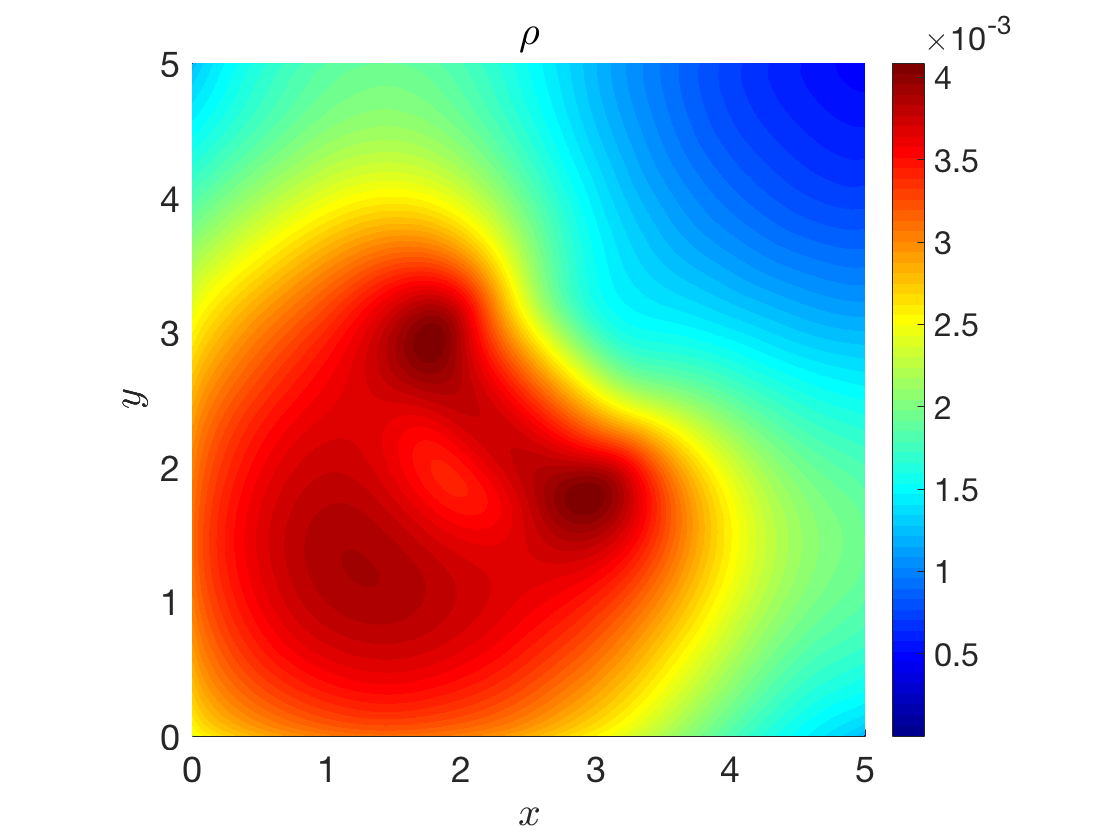}
        \caption{\scriptsize t=7.5}
        \label{ii_t3000}   
     \end{subfigure}
      
     \begin{subfigure}{0.32\textwidth}
        \centering
        \includegraphics[width=\textwidth]{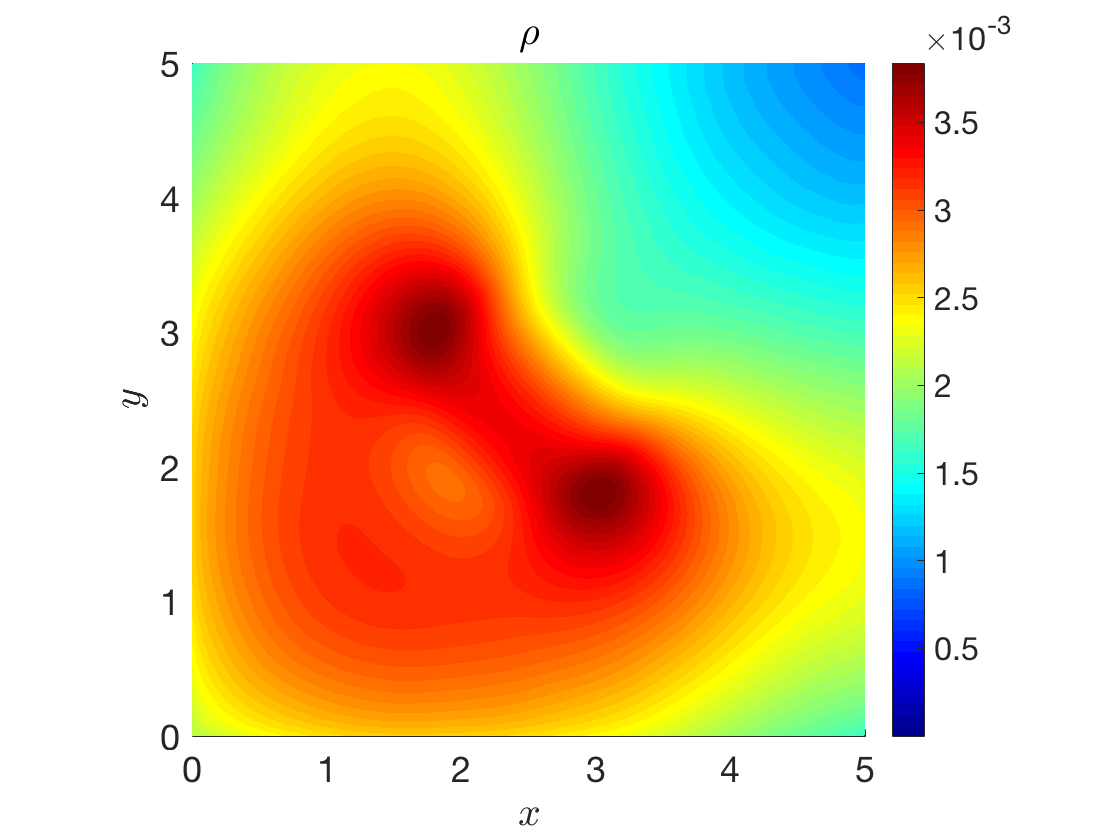}   
        \caption{\scriptsize t=10}
        \label{ii_t4000}
     \end{subfigure}
       \begin{subfigure}{0.32\textwidth}
        \centering
        \includegraphics[width=\textwidth]{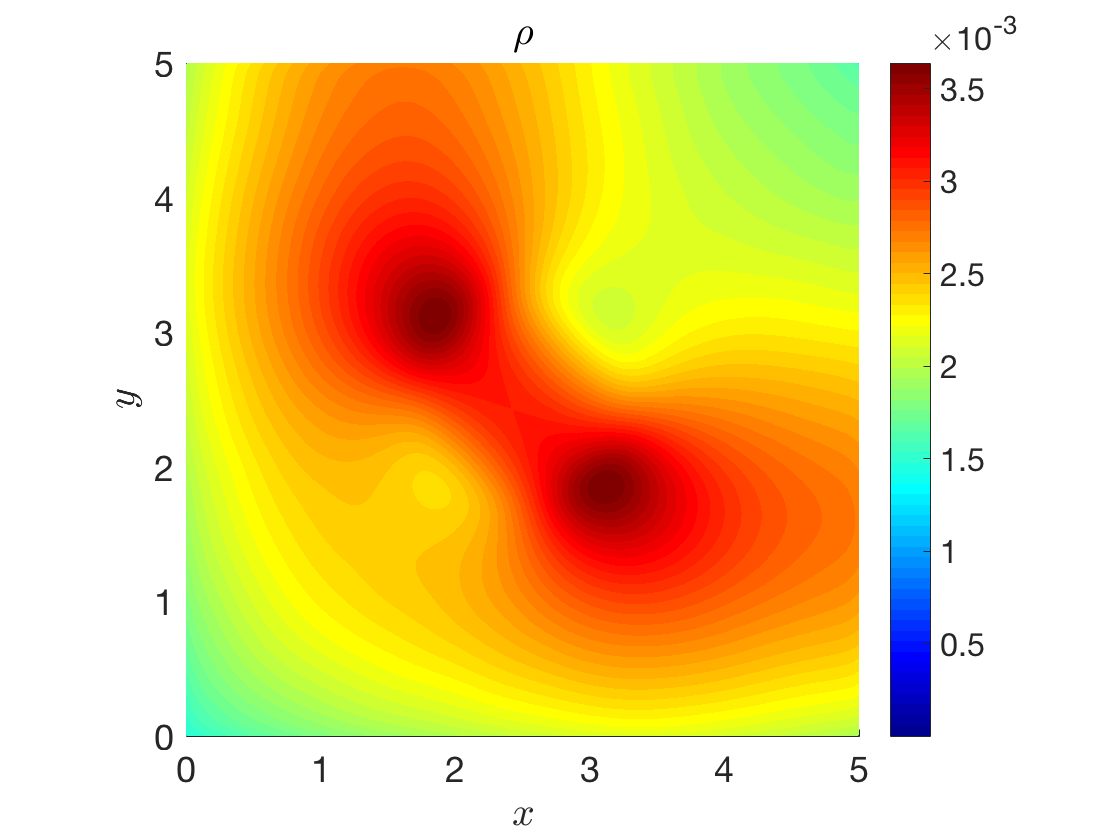}   
        \caption{\scriptsize t=15}
        \label{ii_t6000}
     \end{subfigure}
     \begin{subfigure}{0.32\textwidth}
        \centering
        \includegraphics[width=\textwidth]{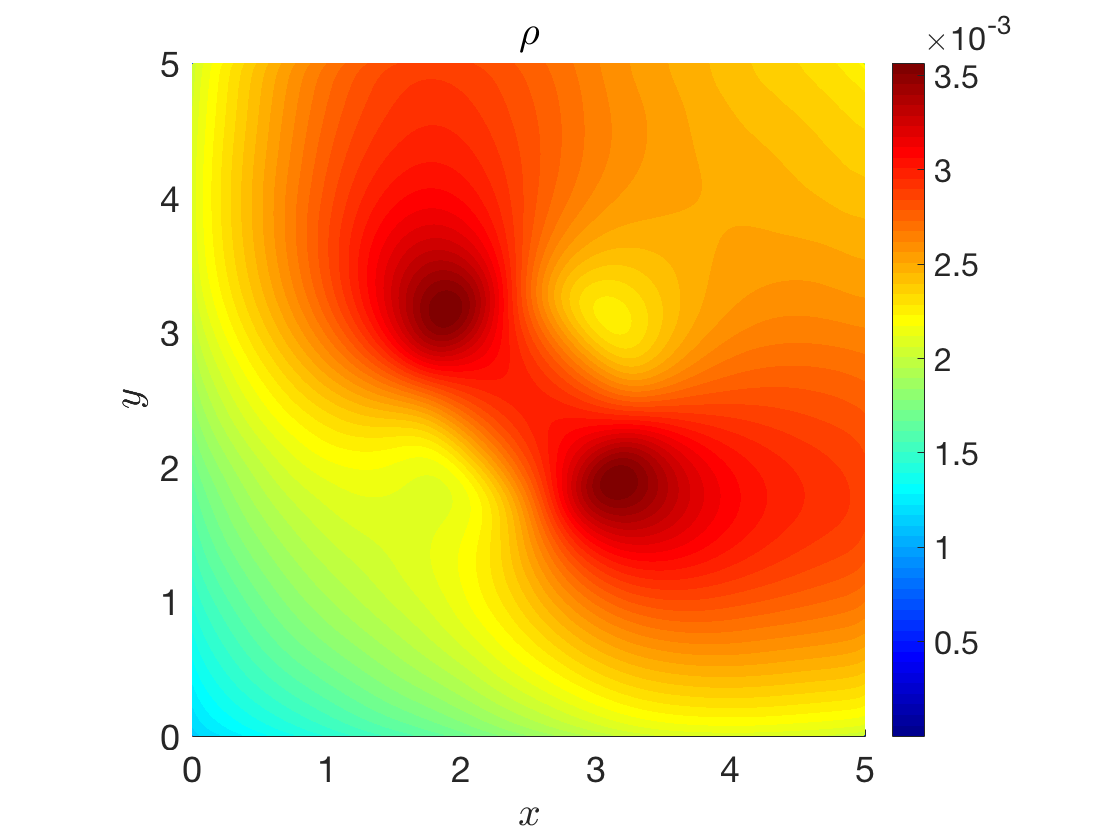}   
        \caption{\scriptsize t=20}
        \label{ii_t8000}
     \end{subfigure}
    \caption{ \textbf{Test 3} Case $ii)$ with non-local $q$ and $\cS$, independent and sensing with $\gamma_q=\gamma_{\scS}=H(R-\lambda)$. $\cS$ is given in Fig. \ref{S025}, that corresponds to $l_\cS=0.055$, while $m_k=10$, so that $l_q \approx0.031$. The sensing radius of the cells is $R=0.02$.}
       \label{Sim_ii}
\end{figure}

\begin{figure}[!h] 
\begin{subfigure}{0.32\textwidth}
        \centering
        \includegraphics[width=\textwidth]{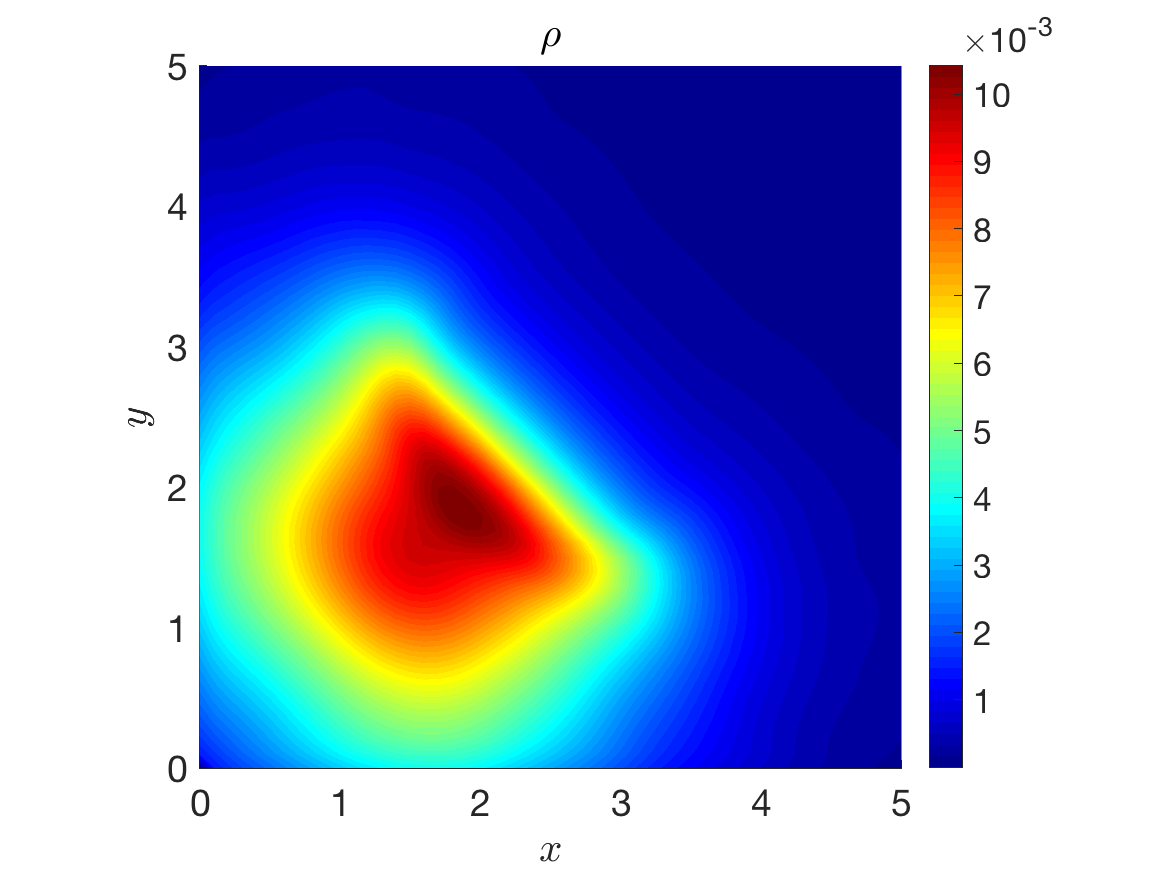}
        \caption{\scriptsize t=2.5}
        \label{iii_t100}
    \end{subfigure}
\begin{subfigure}{0.32\textwidth}
        \centering
        \includegraphics[width=\textwidth]{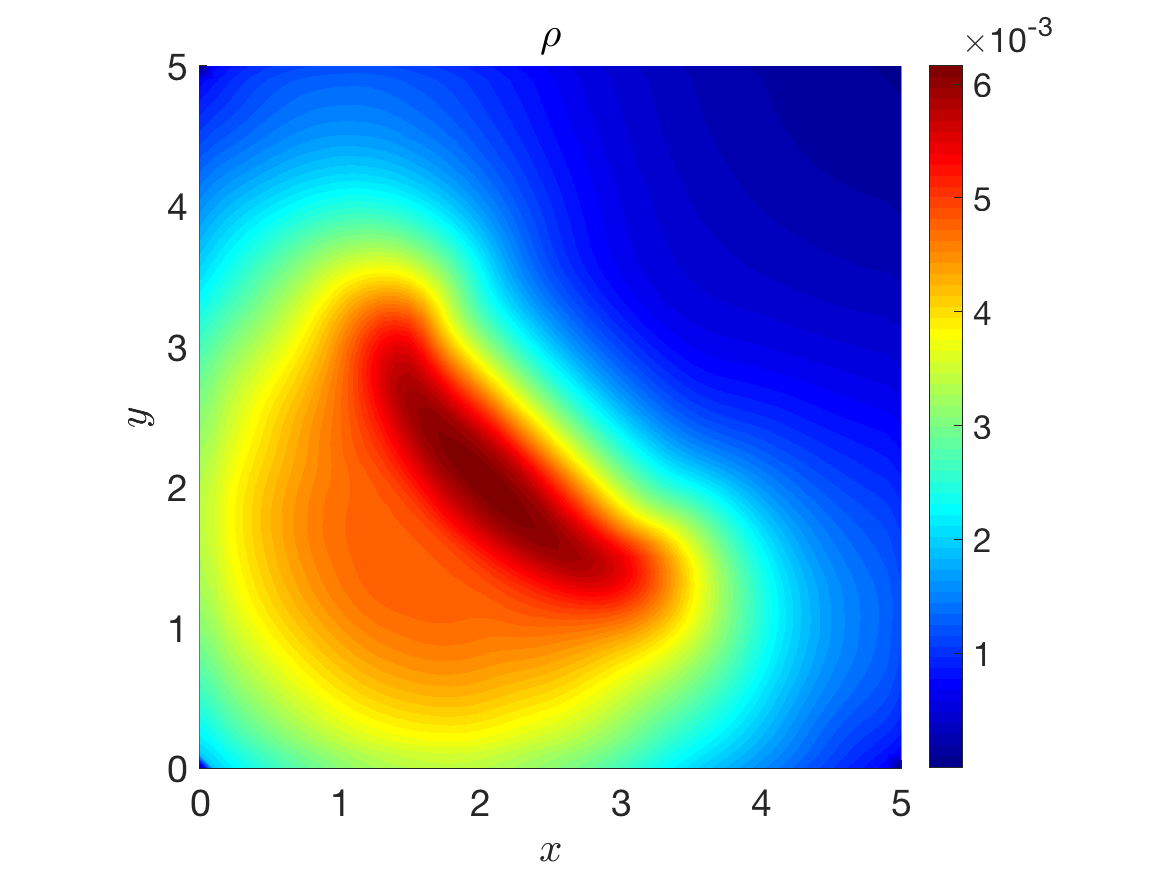}
        \caption{\scriptsize t=5}
        \label{iii_t200}
    \end{subfigure}
    \begin{subfigure}{0.32\textwidth}
        \centering
        \includegraphics[width=\textwidth]{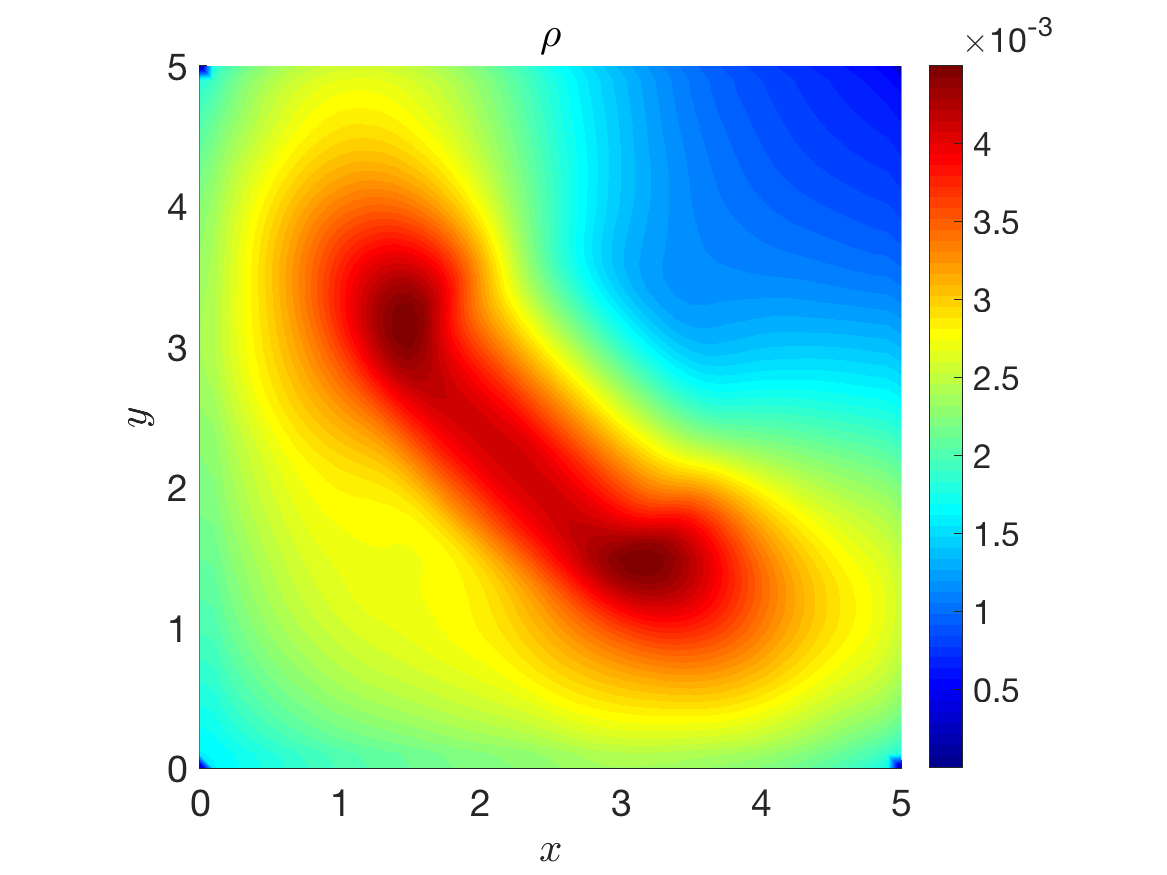}
        \caption{\scriptsize t=10}
        \label{iii_t400}
    \end{subfigure}
    
   \begin{subfigure}{0.32\textwidth}
        \centering
        \includegraphics[width=\textwidth]{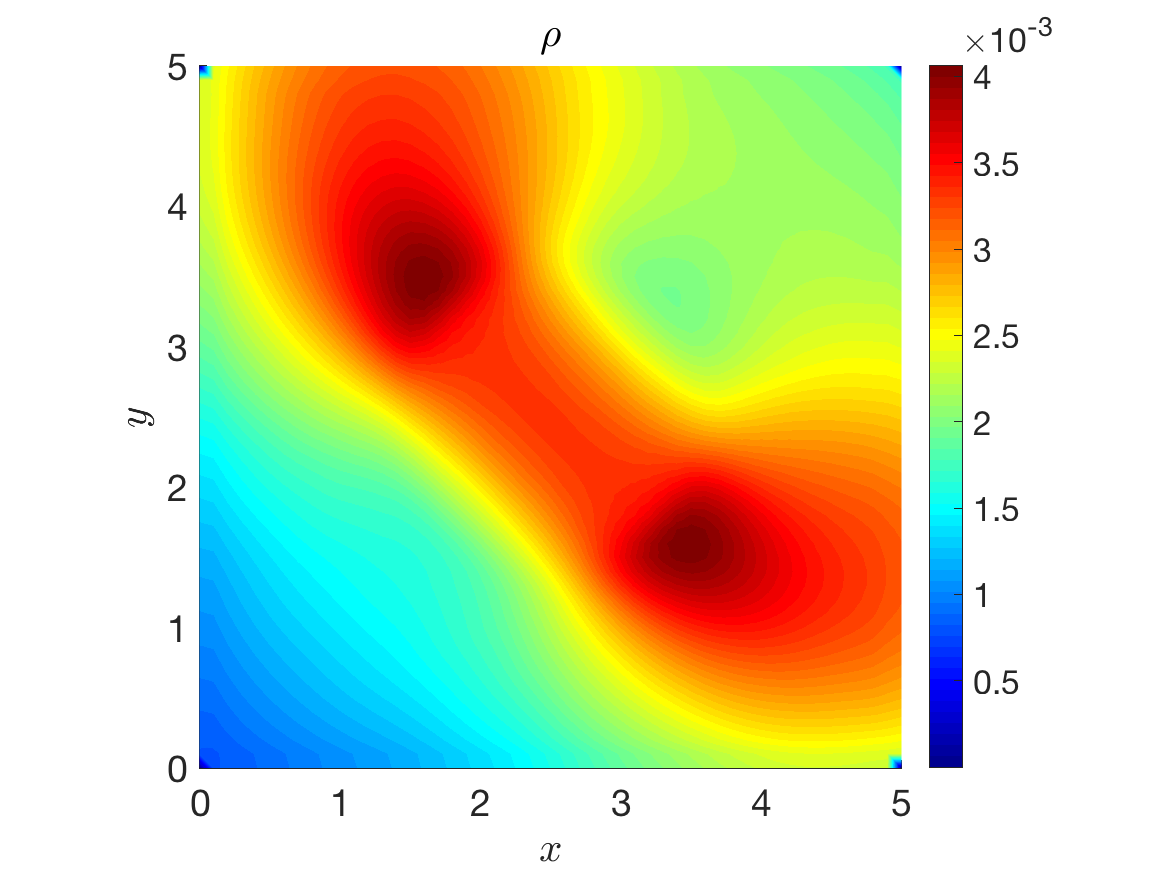}
        \caption{\scriptsize t=20}
        \label{iii_t800}   
     \end{subfigure} 
     \begin{subfigure}{0.32\textwidth}
        \centering
        \includegraphics[width=\textwidth]{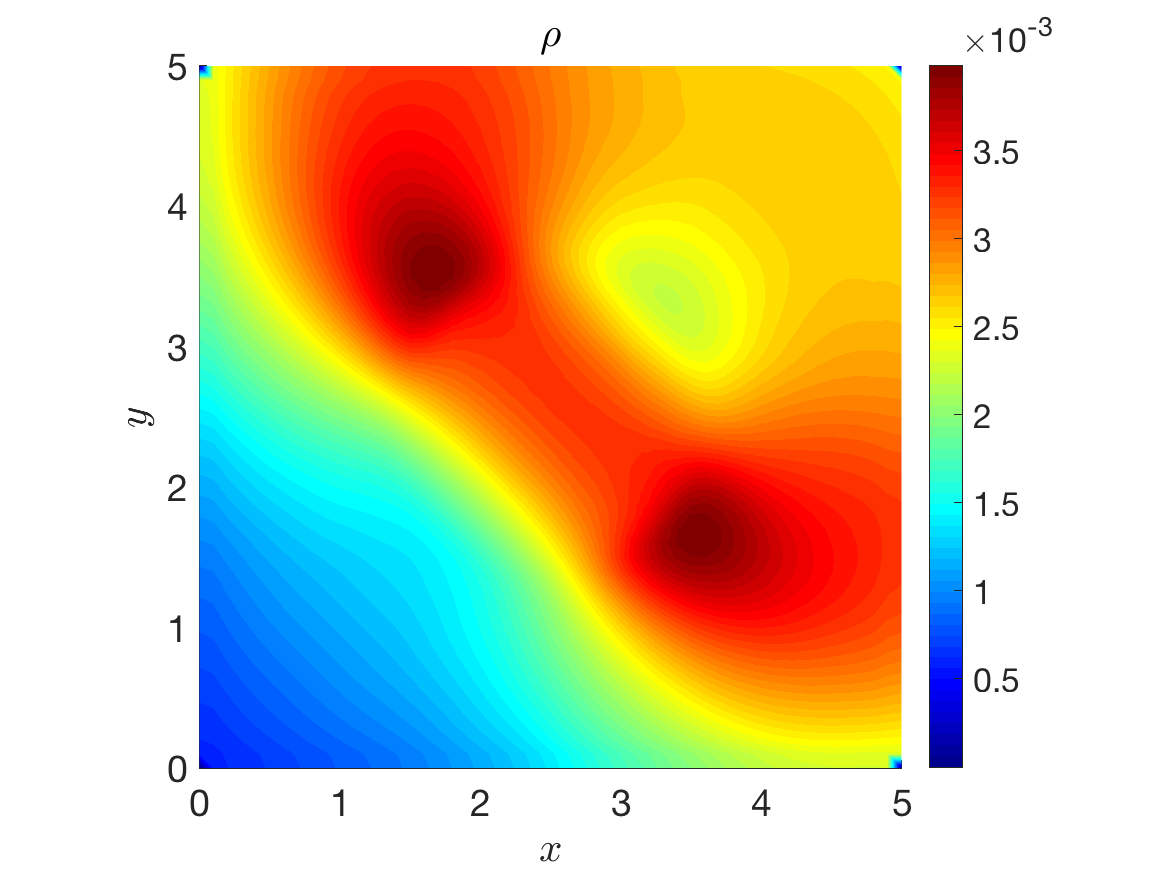}   
        \caption{\scriptsize t=30}
        \label{iii_t1200}
     \end{subfigure}
       \begin{subfigure}{0.32\textwidth}
        \centering
        \includegraphics[width=\textwidth]{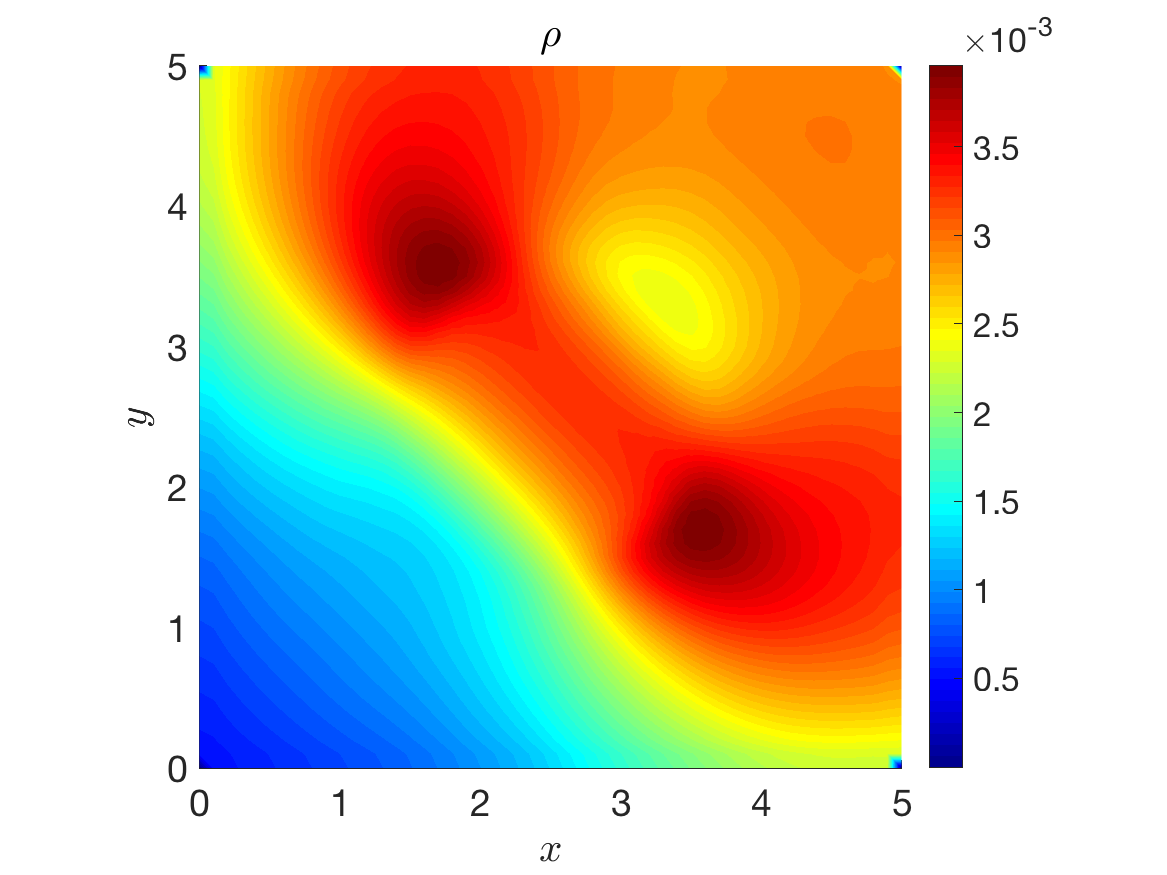}   
        \caption{\scriptsize t=60}
        \label{ii_t2400}
     \end{subfigure}
    \caption{ \textbf{Test 3} Case $iii)$ with non-local $q$ and $\cS$, independent and with sensing function $\gamma_q=\gamma_{\scS}=H(R-\lambda)$. $\cS$ is given in Figure \ref{S18}, so that $l_{\cS}=0.25$, while for the fiber distribution $m_k=100$, corresponding to $l_q\approx0.0031$. The sensing radius of the cells is set to $R=0.02$.}
       \label{Sim_iii}
\end{figure}

\begin{figure}[!h] 
\begin{subfigure}{0.32\textwidth}
        \centering
        \includegraphics[width=\textwidth]{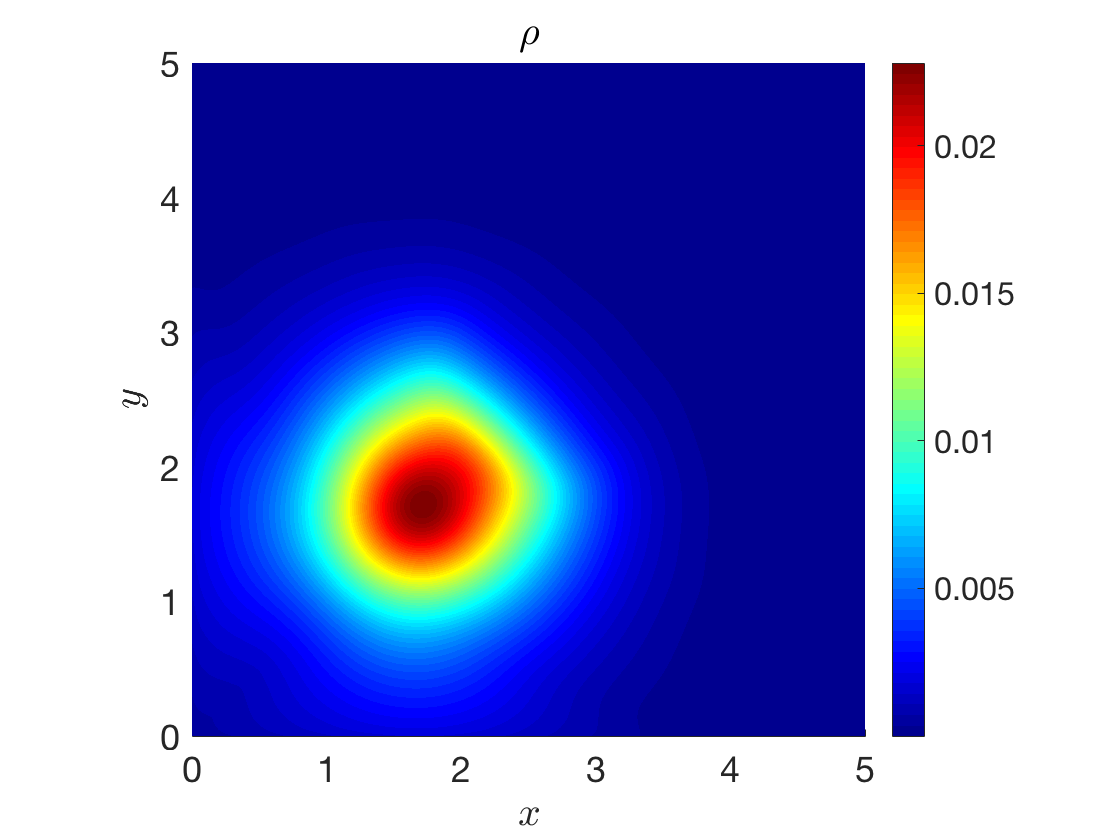}
        \caption{\scriptsize t=1.25}
        \label{iv_t500}
    \end{subfigure}
\begin{subfigure}{0.32\textwidth}
        \centering
        \includegraphics[width=\textwidth]{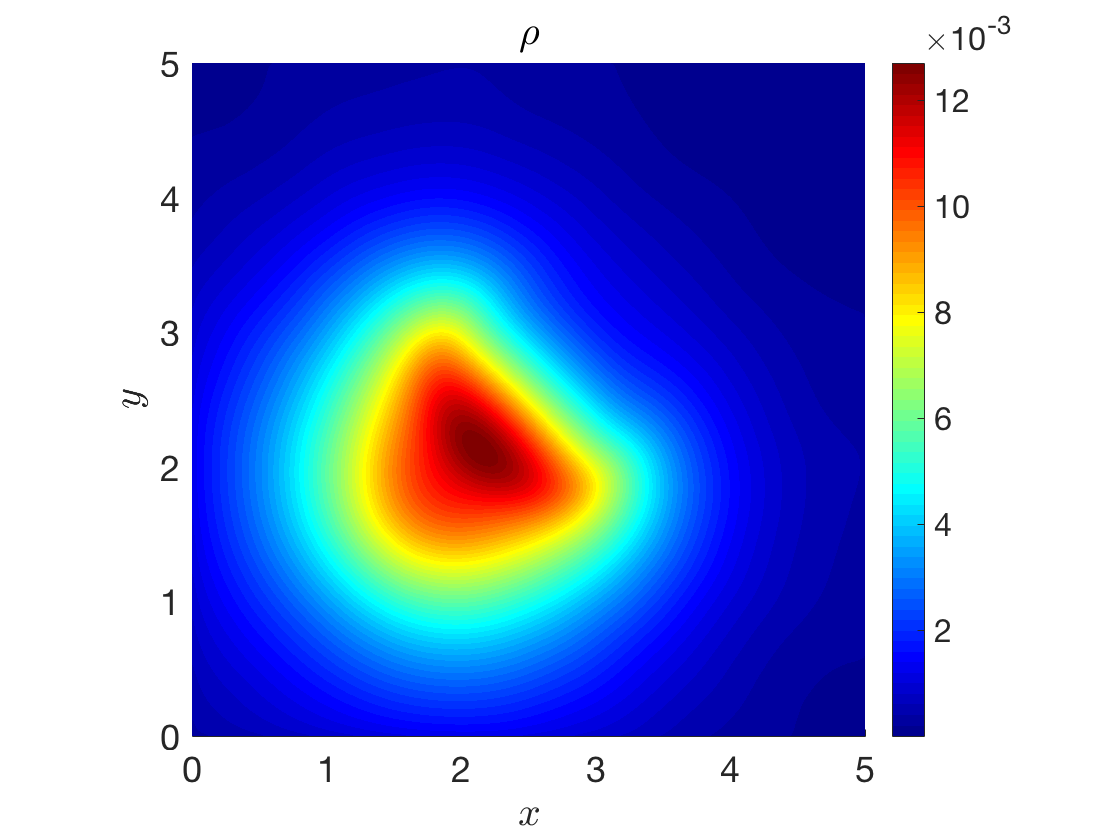}
        \caption{\scriptsize t=2.5}
        \label{iv_t1000}
    \end{subfigure}
    \begin{subfigure}{0.32\textwidth}
        \centering
        \includegraphics[width=\textwidth]{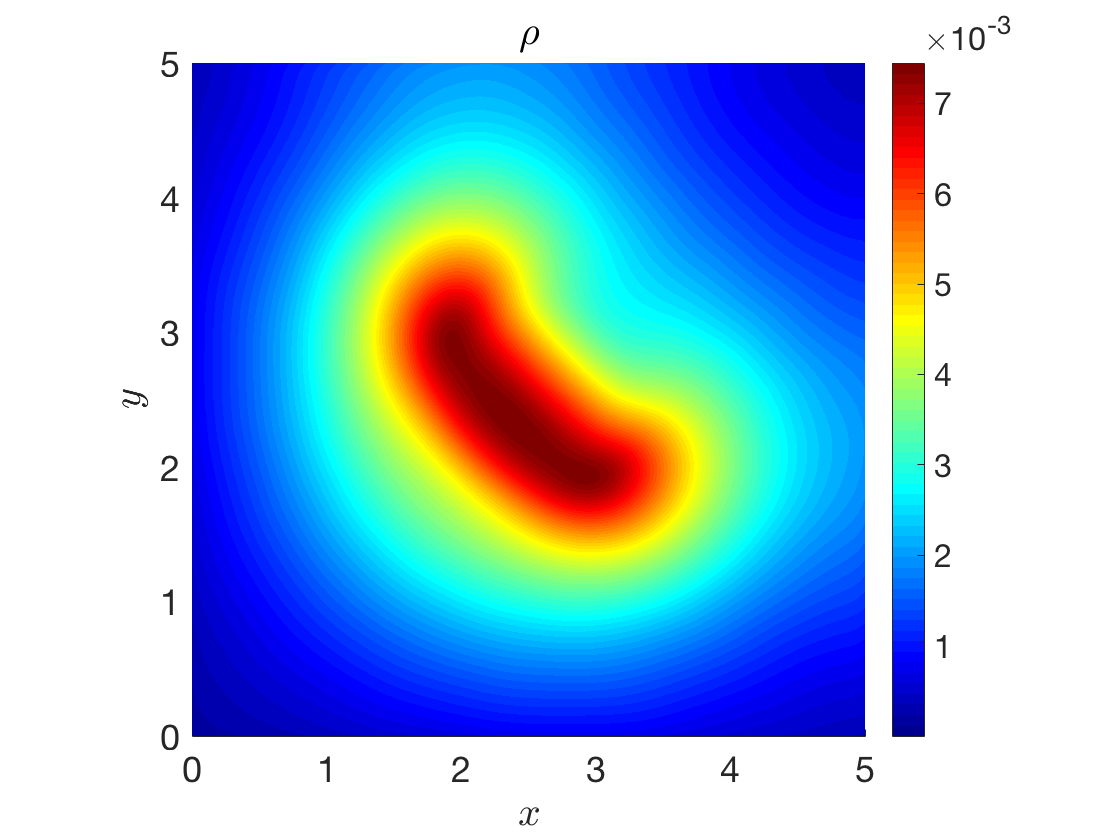}
        \caption{\scriptsize t=5}
        \label{iv_t2000}
    \end{subfigure}
    
   \begin{subfigure}{0.32\textwidth}
        \centering
        \includegraphics[width=\textwidth]{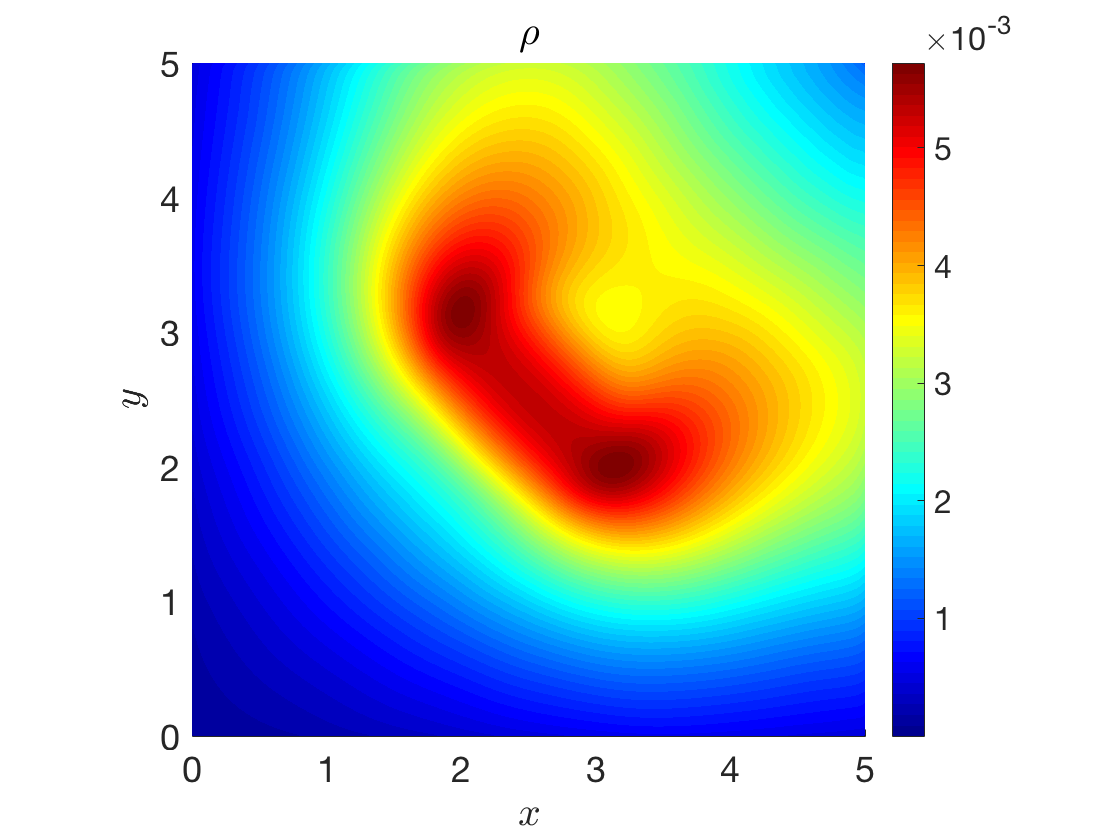}
        \caption{\scriptsize t=7.5}
        \label{iv_t3000}   
     \end{subfigure} 
     \begin{subfigure}{0.32\textwidth}
        \centering
        \includegraphics[width=\textwidth]{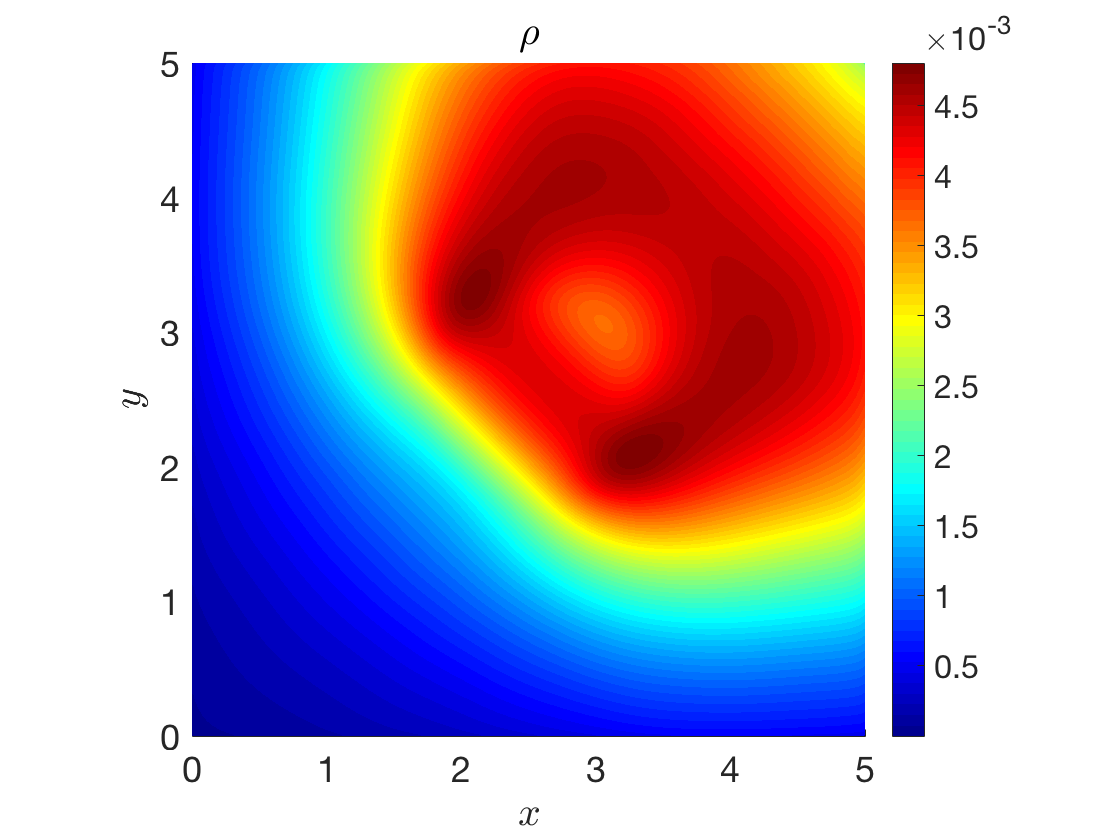}   
        \caption{\scriptsize t=10}
        \label{iv_t4000}
     \end{subfigure}
       \begin{subfigure}{0.32\textwidth}
        \centering
        \includegraphics[width=\textwidth]{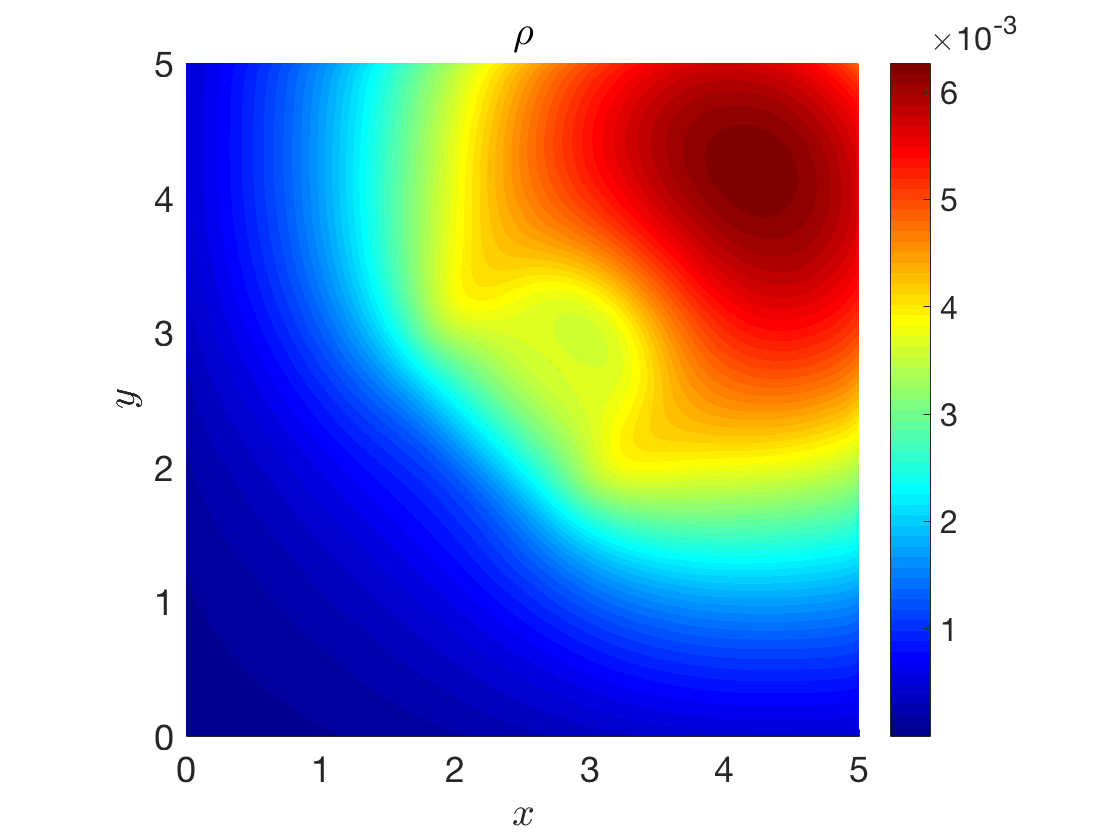}   
        \caption{\scriptsize t=15}
        \label{iv_t5000}
     \end{subfigure}
    \caption{ \textbf{Test 3} Case $iv)$ with non-local $q$ and $\cS$, independent sensing with $\gamma_q=\gamma_{\scS}=H(R-\lambda)$. $\cS$ is given in Fig. \ref{S005}, that corresponds to $l_\cS=0.002$, whilst $m_k=10$, so that $l_q \approx 0.031$. The sensing radius of the cells is $R=0.02$.}
       \label{Sim_iv}
\end{figure}

We shall consider the turning kernel describing contact guidance lead by a $q$ with mean direction $\theta_q(x,y)=3\pi/4$ $\forall (x,y) \in\Omega$ and coefficient $k(x,y)$, modulating the strength of the alignment, given by a gaussian distribution
\begin{equation}\label{k_gaussian}
k(x,y)=m_k e^{-\dfrac{\left( (x,y)-(x_k,y_k)\right)^2}{2\sigma_k^2}}
\end{equation} 
where $(x_k,y_k) = (2.5,2.5)$ and $\sigma_k^2=0.15$ (Fig. \ref{k_gauss}). This mimics the situation of fibers more aligned in the central circular region and uniformly disposed in the rest of the domain. We shall consider different values of $m_k$ in order to obtain different values of $l_q$: $m_k=10$ corresponds to $l_q \approx 0.031$ and $m_k=100$ corresponds to $l_q\approx 0.0031$. Details about the estimation of $l_q$ for a Bimodal Von Mises distribution of fibers $q$ are given in Appendix \ref{est_lq}. The chemoattractant is \eqref{S.gauss} with $(x_{\cS},y_{\cS})=(4.5, 4.5)$ and $m_{\cS}=10$. In the simulations, we shall consider three different values for the variance of the chemoattractant $\sigma_{\cS}^2$ in order to obtain different values of $l_{\cS}$: $\sigma_{\cS}^2=0.05$ that corresponds to $l_{\cS}=0.002$ in Fig. \ref{S005}, $\sigma_{\cS}^2=0.25$ that corresponds to $l_{\cS}=0.055$ in Fig. \ref{S025} and $\sigma_{\cS}^2=1.8$ that corresponds to $l_{\cS}=0.25$ in Fig. \ref{S18}. The initial distribution of cells for all the tests presented in Figs. \ref{Sim_i_varS005}, \ref{Sim_i_varS18}, \ref{Sim_ii}, \ref{Sim_iii} and \ref{Sim_iv} is given by \eqref{ci_cells_gaussian} with $(x_0,y_0)=(1.5,1.5)$, $r_0=0.1$, $\sigma_0^2=0.1$. In particular, we present five sets of simulations that are summarized in Table \ref{Table_test3}.
\begin{table}[!htbp]
\footnotesize
\begin{center}
\begin{tabular}{|c|c|c|c|c|c|}
\hline
$l_{\cS}$ & $l_{q}$ & $R$ & Case &$\eta$ & Figure \\[1.5ex]
\hline\hline
$0.002$ & $0.0031$ & $0.7$ & $i)$ & $<1$ &\ref{Sim_i_varS005}\\[1.5ex]
\hline
$0.25$ & $0.0031$ & $0.7$ & $i)$ & $\gg 1$ &\ref{Sim_i_varS18}\\[1.5ex]
\hline
$0.055$ & $0.031$ & $0.02$ & $ii)$ & $>1$ &\ref{Sim_ii}\\[1.5ex]
\hline
$0.25$ & $0.0031$ & $0.02$ & $iii)$ & $\gg 1$ &\ref{Sim_iii} \\[1.5ex]
\hline
$0.002$ & $0.031$ & $0.02$ & $iv)$ & $<1$ &\ref{Sim_iv}\\
\hline
\end{tabular}
\caption{Summary of the simulations presented in Test 3.}
\label{Table_test3}
\end{center}
\end{table}

\indent In Fig. \ref{Sim_i_varS005}, we consider the case in which $\eta_\cS,\eta_q\gg1$, $\ie$, we are dealing with case $i)$. The macroscopic behavior is strongly hyperbolic with macroscopic velocity given by \eqref{U_T_indip}. In fact, in Fig. \ref{Sim_i_varS005} we can observe that the behavior is not diffusive and the cluster of cells is quite compact. Moreover, when cells reach the region in which fibers are strongly aligned in the direction $3\pi/4$ (as shown in Fig. \ref{k_gauss}), that is perpendicular to the favorable direction $\pi/4$ induced by the chemoattractant, they surround that region inducing strong alignment and go over towards the chemoattractant. In this setting, the parameter defined in \eqref{eta} is slightly smaller then 1 and, in fact, chemotaxis prevails in the overall dynamics, as the stationary state is clearly peaked on the chemoattractant profile, but the fibers structure influences the transient.

In Fig. \ref{Sim_i_varS18}, we shall consider $\cS$ with $\sigma_{\cS}^2=1.8$ and, consequently, $l_{\cS}=0.25$ (see Fig. \ref{S18}). Concerning the fibers, we have $m_k=100$, so that $l_q \approx 0.0031$, and the sensing radius is $R=0.7$. This setting falls again in case $i)$, but the behavior is different with respect to the previous simulation in Fig. \ref{Sim_i_varS005}. The chemoattractant in Fig. \ref{S18}, in fact, is spread over the whole domain and, actually, the quantity $l_{\cS}$ is almost $10^2$ times the $l_{\cS}$ considered in Fig. \ref{S005} and used for the simulation in Fig. \ref{Sim_i_varS005}. Even though we are still in a strongly hyperbolic case and cells are guided by the strong drift \eqref{U_T_indip}, as $R$ is slightly larger then $l_{\cS}$ and $l_{\cS}$ is large, the cell cluster diffuses a bit more in the domain. When it reaches the region of strongly aligned fibers, it starts to surround that region (see Figs. \ref{i_t100_S18}-\ref{i_t400_S18}), but, as $\eta_{\cS}= 2.8=\mathcal{O}(1)$, some cells, that do not surround the region, are slowed down and partially tend to align along the fibers. In Fig. \ref{i_t600_S18}, for instance, we have a high density of cells both in the strongly aligned fiber region and in the region of high density of chemoattractant. Eventually, cells manage to overcome the area of highly aligned fibers and they tend to converge to the chemoattractant profile (see Figs. \ref{i_t900_S18}-\ref{i_t1100_S18}). Now, the the overall dynamics is greatly affected by the fibers and, in fact, $\eta \gg 1$. 

The second scenario, illustrated in Fig. \ref{Sim_ii}, refers to the case $ii)$, since the sensing radius $R=0.02$ is smaller than both $l_\cS=0.055$ and $l_q\approx 0.031$. 
At the macroscopic level, the behavior of the system is described by the diffusion-advection equation \eqref{macro_indip.ii} with macroscopic velocity \eqref{UT_ind.ii}. Actually, in Fig. \ref{Sim_ii} we can observe a highly diffusive behavior, as the macroscopic density of cells has invaded almost the half of the domain before even starting to be influenced by the fibers. If we compare the same time step in Figs. \ref{ii_t2000} and \ref{i_t200_S18}, we see that the cells are in both cases reaching the fibers and feeling the region in which fibers are aligned the most. However, in Fig. \ref{i_t200_S18} the cell cluster is much more compact than in Fig. \ref{ii_t2000}, where, instead, cells already occupied half of the domain, because of diffusion, and we have high density of cells both closely to the strongly aligned fiber region and around the initial position. Therefore, cells start surrounding the central region of strongly aligned fibers, because they already sense the chemoattractant, and, once overcome this area, they tend to the chemoattractant profile (see Figs. \ref{ii_t3000}-\ref{ii_t8000}). In particular, in the transient time, cells accumulate the most at the sides of the region with highly aligned fibers. In this specific setting, $\eta >1$ and, in fact, contact guidance highly affects the dynamics.

The third scenario, illustrated in Fig. \ref{Sim_iii}, refers to the case $iii)$, since the sensing radius $R=0.02$ is smaller than $l_{\cS}=0.25$ but it is larger then $l_q\approx 0.0031$. The macroscopic setting is described by a diffusion-advection equation with diffusion tensor and drift velocity given by \eqref{DT_0_indip.iii} and \eqref{UT_indip.iii}, respectively. As $\eta_{\cS}<1$, we have that the chemoattractant induces a strong diffusivity, but being $\eta_q >1$, the alignment of fibers strongly affects the dynamics (see Figs. \ref{iii_t400}-\ref{iii_t800}). Comparing, in addition, Figs. \ref{ii_t2000} and \ref{iii_t200}, we have now that the highest cell concentration is in the mean fiber direction $\theta_q=3\pi/4$ in the region surrounding the center of the domain, where the fibers are aligned with a higher degree. As already observe in section 3, this scenario prescribes $\eta \gg 1$ and, in fact, contact guidance dominates again the dynamics.

Eventually, for a sensing radius $R=0.02$ smaller than $l_q \approx 0.031$, but larger than $l_\cS=0.002$, the macroscopic behavior is approximated by an hyperbolic equation with drift velocity given in \eqref{UT_0:indip.iv}. Results of the simulation are presented in Fig. \ref{Sim_iv}. Here, the chemoattractant has the profile shown in Fig. \ref{S005}. Cells diffuse in the domain because $\eta_q$ is smaller than 1, and they start moving in a region with randomly disposed fibers (see Fig. \ref{iv_t500}). Then, they mainly follow the preferential direction $\pi/4$ thanks to the presence of the chemoattractant. In fact, it induces a strong drift because of the high non-locality, determining $\eta_{\cS}\gg1$. Here chemotaxis is slightly dominating the dynamics and, in fact, $\eta <1$.

\subsection{Test 4: heterogeneous ECM environment}
\begin{figure}[!h] 
\begin{subfigure}{0.32\textwidth}
        \centering
        \includegraphics[width=0.75\textwidth]{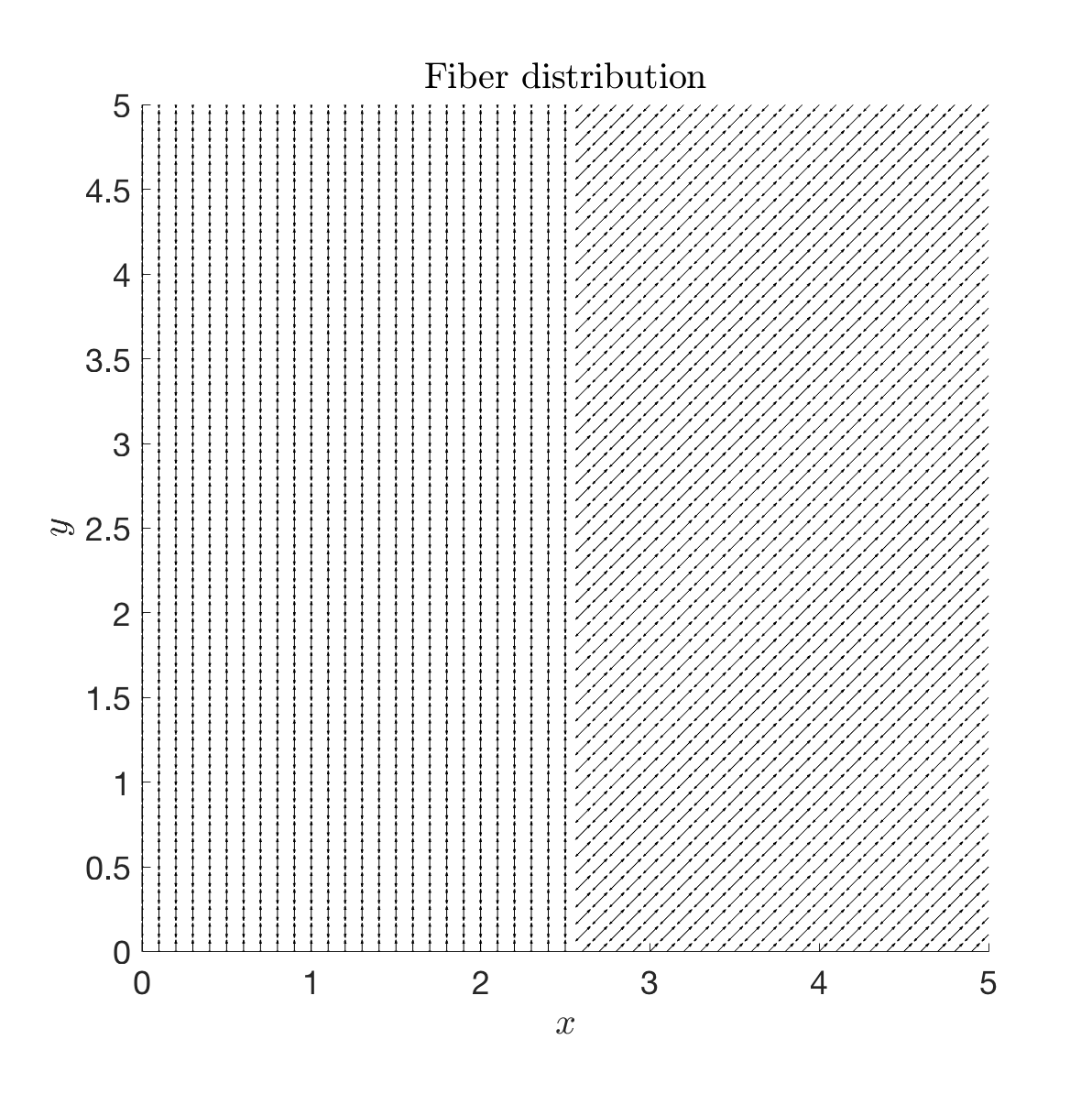}
        \caption{\scriptsize Fibers distribution.}
        \label{Fib2side}
    \end{subfigure}
    \begin{subfigure}{0.32\textwidth}
        \centering
        \includegraphics[width=\textwidth]{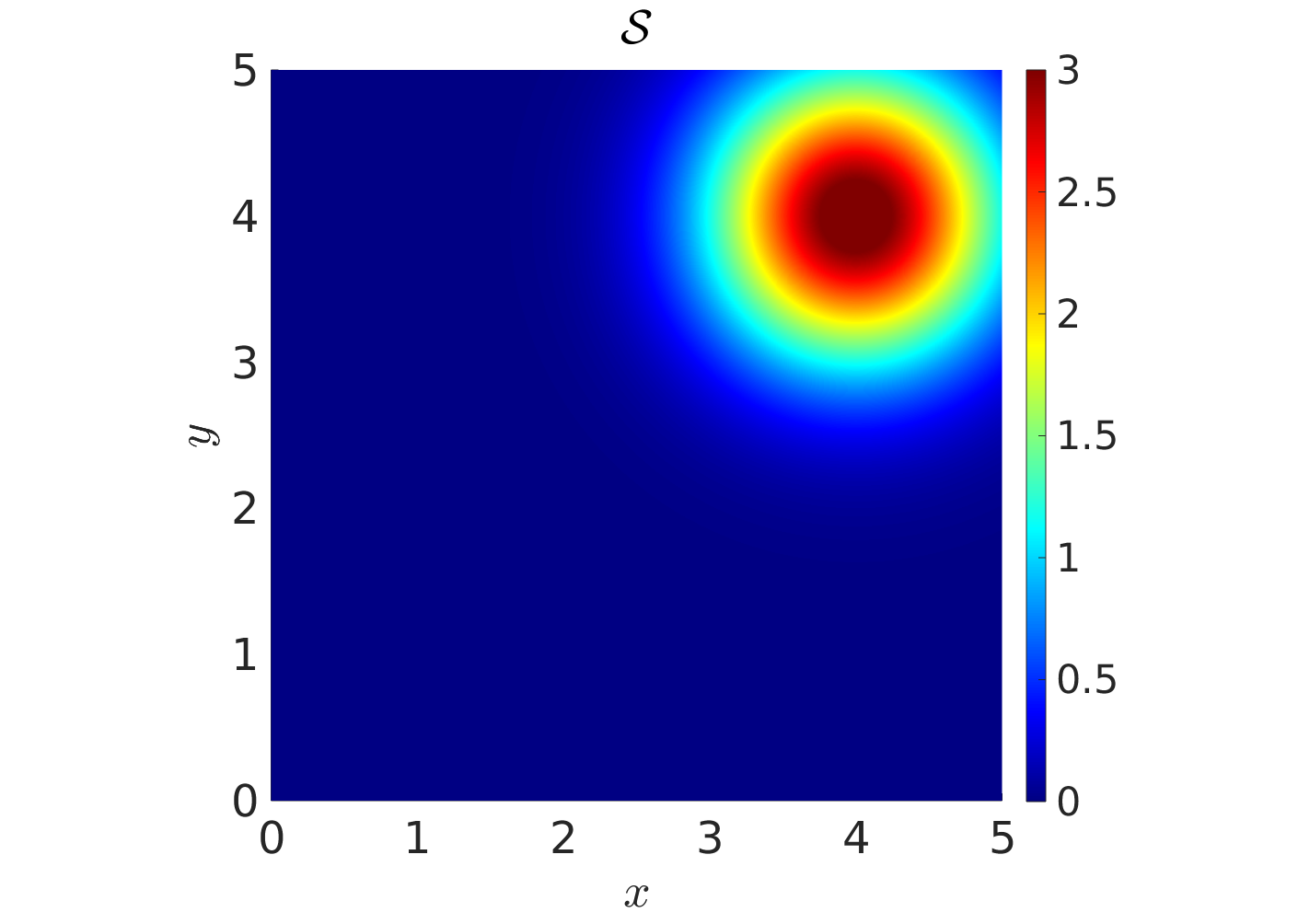}
        \caption{\scriptsize Chemoattractant $\cS$.}
        \label{InS_Fib2side}
    \end{subfigure}
\begin{subfigure}{0.32\textwidth}
        \centering
        \includegraphics[width=\textwidth]{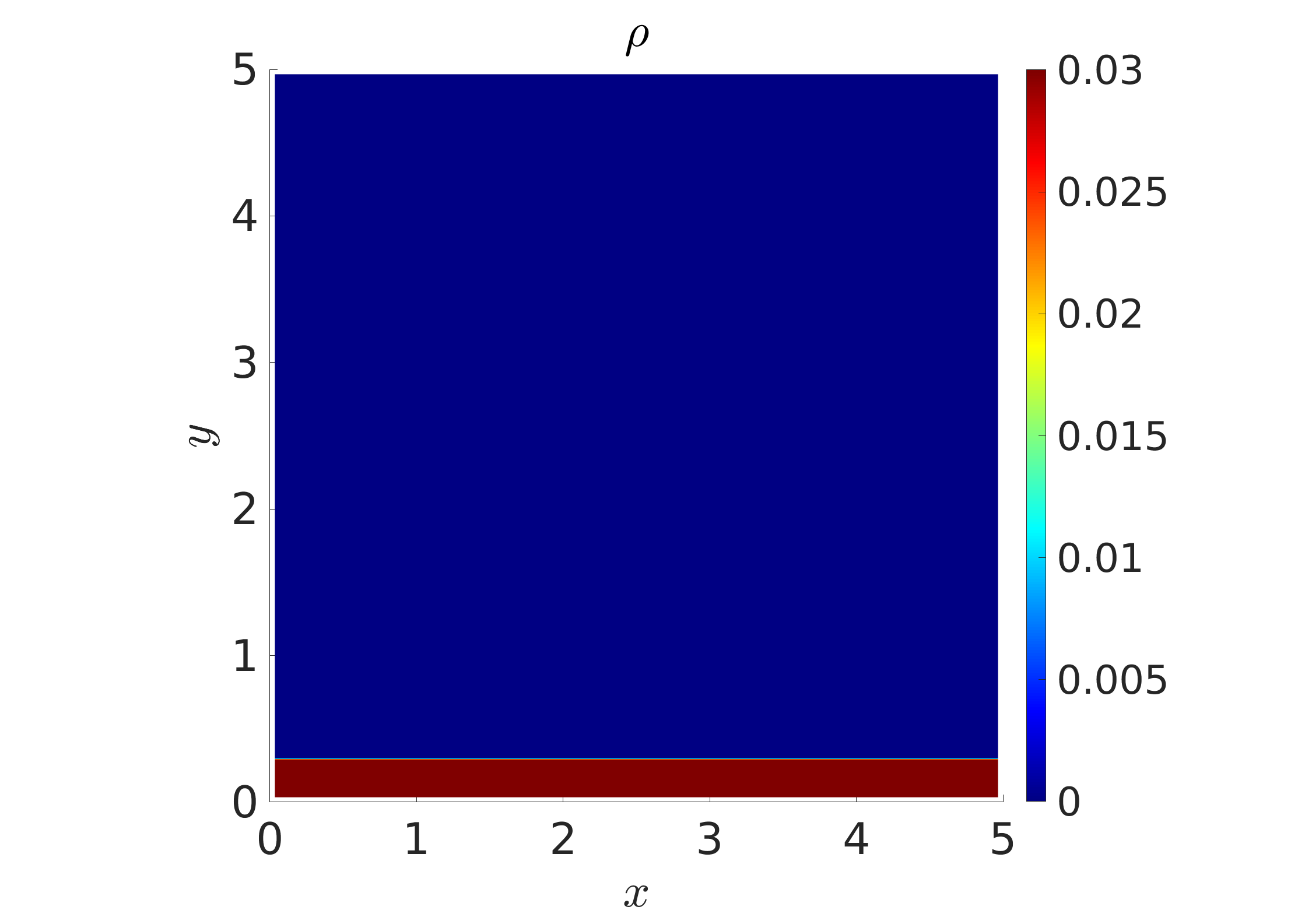}
        \caption{\scriptsize Initial condition for the cell.}
        \label{InCell_Fib2side}
    \end{subfigure}\\
    
    \begin{subfigure}{0.32\textwidth}
        \centering
        \includegraphics[width=\textwidth]{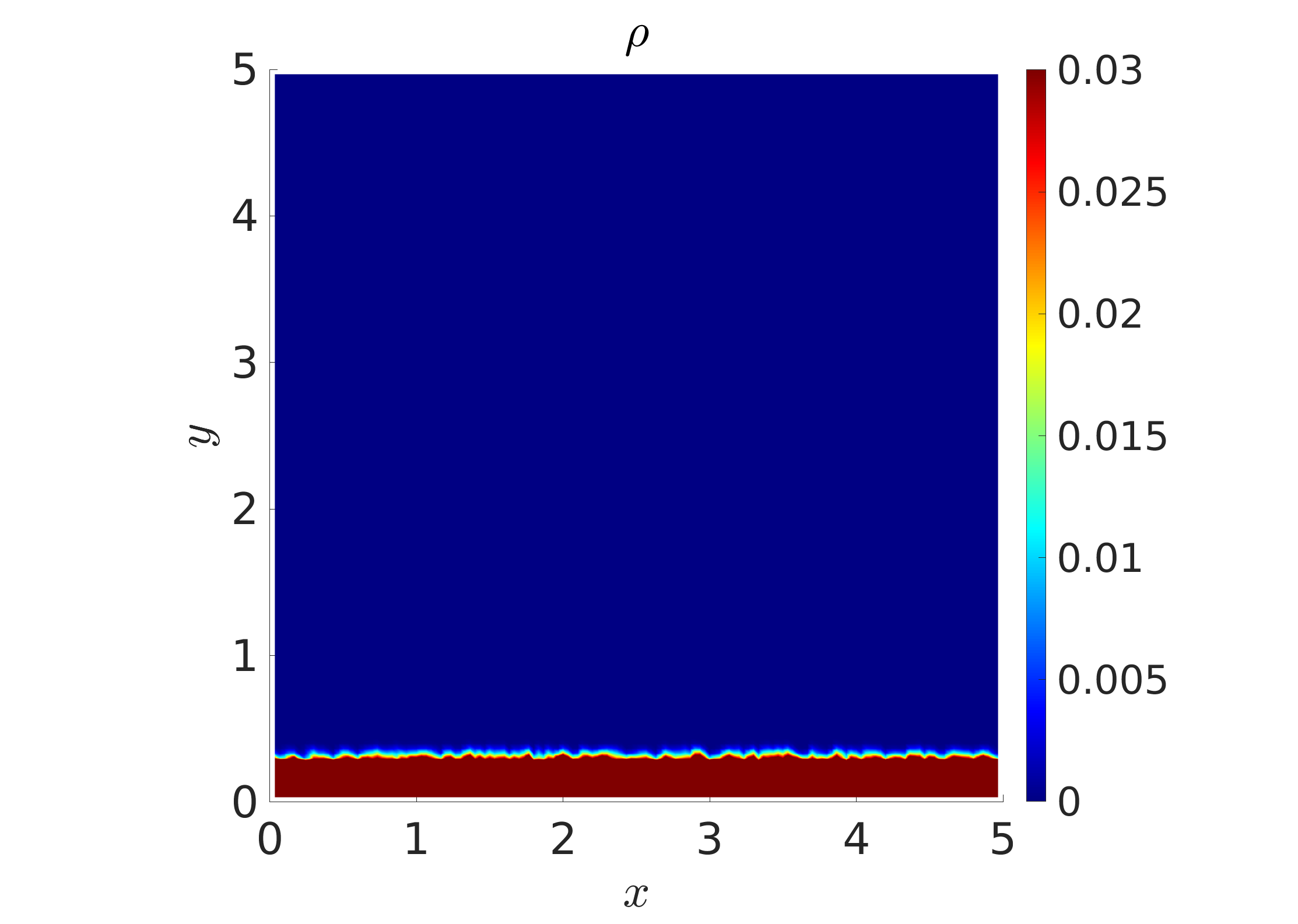}
        \caption{\scriptsize t=0.04}
        \label{MD1_t5}
    \end{subfigure}
\begin{subfigure}{0.32\textwidth}
        \centering
        \includegraphics[width=\textwidth]{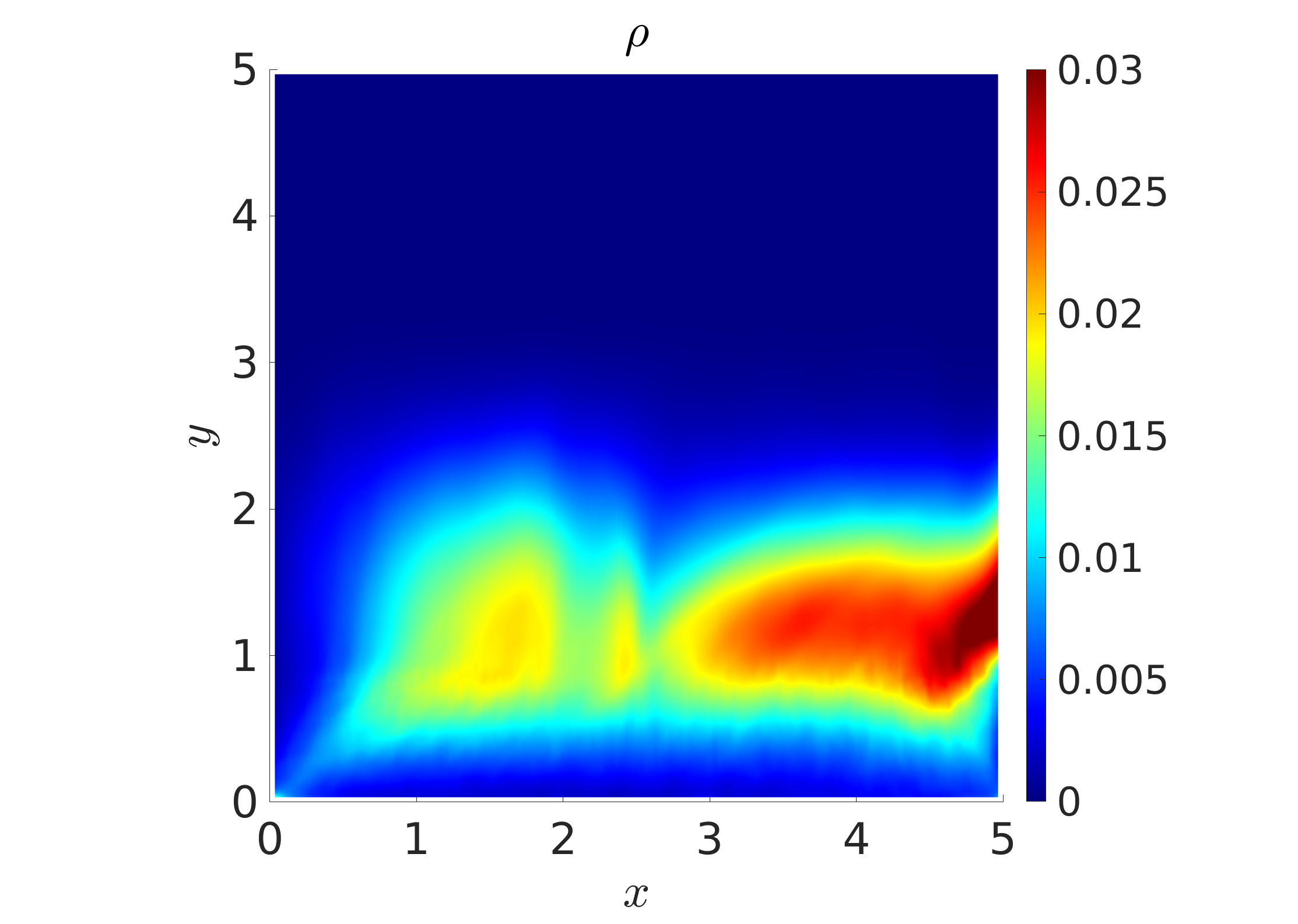}
        \caption{\scriptsize t=1.6}
        \label{MD1_t200}
    \end{subfigure}
    \begin{subfigure}{0.32\textwidth}
        \centering
        \includegraphics[width=\textwidth]{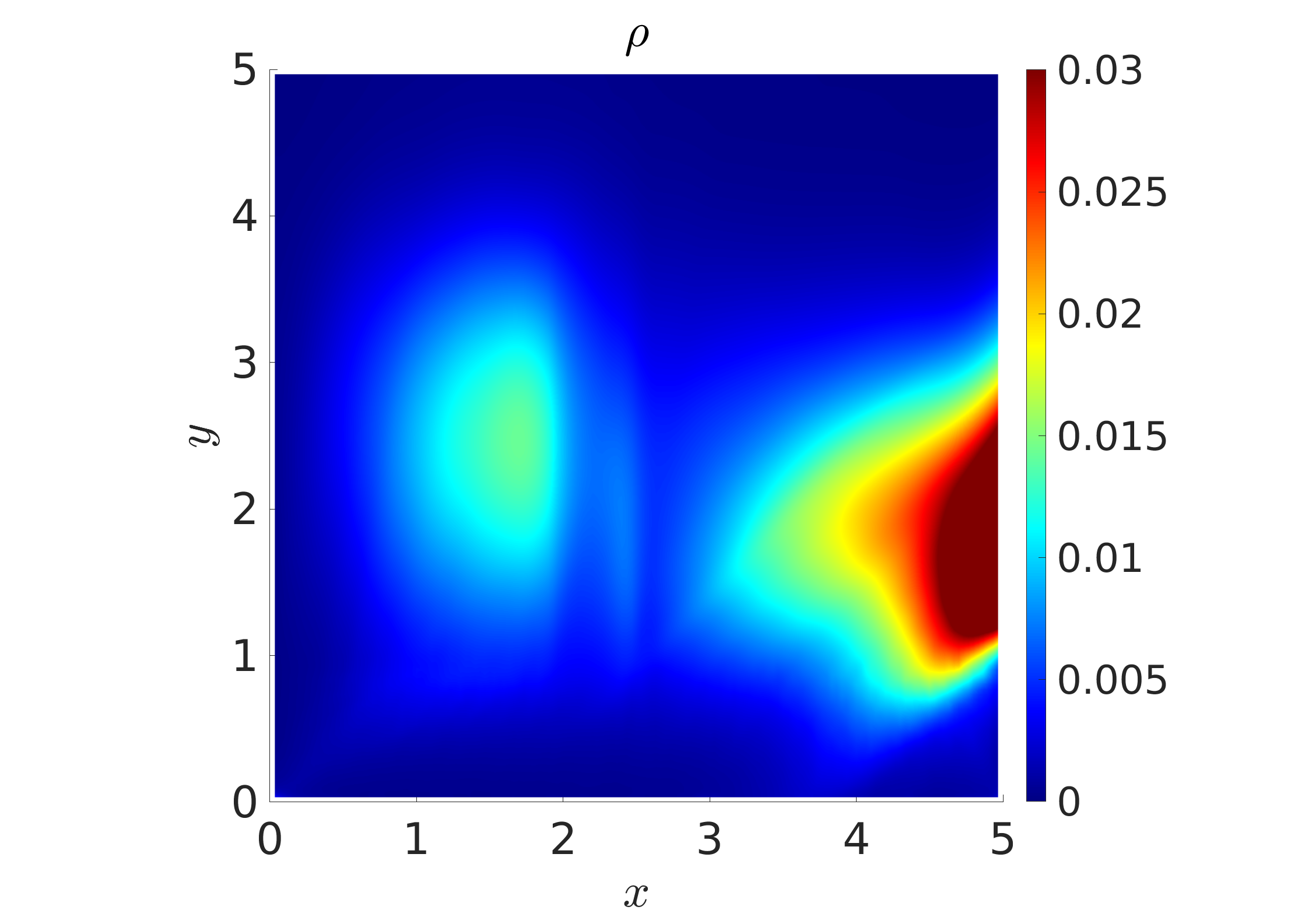}
        \caption{\scriptsize t=2.904}
        \label{MD1_t363}
    \end{subfigure}\\
    
   \begin{subfigure}{0.32\textwidth}
        \centering
        \includegraphics[width=\textwidth]{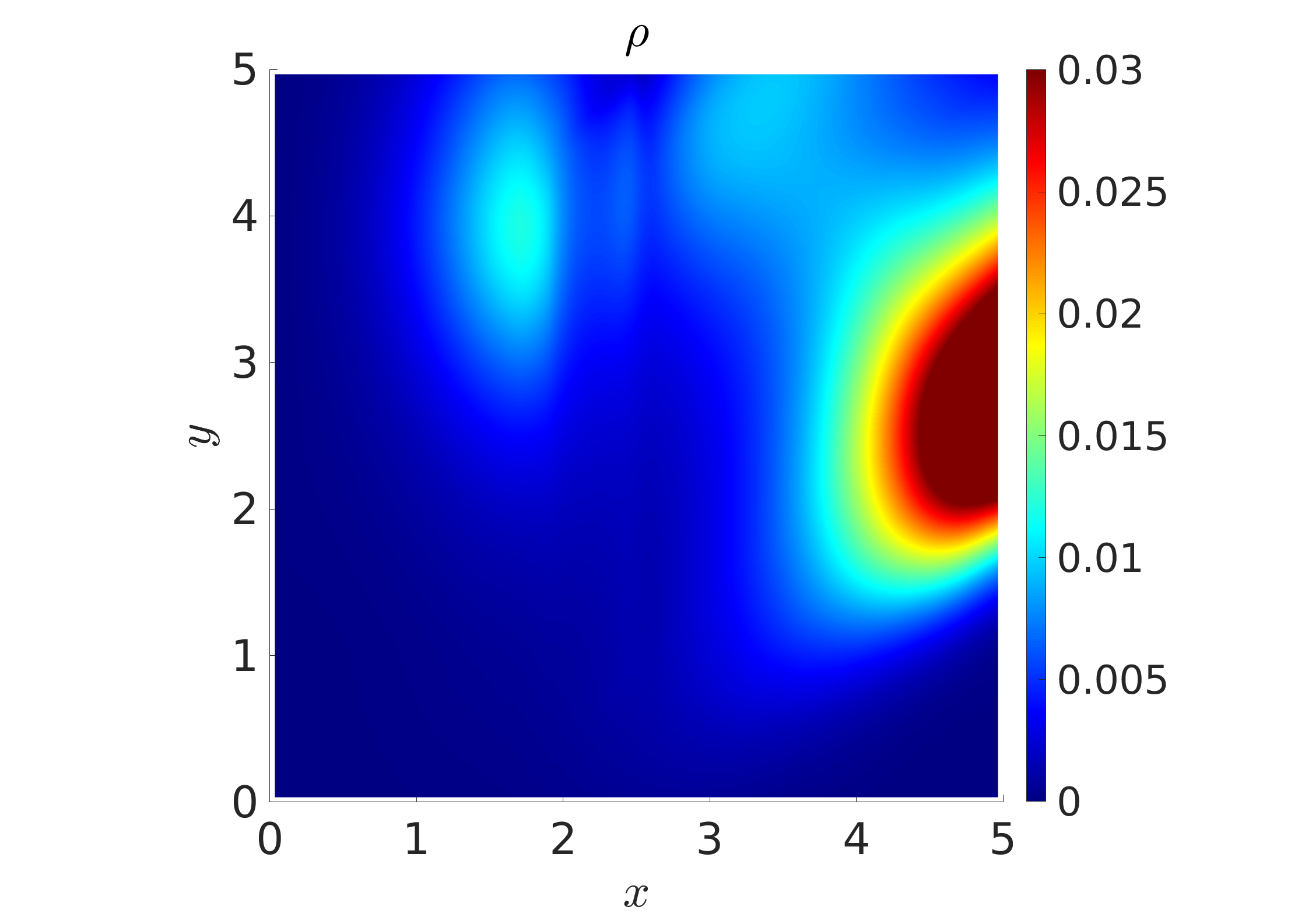}
        \caption{\scriptsize t=18.4}
        \label{MD1_t2300}   
     \end{subfigure} 
     \begin{subfigure}{0.32\textwidth}
        \centering
        \includegraphics[width=\textwidth]{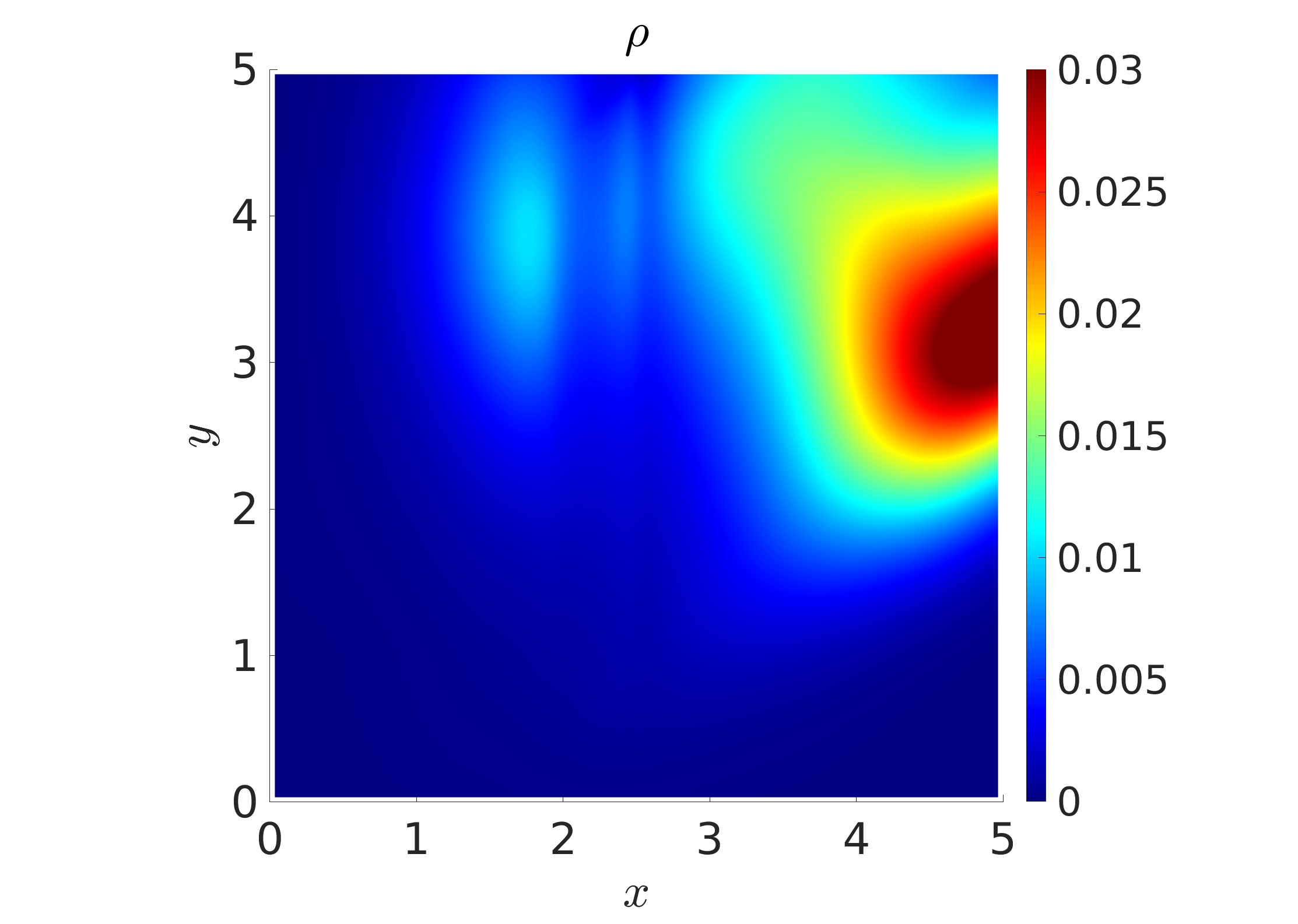}   
        \caption{\scriptsize t=44}
        \label{MD1_t5500}
     \end{subfigure}
       \begin{subfigure}{0.32\textwidth}
        \centering
        \includegraphics[width=\textwidth]{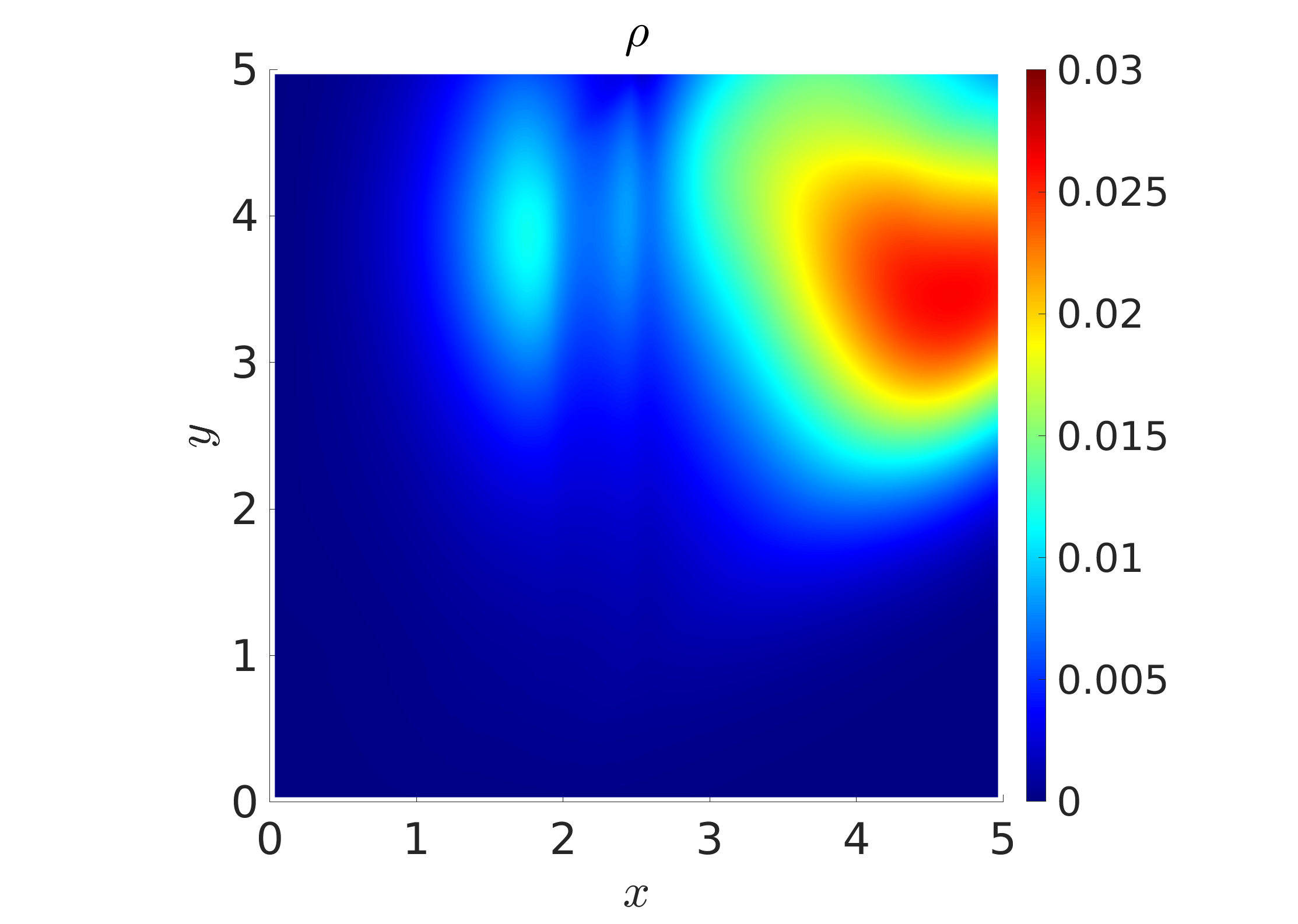}   
        \caption{\scriptsize t=67.2}
        \label{MD1_t8400}
     \end{subfigure}
    \caption{ \textbf{Test 4} Migration of cells in an heterogenous domain as illustrated in (a).  The sensing radius of the cells is $R=0.8$. The chemoattractant (b) is \eqref{S.gauss} with $m_{\cS}=10$ and $\sigma_{\cS}^2=0.5$. The initial cell profile (c) evolves in time as illustrated in (d)-(i).}
       \label{4.Mix_domain.1}
\end{figure}

We now consider a domain $\Omega$ divided in several regions, each of them characterized by a different average direction of the fibers. In particular, we shall do this in the case of independent sensing model 
with $\gamma_q=\gamma_{\scS}=H(R-\lambda)$, as for Test 3; the independence of the two sensings, in fact, allows to visualize more efficiently the two distinct directional effects. 
 As first scenario, we shall consider the domain schematized in Fig. \ref{Fib2side}; in each subdomain we have $k(x,y)=50$, that corresponds to $D_q=0.005$. The initial condition of the cells is represented in Fig. \ref{InCell_Fib2side}, with initial density $r_0=0.1$, while the chemoattractant has a gaussian profile \eqref{S.gauss} centered in $(x_{\cS},y_{\cS})= (4,4)$, with $m_{\cS}=10$ and $\sigma_{\cS}^2= 0.5$, as shown in Fig. \ref{InS_Fib2side}. We observe that cells do not migrate collectively towards the chemoattractant, but they divide into two main separated clusters (see Figs. \ref{MD1_t363} - \ref{MD1_t5500}): in fact, although the sensing radius $R=0.8$ is quite large, the cells that are closer to the left boundary remain trapped in the first subdomain, showing a loss of adhesion with the rest of the cell population. As shown in Fig. \ref{MD1_t8400}, even though the cells that are in the left subdomain horizontally align to the chemoattractant, the high degree of alignment of the fiber does not allow them to escape this region, even for large times.
 
\begin{figure}[!h] 
\begin{subfigure}{0.32\textwidth}
        \centering
        \includegraphics[width=0.75\textwidth]{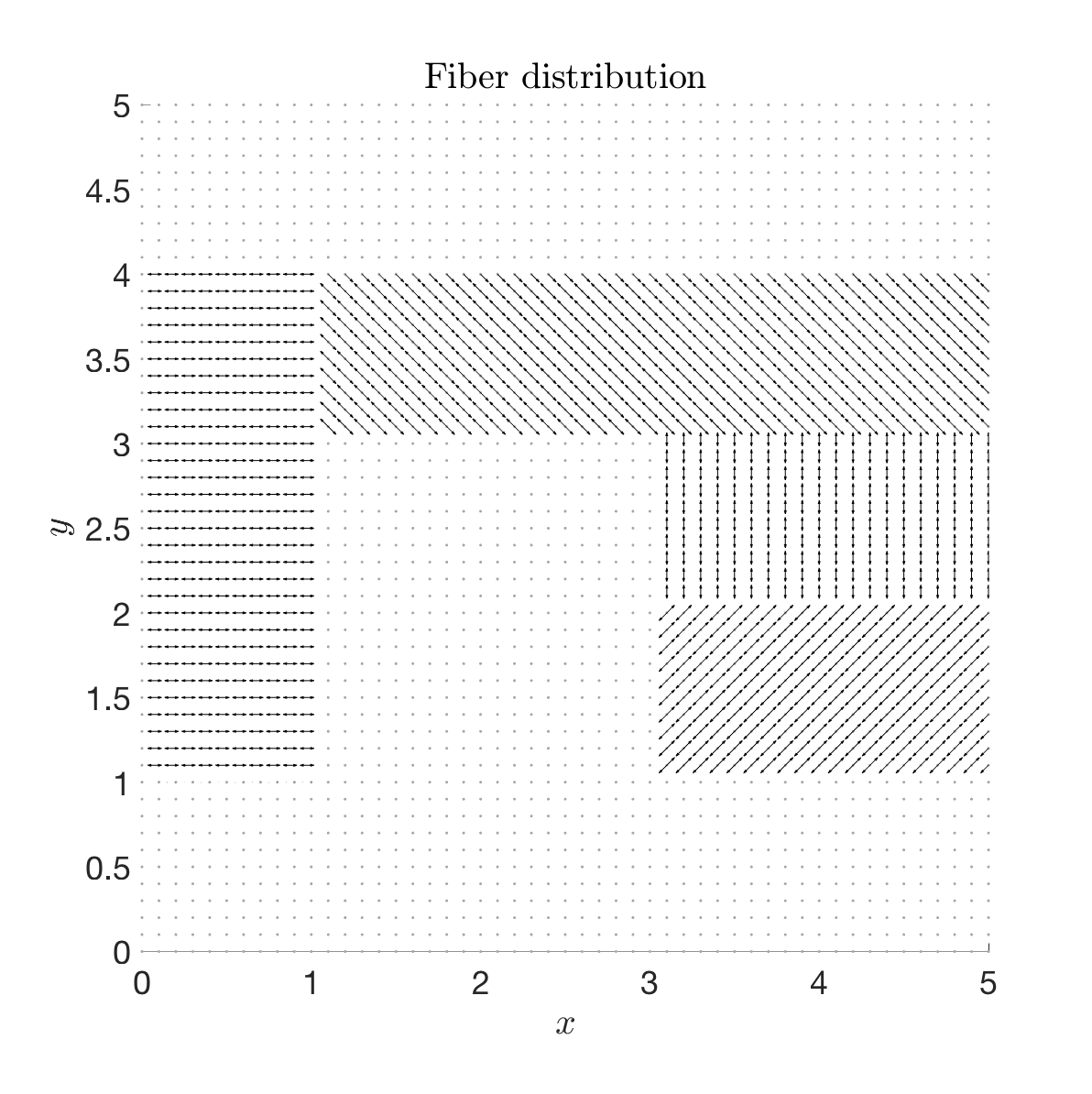}
        \caption{\scriptsize Fibers distribution.}
        \label{FibMix}
    \end{subfigure}
    \begin{subfigure}{0.32\textwidth}
        \centering
        \includegraphics[width=\textwidth]{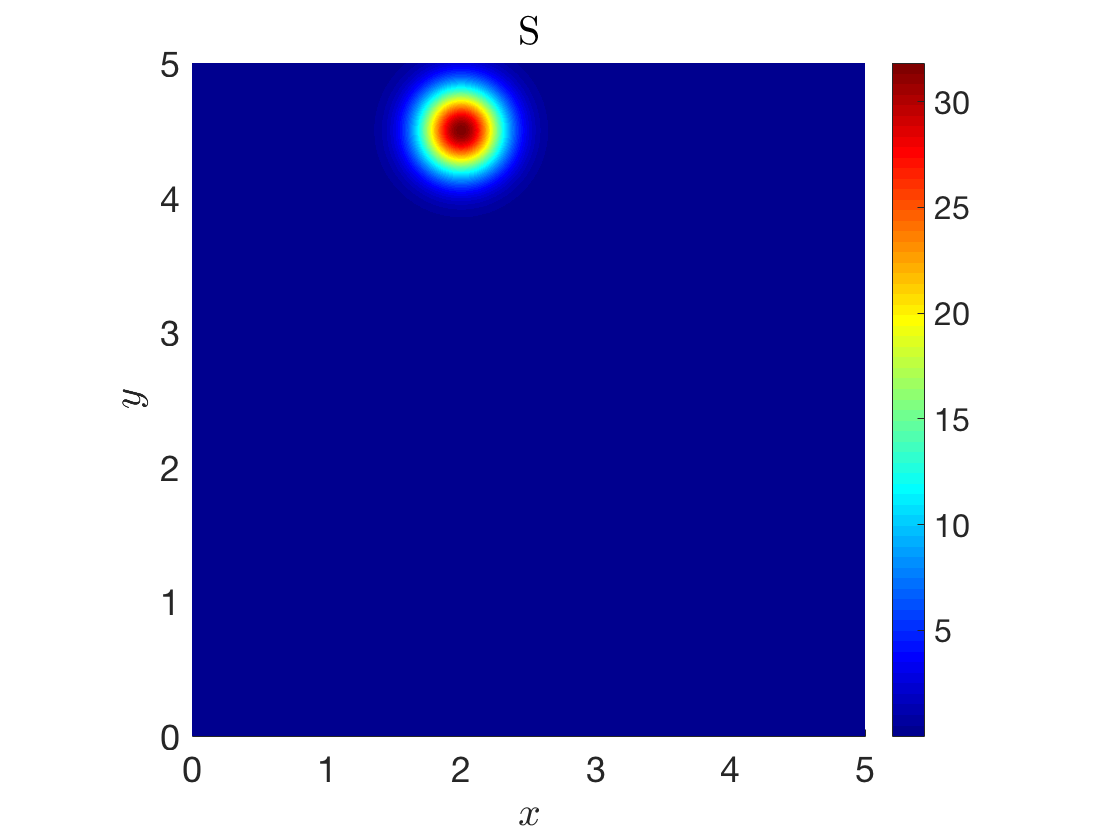}
        \caption{\scriptsize Chemoattractant $\cS$.}
        \label{InS_FibMix}
    \end{subfigure}
\begin{subfigure}{0.32\textwidth}
        \centering
        \includegraphics[width=\textwidth]{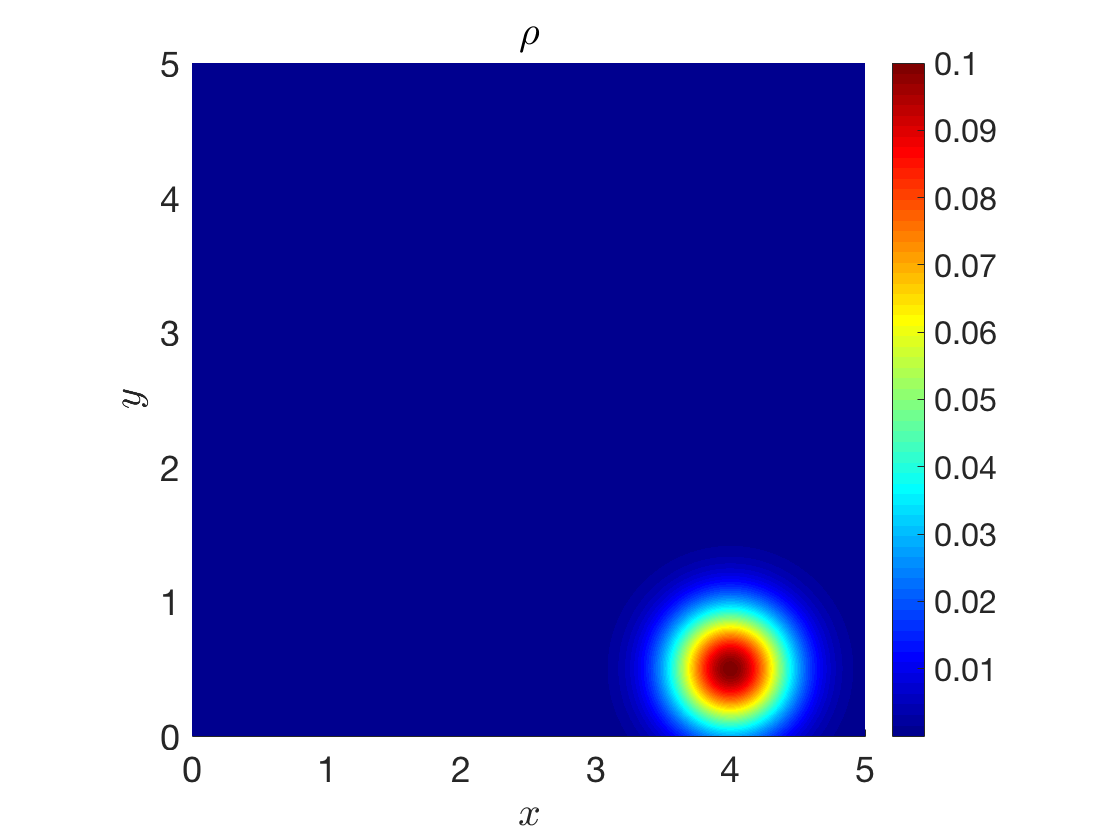}
        \caption{\scriptsize Cells initial condition.}
        \label{InCell_FibMix}
    \end{subfigure}\\
    
    \begin{subfigure}{0.32\textwidth}
        \centering
        \includegraphics[width=\textwidth]{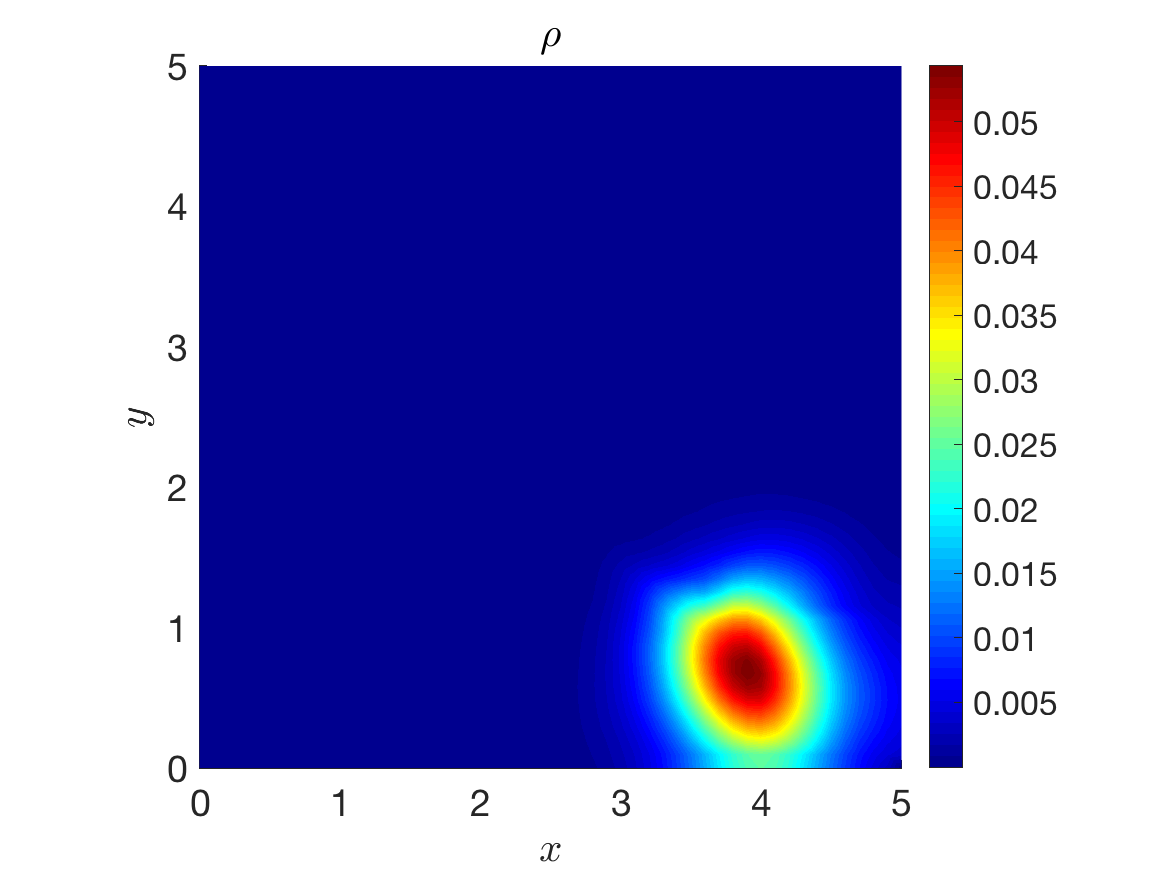}
        \caption{\scriptsize t=0.5}
        \label{MD_t120}
    \end{subfigure}
\begin{subfigure}{0.32\textwidth}
        \centering
        \includegraphics[width=\textwidth]{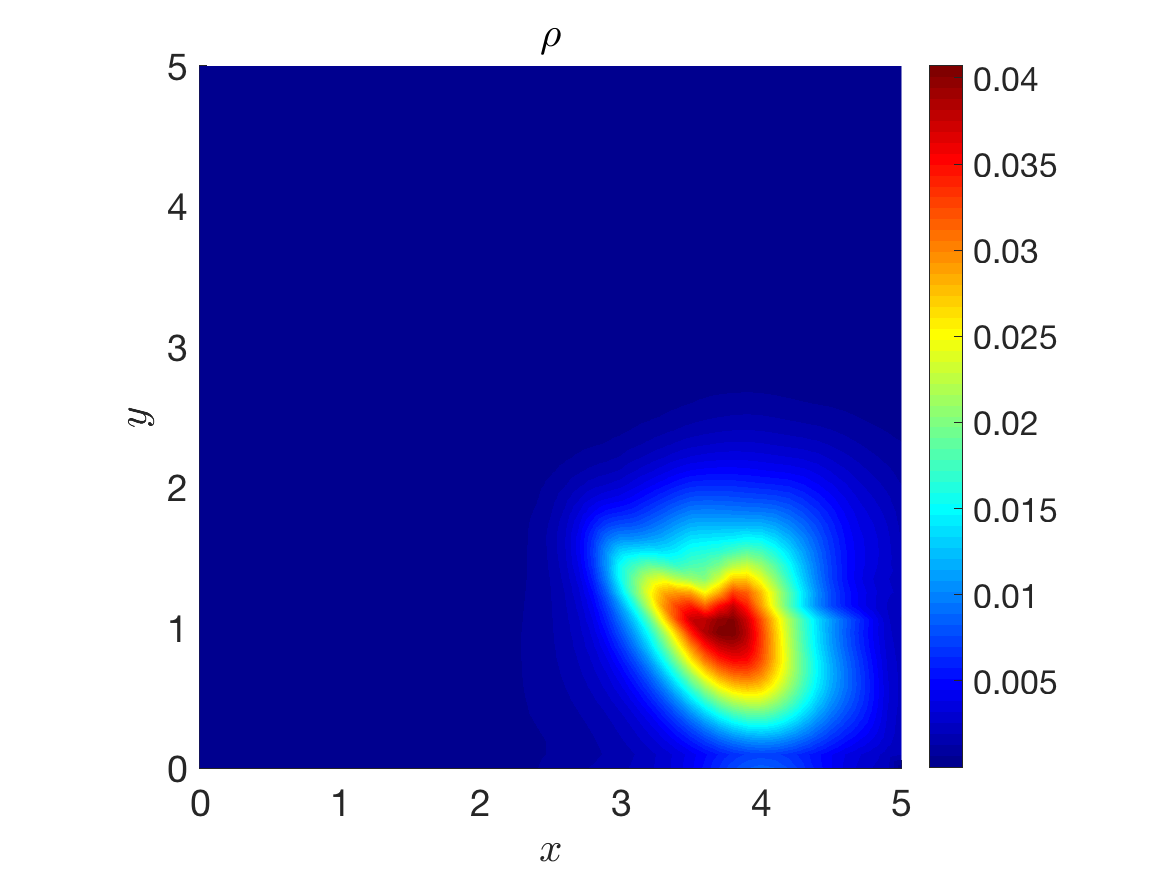}
        \caption{\scriptsize t=1}
        \label{MD_t40}
    \end{subfigure}
    \begin{subfigure}{0.32\textwidth}
        \centering
        \includegraphics[width=\textwidth]{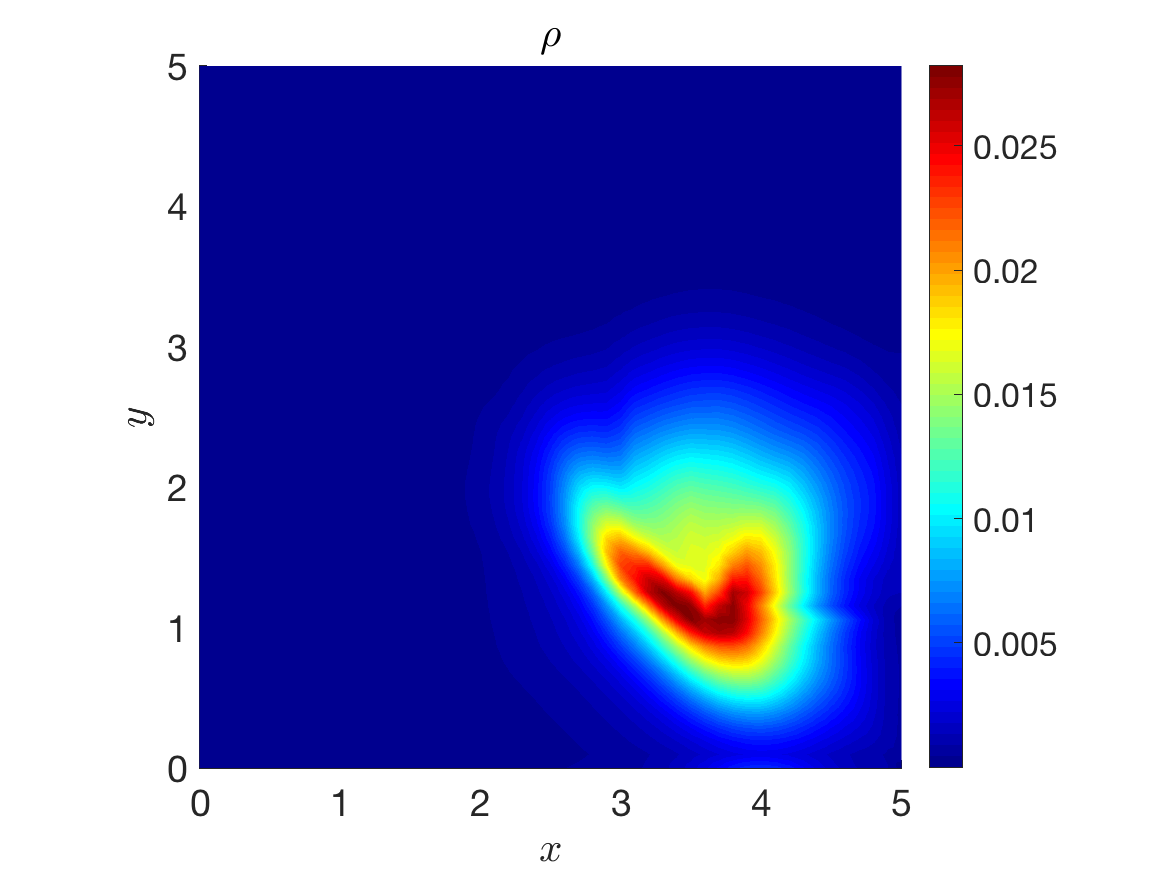}
        \caption{t=1.5}
        \label{MD_t160}
    \end{subfigure}\\
    
   \begin{subfigure}{0.32\textwidth}
        \centering
        \includegraphics[width=\textwidth]{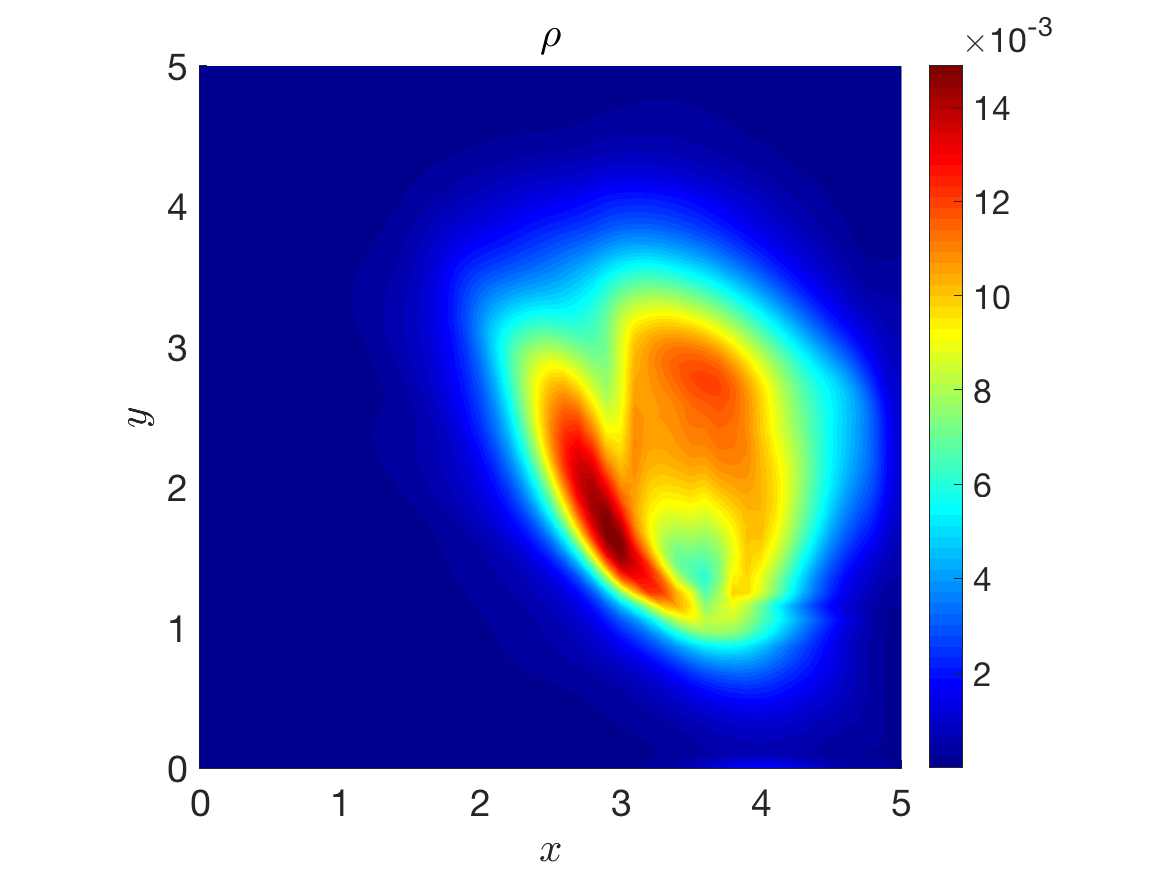}
        \caption{\scriptsize t=2.5}
        \label{MD_t100}   
     \end{subfigure} 
     \begin{subfigure}{0.32\textwidth}
        \centering
        \includegraphics[width=\textwidth]{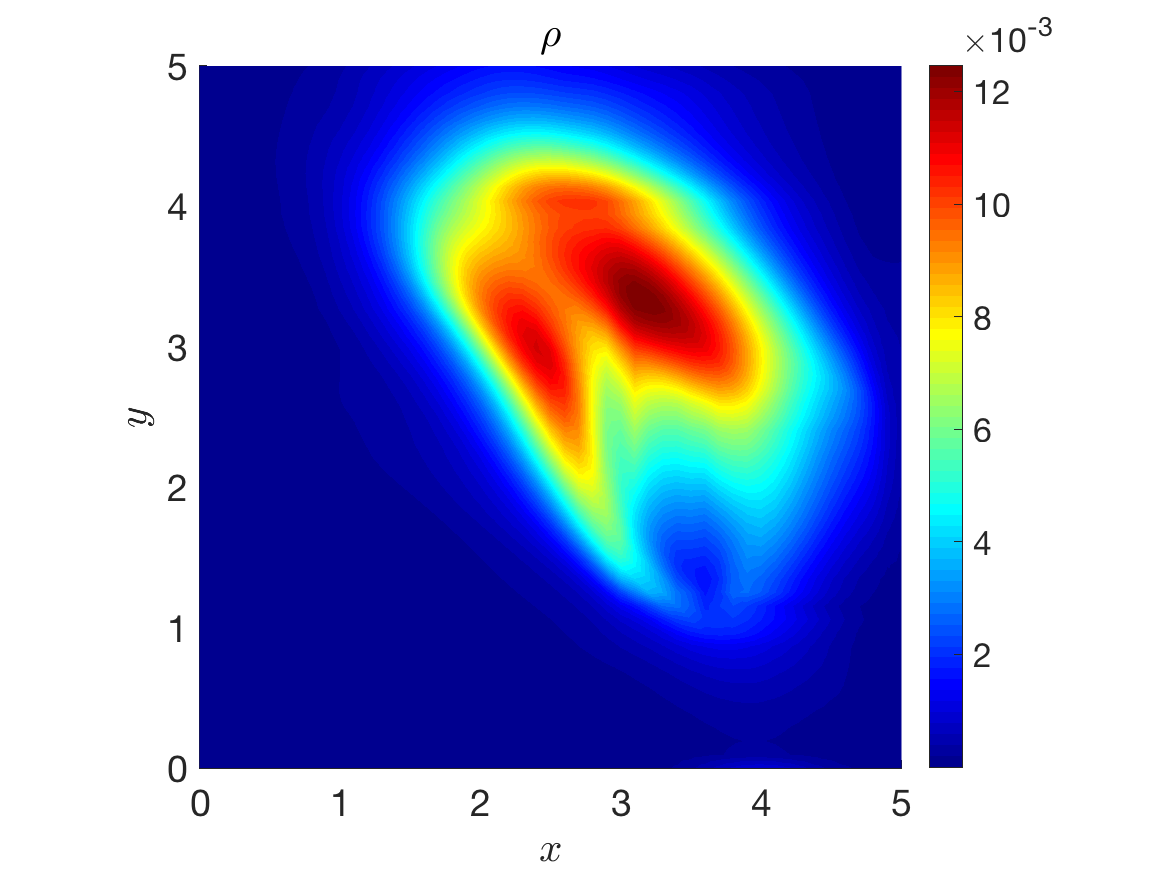}   
        \caption{\scriptsize t=3.5}
        \label{MD_t140}
     \end{subfigure}
       \begin{subfigure}{0.32\textwidth}
        \centering
        \includegraphics[width=\textwidth]{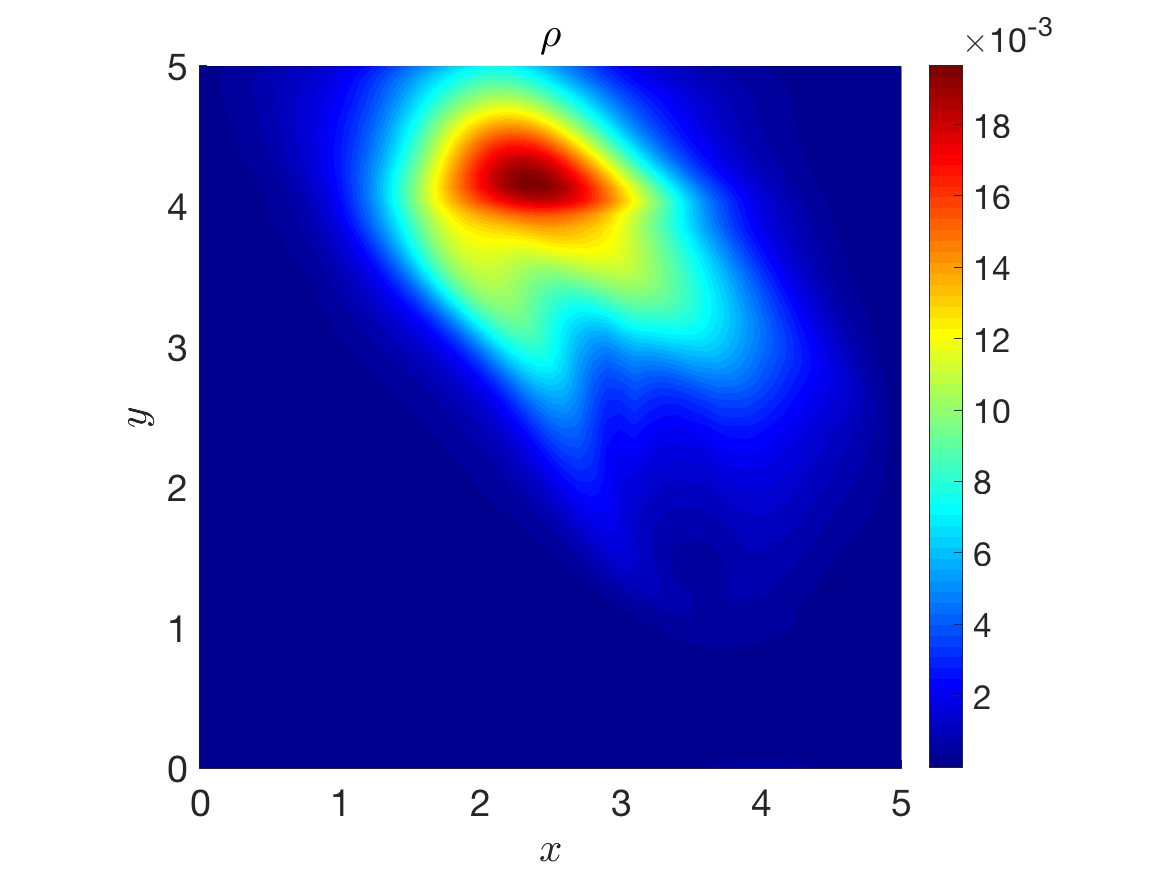}   
        \caption{\scriptsize t=4.5}
        \label{MD_t180}
     \end{subfigure}
    \caption{ \textbf{Test 4} Migration of cell in an heterogenous domain as illustrated in (a).  The sensing radius of the cells is $R=0.7$. The chemoattractant (b) is \eqref{S.gauss} with $m_{\cS}=10$ and $\sigma_{\cS}^2=0.05$. The initial cell profile (c) evolves in time as illustrated in (d)-(i).}
       \label{4.Mix_domain.2}
\end{figure}

 As second scenario, we shall consider the domain represented in Fig. \ref{FibMix}; in each subdomain, the parameter $k(x,y)=50$. The initial condition of the cell population is \eqref{ci_cells_gaussian} with $(x_0,y_0)=(4,0.5)$ and $r_0=0.1$, while the chemoattractant has a gaussian profile \eqref{S.gauss} centered in $(x_{\cS},y_{\cS})= (2,4.5)$ with $m_{\cS}=10$ and $\sigma_{\cS}^2= 0.05$, as shown in Fig. \ref{InCell_FibMix} and \ref{InS_FibMix}, respectively. We observe that cells do not migrate directly towards the chemoattractant, as they sense the heterogeneous fibrous environment and, consequently, adapt their migration to it. In particular, cells that are able to reach and sense the isotropic subdomain where the fibers are uniformly distributed (defined by $1\le x\le3$ and $0\le y\le3$), go in this direction imposed by the gradient of the chemoattractant. On the other hand, in the subdomain $3\le x\le5$ and $1\le y\le2$, they follow the direction of fiber alignment, that is $\pi/4$, perpendicular to the favorable direction imposed by $\cS$. However, the sensing radius $R=0.7$ allows the cells that are closer to the right boundary to escape quite fast the disadvantageous (in terms of preferential direction) subdomains and, following firstly the direction $\pi/2$ in $2\le y\le3$ and, then, $3\pi/4$ in $3\le y\le4$, to reach the chemoattractant. 

\section{Conclusion}
We have proposed a kinetic model for describing cell migration in a multi-cue environment. In particular, in the same spirit as \cite{loy2019JMB}, we have considered that cells, as they can extend protrusions up to several cell diameters, perform a non-local sensing of the environment up to a distance $R$ (named the sensing radius) from its nucleus. In the present model, there are two guidance cues affecting the polarization, and, therefore, the direction of motion of the cells: contact guidance, that is a bi-directional cue, and a chemical gradient, that is a mono-directional cue. We remark that for the first time in this work a non-local sensing in the physical space of the mesoscopic distribution of fibers is considered. In particular, we introduced two classes of models: in the first one, the cells perform an independent sensing of the fibers and of the chemical in its neighborhood, while in the second class of models the cells average the chemical and the fibers with the same sensing kernel.

In the two cases, a particular attention was devoted to the identification of the proper macroscopic limit according to the properties of the turning operator.  We detected two parameters, $\eta_q$ and $\eta_{\cS}$, that measure the relation between the cell sensing radius and the characteristic lengths of variation $-$ $l_{\cS}$ and $l_q$ $-$ of the two cues, and discriminate between a diffusion-driven regime with an advective correction and a drift-driven regime. In particular, when the sensing radius does not exceed the characteristic length of the chemoattractant, the bi-directional nature of the fibers allows for a diffusive regime; otherwise the hyperbolic scaling leads to macroscopic drift.  A common feature in the different cases is the dependency of the macroscopic velocity on both the fibers network and the chemoattractant. This aspect enhances the non-trivial influence of contact guidance on the cell drift, although we considered a non polarized fibers network. This interdependence is in accordance with the model proposed in \cite{wagle2000}. Moreover, in absence of a chemoattractant, this impact on the drift term could persist for spatial heterogenous fiber distributions. This is in accordance to what is observed in \cite{Hillen.05} and it represents a step forward with respect to \cite{wagle2000}, in which the drift is a function of contact guidance only through to the presence of a chemical gradient, $\ie$, without chemoattractant there will be no drift. 

The numerical simulations of the transport model pointed out the main features characterizing the two classes of models and the possible scenarios that they are able to capture. We observed that the presence of two cues influencing cell polarization, even when the fibers are sensed locally, ensures a preferential sense of motion for cells laying on regions of highly aligned non-oriented fibers. Test 3 allowed to show the importance of deriving the macroscopic equations from an underlying microscopic dynamics and in the appropriate regime: a directly postulated drift-diffusion equation would not capture the exact dynamics in all the possible regimes. The competitive or collaborative effects of the cues depend, in a first instance, on the angle between their relative orientations, $\ie$, the direction of fiber alignment $\theta_q$ and the gradient of the chemoattractant. Moreover, especially for the cases of competitive cues, determining which one is the dominant cue depends on their relative strengths, in terms of both concentration and intensity (degree of alignment of the fiber $k(\x)$ or steepness of the chemoattractive gradient). 
We introduced the parameter $\eta=l_\cS/l_q$ that, independently on the cell size or its sensing capability, quantifies the relative contribution of guidance to chemotaxis and provide a first separation between the cases of fiber-dominating and chemotaxis-dominating dynamics ($\eta\gg1$ or $\eta \ll 1$, respectively). The presented framework also allows for the direct calculation of parameters that can be used to quantify directed cell migration and to set its efficiency, like, for instance, mean square displacement, persistence time, directional persistence and mean speed
 \cite{Alt.88}. 

\noindent Additionally, the non locality brings an further level of detail to the model, allowing to obtain different macroscopic behaviour depending on the characteristics of the two sensing. In fact, we did not observe strong differences between the independent and the dependent sensing models, when we assume in the former the same sensing kernel for fibers and chemoattractant, $\ie$, when $\gamma_q=\gamma_{\scS}$. However, if there are biological observations sustaining the possibility that a cell might implement different strategies for sensing the underlying fibers network and the chemoattractant, it would be possible to use the proposed model, in its independent sensing version, to investigate this scenario and to compare the possible outcomes of this sensing approach with the case of a unique and common sensing strategy.

Potentially, the case of competitive cues, combined with the non-local aspect of the model, could lead to interesting further analysis. As observed in the last numerical tests, the combination of heterogenous landscapes of fiber with chemoattractive agents show how the cell density can divide and cross the domain using different migration strategies. This leads to natural questions about the deeper mechanisms leading the competition between the two cues, considering, for instance, the possible role of cell adhesion in recovering collective migration.

We remark that, even if simulations were performed in a two dimensional setting, the transport model (and its macroscopic limits, as a consequence) is formulated in a general d-dimensional setting. Hence, a possible future development is to perform simulations in the three dimensional case, that would be much more realistic for mimicking in-vivo migration of cells in the extracellular matrix. Moreover, the model that we proposed may be adapted to describe other directional cues that might describe, among others, haptotactic, durotactic or electrotactic mechanisms. Furthermore, in the same spirit as in \cite{Loy_Preziosi2} we could enrich this model with a non-constant sensing-radius, as it may vary according to the spatial and directional variability of the external guidance cues. Lastly, this study was restricted to the case in which the cues affect only cell polarization, considering a uniform distribution of the speeds. However, similarly to what is done in \cite{loy2019JMB, Loy_Preziosi2}, it may be modified to model a multi-cue environment in which one of the signals affects also the speed of the cells.

 \appendix 
 
\section{Estimation of $l_q$}
\label{est_lq}
Let us consider the fiber density distribution $q(\x,\hv)$ defined by a bimodal Von Mises Fisher 
\[
q(\x,\hv)=\dfrac{1}{4\pi I_0(k(\x))}\left(e^{k(\x)\,\ub\cdot \hv}+e^{-k(\x)\,\ub\cdot \hv}\right)\,,
\]
where $k(\x)\in \mathcal{C}^1(\Omega)$ and $I_\nu(k(\x))$ denotes the modified Bessel function of first kind of order $\nu$. 

We now want to give an estimation for the range of variability of the characteristic length $l_q$, defined as:
\[
l_q:=\dfrac{1}{\max\limits_{\x \in \Omega}\,\max\limits_{\hv \in \mathbb{S}^{d-1}}\frac{|\nabla q \cdot \hv|}{q}}\,.
\]
Since $\dfrac{\partial I_0}{\partial k}=\dfrac{I_1(k)}{I_0(k)}$, we have that 
\begin{equation*}
\begin{split}
\nabla q&=\dfrac{\left(e^{k(\x)\,\ub\cdot \hv}-e^{-k(\x)\,\ub\cdot \hv}\right)}{4 \pi I_0(k(\x))}\,\nabla k\,(\ub\cdot\hv)-\dfrac{\left(e^{k(\x)\,\ub\cdot \hv}+e^{-k(\x)\,\ub\cdot \hv}\right)}{4 \pi I_0^2(k(\x))}\,\dfrac{\partial I_0}{\partial k}\,\nabla k=\\[0.3cm]
&=\dfrac{\left(e^{k(\x)\,\ub\cdot \hv}-e^{-k(\x)\,\ub\cdot \hv}\right)}{4 \pi I_0(k(\x))}\,\nabla k\,(\ub\cdot\hv)-\dfrac{\left(e^{k(\x)\,\ub\cdot \hv}+e^{-k(\x)\,\ub\cdot \hv}\right)}{4 \pi I_0(k(\x))}\,\dfrac{ I_1(k(\x))}{I_0(k(\x))}\,\nabla k
\end{split}
\end{equation*}
\noindent Since $q(\x,\hv)>0$, we have:
\begin{equation*}
\begin{split}
\dfrac{\nabla q \cdot \hv}{q}
=\left|\dfrac{\left(e^{k(\x)\,\ub\cdot \hv}-e^{-k(\x)\,\ub\cdot \hv}\right)}{\left(e^{k(\x)\,\ub\cdot \hv}+e^{-k(\x)\,\ub\cdot \hv}\right)}\,(\ub\cdot\hv)-\,\dfrac{ I_1(k(\x))}{I_0(k(\x))}\right|||\nabla k|| \cos(\nabla k \cdot \hv)
\end{split}
\end{equation*}
where $||\cdot ||$ denotes the $L_2$-norm and we use the fact that $||\hv||=1$.
Therefore,
\begin{equation*}
\begin{split}
\left|\dfrac{\nabla q \cdot \hv}{q}\right|
=\left|\dfrac{\left(e^{k(\x)\,\ub\cdot \hv}-e^{-k(\x)\,\ub\cdot \hv}\right)}{\left(e^{k(\x)\,\ub\cdot \hv}+e^{-k(\x)\,\ub\cdot \hv}\right)}\,(\ub\cdot\hv)-\,\dfrac{ I_1(k(\x))}{I_0(k(\x))}\right|||\nabla k|| \left|\cos(\nabla k \cdot \hv)\right|
\end{split}
\end{equation*}
Recalling that $|a-b|\le |a|+|b|$, $-1\le \dfrac{\left(e^{k(\x)\,\ub\cdot \hv}-e^{-k(\x)\,\ub\cdot \hv}\right)}{\left(e^{k(\x)\,\ub\cdot \hv}+e^{-k(\x)\,\ub\cdot \hv}\right)}\le 1$ and $|\cos\,(\cdot) |\le 1$,
we get
\[
\left|\dfrac{\nabla q\cdot \hv}{q}\right|\le\left(1+\left|\dfrac{I_1(k(\x))}{I_0(k(\x))}\right|\right)||\nabla k||\,.
\]
Considering Eq. (1.12) in \cite{Natalini} for $\nu=1$, we obtain that $\left|\dfrac{I_1}{I_0}\right|<1$, and, therefore,
\[
\left|\dfrac{\nabla q\cdot \hv}{q}\right|< 2||\nabla k||
\]
that implies
\[
\max_{\x \in \Omega} \max_{\hv \in \mathbb{S}^{d-1}}\left|\dfrac{\nabla q\cdot \hv}{q}\right|< 2 \max_{\x \in \Omega} ||\nabla k||.
\]
This translates into 
\begin{equation}\label{lq_sx}
l_q\ge\dfrac{1}{2\max\limits_{\x \in \Omega}||\nabla k||}\,.
\end{equation}
In particular, if there exists $\x$ such that $\nabla k(\x) \cdot \hv =1$ and, at the same time, also satisfies $\nabla k(\x) \parallel \ub$, then
\eqref{lq_sx} is true with the equal sign. In particular, for the symmetry of \eqref{k_gaussian} and \eqref{S.gauss} we shall consider
\[
l_q\approx \dfrac{1}{2\max\limits_{\x \in \Omega}||\nabla k||}\,.
\]

\textbf{Acknowledgments}
The authors would like to thank Prof. Luigi Preziosi for fruitful discussions and valuable comments. 
This work was partially supported by Istituto Nazionale di Alta Matematica, Ministry of Education, Universities and Research, through the MIUR grant Dipartimento di Eccellenza 2018-2022, Project no. E11G18000350001, and the Scientific Reseach Programmes of Relevant National Interest project n. 2017KL4EF3. NL also acknowledges Compagnia di San Paolo. This research was also partially supported by the Basque Government through the BERC 2018- 2021 program and by the Spanish State Research Agency through BCAM Severo Ochoa excellence accreditation SEV-2017-0718. MC has received funding from the European Union’s Horizon 2020 research and innovation programme under the Marie Skłodowska- Curie grant agreement No. 713673. The project that gave rise to these results received the support of a fellowship from ”la Caixa” Foundation (ID 100010434). The fellowship code is LCF/BQ/IN17/11620056.

\bibliography{references}

\begin{thebibliography}{10}

\bibitem{Azimzade2019PRE}
Y.~Azimzade, A.~A. Saberi, and M.~Sahimi.
\newblock Regulation of migration of chemotactic tumor cells by the spatial
  distribution of collagen fiber orientation.
\newblock {\em Phys. Rev. E}, 99:062414, 2019.

\bibitem{bellomo2007}
N.~Bellomo, A.~Bellouquid, J.~Nieto, and J.~Soler.
\newblock Multicellular biological growing systems: Hyperbolic limits towards
  macroscopic description.
\newblock {\em Math. Mod. Meth. Appl. S.}, 17(supp01):1675--1692, 2007.

\bibitem{bellomo2015}
N.~Bellomo, A.~Bellouquid, Y.~Tao, and M.~Winkler.
\newblock Toward a mathematical theory of keller--segel models of pattern
  formation in biological tissues.
\newblock {\em Math. Mod. Meth. Appl. S.}, 25(09):1663--1763, 2015.

\bibitem{Berg}
H.~C. Berg.
\newblock {\em Random Walks in Biology}.
\newblock Princeton University Press, revised edition, 1983.

\bibitem{Berg_Purcell.77}
H.~C. Berg and E.~M. Purcell.
\newblock Physics of chemoreception.
\newblock {\em Biophys. J.}, 20(2):193--219, 1977.

\bibitem{Bisi.Carrillo.Lods}
M.~Bisi, J.~A. Carrillo, and B.~Lods.
\newblock Equilibrium solution to the inelastic boltzmann equation driven by a
  particle bath.
\newblock {\em J. Stat. Phys.}, 133(5):841--870, 2008.

\bibitem{Berg_Block_Segall}
S.~M. Block, J.~E. Segall, and H.~C. Berg.
\newblock Adaptation kinetics in bacterial chemotaxis.
\newblock {\em J. Bacteriol. Res.}, 154(1):312--323, 1983.

\bibitem{BROMBEREK2002ECR}
B.~A. Bromberek, P.~A.~J. Enever, D.~I. Shreiber, M.~D. Caldwell, and R.~T.
  Tranquillo.
\newblock Macrophages influence a competition of contact guidance and
  chemotaxis for fibroblast alignment in a fibrin gel coculture assay.
\newblock {\em Exp. Cell Res.}, 275(2):230--242, 2002.

\bibitem{Chalub_Markowich_Perthame_Schmeiser.04}
F.~A. C.~C. Chalub, P.~A. Markowich, B.~Perthame, and C.~Schmeiser.
\newblock Kinetic models for chemotaxis and their drift-diffusion limits.
\newblock {\em Monatsh. Math.}, 142(1):123--141, 2004.

\bibitem{Chauviere_Hillen_Preziosi.07}
A.~Chauviere, T.~Hillen, and L.~Preziosi.
\newblock Modeling cell movement in anisotropic and heterogeneous network
  tissues.
\newblock {\em Netw. Heterog. Media}, 2(2):333--351, 2007.

\bibitem{Chauviere_Hillen_Preziosi.08}
A.~Chauviere, T.~Hillen, and L.~Preziosi.
\newblock Modeling the motion of a cell population in the extracellular matrix.
\newblock {\em Discrete Cont. Dyn.-B}, 2007(Supplemental volume):250--259,
  2007.

\bibitem{chen2019}
L.~Chen, K.~J. Painter, C.~Surulescu, and A.~Zhigun.
\newblock Mathematical models for cell migration: a nonlocal perspective.
\newblock {\em arXiv preprint arXiv:1911.05200}, 2019.

\bibitem{Col_Sci_Prez.17}
A.~Colombi, M.~Scianna, and L.~Preziosi.
\newblock Coherent modelling switch between pointwise and distributed
  representations of cell aggregates.
\newblock {\em J. Math. Biol.}, 74(4):783--808, 2017.

\bibitem{Col_Sci_Tos.15}
A.~Colombi, M.~Scianna, and A.~Tosin.
\newblock Differentiated cell behavior: a multiscale approach using measure
  theory.
\newblock {\em J. Math. Biol.}, 71:1049--1079, 2015.

\bibitem{Conte_Groppi_Gerardo}
M.~Conte, L.~Gerardo-Giorda, and M.~Groppi.
\newblock Glioma invasion and its interplay with nervous tissue and therapy: A
  multiscale model.
\newblock {\em J. Theo. Biol.}, 486:110088, 2020.

\bibitem{Menci}
E.~Di~Costanzo, M.~Menci, E.~Messina, R.~Natalini, and A.~Vecchio.
\newblock A hybrid model of collective motion of discrete particles under
  alignment and continuum chemotaxis.
\newblock {\em Discrete Cont. Dyn.-B}, 25:443--472, 2020.

\bibitem{Dickinson_Tranquillo.93}
R.~Dickinson and R.~T. Tranquillo.
\newblock Stochastic model of biased cell migration based on binding
  fluctuations of adhesion receptors.
\newblock {\em J. Math. Biol.}, 19:563--600, 1991.

\bibitem{Dickinson}
R.~B. Dickinson.
\newblock A generalized transport model for biased cell migration in an
  anisotropic environment.
\newblock {\em J. Math. Biol.}, 40(2):97--135, 2000.

\bibitem{Eftimie2}
R.~Eftimie.
\newblock Hyperbolic and kinetic models for self-organized biological
  aggregations and movement: a brief review.
\newblock {\em J. Math. Biol.}, 65(1):35--75, 2012.

\bibitem{Engwer_Hillen_Surulescu_15}
C.~Engwer, T.~Hillen, M.~Knappitsch, and C.~Surulescu.
\newblock Glioma follow white matter tracts: a multiscale dti-based model.
\newblock {\em J. Math. Biol.}, 71(3):551--582, 2015.

\bibitem{Engwer_Knappitsch_Surulescu.16}
C.~Engwer, M.~Knappitsch, and C.~Surulescu.
\newblock A multiscale model for glioma spread including cell-tissue
  interactions and proliferation.
\newblock {\em Math. Biosci. Eng.}, 13:443--460, 2016.

\bibitem{Engwer_Stinner_Surulescu.08}
C.~Engwer, C.~Stinner, and C.~Surulescu.
\newblock On a structured multiscale model for acid-mediated tumor invasion:
  The effects of adhesion and proliferation.
\newblock {\em Math. Mod. Meth. Appl. S.}, 27:1355--1390, 2017.

\bibitem{filbet2005}
F.~Filbet, P.~Lauren{\c{c}}ot, and B.~Perthame.
\newblock Derivation of hyperbolic models for chemosensitive movement.
\newblock {\em J. Math. Biol.}, 50(2):189--207, 2005.

\bibitem{Friedl.04}
P.~Friedl.
\newblock Prespecification and plasticity: shifting mechanisms of cell
  migration.
\newblock {\em Curr. Opin. Cell Biol.}, 16:14–23, 2004.

\bibitem{Friedl_Brocker.00}
P.~Friedl and E.-B. Brocker.
\newblock The biology of cell locomotion within three dimensional extracellular
  matrix.
\newblock {\em Cell Mol Life Sci.}, 57:41--64, 2000.

\bibitem{Gininait2019ModellingCC}
R.~Giniūnaitė, R.~E. Baker, P.~M. Kulesa, and P.~K. Maini.
\newblock Modelling collective cell migration: neural crest as a model
  paradigm.
\newblock {\em J. Math. Biol.}, 80:481--504, 2019.

\bibitem{Hillen.05}
T.~Hillen.
\newblock M5 mesoscopic and macroscopic models for mesenchymal motion.
\newblock {\em J. Math. Biol.}, 53(4):585--616, 2006.

\bibitem{HMPS2017MBE}
T.~Hillen, A.~Murtha, K.~J. Painter, and A.~Swan.
\newblock Moments of the von mises and fischer distributions and applications.
\newblock {\em Math. Biosci. Eng.}, 14(3):673--694, 2017.

\bibitem{Othmer_Hillen.00}
T.~Hillen and H.~G. Othmer.
\newblock The diffusion limit of transport equations derived from velocity-jump
  processes.
\newblock {\em SIAM J. Appl. Math.}, 61:751--775, 2000.

\bibitem{Hillen_Painter.08}
T.~Hillen and K.~J. Painter.
\newblock A user's guide to pde models for chemotaxis.
\newblock {\em J. Math. Biol.}, 58(1):183--217, 2008.

\bibitem{Chiocca}
J.~Johnson, M.~O. Nowicki, C.~H. Lee, E.~A. Chiocca, M.~S. Viapiano, S.~E.
  Lawler, and J.~J Lannutti.
\newblock Quantitative analysis of complex glioma cell migration on electrospun
  polycaprolactone using time-lapse microscopy.
\newblock {\em Tissue Eng. Part C-Me}, 15(4):531--540, 2009.

\bibitem{Keller_Segel}
E.~F. Keller and L.~A. Segel.
\newblock Initiation of slime mold aggregation viewed as an instability.
\newblock {\em J. Theo. Biol.}, 26(3):399--415, 1970.

\bibitem{PAINTER2019JTB}
P.~J. Kevin.
\newblock Mathematical models for chemotaxis and their applications in
  self-organisation phenomena.
\newblock {\em J. Theor. Biol.}, 481:162--182, 2019.

\bibitem{Painter1999}
P.~J. Kevin, P.~K. Maini, and H.~G. Othmer.
\newblock Development and applications of a model for cellular response to
  multiple chemotactic cues.
\newblock {\em J. Math. Biol.}, 41(4):285--314, 2000.

\bibitem{kolbe2020modeling}
N.~Kolbe, N.~Sfakianakis, C.~Stinner, C.~Surulescu, and J.~Lenz.
\newblock Modeling multiple taxis: tumor invasion with phenotypic
  heterogeneity, haptotaxis, and unilateral interspecies repellence.
\newblock {\em arXiv preprint arXiv:2005.01444}, 2020.

\bibitem{Natalini}
A.~Laforgia and P.~Natalini.
\newblock Some inequalities for modified bessel functions.
\newblock {\em J. Inequal. Appl.}, 2010(1):253035, 2010.

\bibitem{Lara2013IB}
L.~Lara and I.~Schneider.
\newblock Directed cell migration in multi-cue environments.
\newblock {\em Integr. Biol.}, 5(11):1306--1323, 2013.

\bibitem{Lods}
B.~Lods.
\newblock Semigroup generation propertiesof streaming operators with
  noncontractive boundary conditions.
\newblock {\em Math. Comput. Model.}, 42:1441--1462, 2005.

\bibitem{loy2019JMB}
N.~Loy and L.~Preziosi.
\newblock Kinetic models with non-local sensing determining cell polarization
  and speed according to independent cues.
\newblock {\em J. Math. Biol.}, 80:373--421, 2019.

\bibitem{Loy_Preziosi2}
N.~Loy and L.~Preziosi.
\newblock Modelling physical limits of migration by a kinetic model with
  non-local sensing.
\newblock {\em J. Math. Biol.}, 2019.
\newblock In Press.

\bibitem{MAHESHWARI1999BioP}
G.~Maheshwari, A.~Wells, L.~G. Griffith, and D.~A. Lauffenburger.
\newblock Biophysical integration of effects of epidermal growth factor and
  fibronectin on fibroblast migration.
\newblock {\em Biophys. J.}, 76(5):2814--2823, 1999.

\bibitem{mardia2009}
K.~V. Mardia and P.~E. Jupp.
\newblock {\em Directional statistics}, volume 494.
\newblock John Wiley \& Sons, 2009.

\bibitem{Othmer_Hillen.02}
H.~Othmer and T.~Hillen.
\newblock The diffusion limit of transport equations ii: Chemotaxis equations.
\newblock {\em SIAM J. Appl. Math.}, 62:1222--1250, 2002.

\bibitem{Othmer_Stevens.97}
H.~Othmer and A.~Stevens.
\newblock Aggregation, blowup, and collapse: The {ABC}'s of taxis in reinforced
  random walks.
\newblock {\em SIAM J. Appl. Math.}, 57:1044--1081, 2001.

\bibitem{Alt.88}
H.~G. Othmer, S.~R. Dunbar, and W.~Alt.
\newblock Models of dispersal in biological systems.
\newblock {\em J. Math. Biol.}, 26(3):263--298, 1988.

\bibitem{Painter2008}
K.~J. Painter.
\newblock Modelling cell migration strategies in the extracellular matrix.
\newblock {\em J. Math. Biol.}, 58(4):511--543, 2008.

\bibitem{Hillen_Painter.13}
K.~J. Painter and T.~Hillen.
\newblock {\em Transport and anisotropic diffusion models for movement in
  oriented habitats}, volume 2071, pages 177--222.
\newblock Lect. Notes Math., Springer - verlag -, 2013.

\bibitem{Palc}
A.~Palcewski.
\newblock {\em Velocity averaging for boundary value problems}, pages 1--284.
\newblock Ser. Adv. Math. Appl. Sci. World Scientific Publishing Company, 1992.

\bibitem{Petterson}
R.~Pettersson.
\newblock On solutions to the {L}inear {B}oltzmann equation for granular gases.
\newblock {\em Transport Theor. Stat.}, 33(5-7):527--543, 2004.

\bibitem{Plaza}
R.~G. Plaza.
\newblock Derivation of a bacterial nutrient-taxis system with doubly
  degenerate cross-diffusion as the parabolic limit of a velocity-jump process.
\newblock {\em J. Math. Biol.}, 78(6):1681--1711, 2019.

\bibitem{Pourf2018APLB}
K.~E. Pourfarhangi, E.~Hoz, A.~Cohen, and B.~Gligorijevic.
\newblock Contact guidance is cell cycle-dependent.
\newblock {\em APL Bioeng.}, 2:031904, 2018.

\bibitem{Provenzano2006BMCMed}
P.~P. Provenzano, K.~W. Eliceiri, J.~M. Campbell, and et~al.
\newblock Collagen reorganization at the tumor-stromal interface facilitates
  local invasion.
\newblock {\em BMC Med.}, 4(1):38, 2006.

\bibitem{Provenzano2009TCB}
P.~P. Provenzano, K.~W. Eliceiri, and P.~J. Keely.
\newblock Shining new light on 3d cell motility and the metastatic process.
\newblock {\em Trends Cell Biol.}, 19(11):638--648, 2009.

\bibitem{RAJNICEK2007DB}
A.~M. Rajnicek, L.~E. Foubister, and C.~D. McCaig.
\newblock Prioritising guidance cues: Directional migration induced by
  substratum contours and electrical gradients is controlled by a rho/cdc42
  switch.
\newblock {\em Dev. Biol.}, 312(1):448--460, 2007.

\bibitem{Woo2005LoC}
S.~W. Rhee, A.~M. Taylor, C.~H. Tu, D.~H. Cribbs, C.~Cotman, and N.~Li Jeon.
\newblock Patterned cell culture inside microfluidic devices.
\newblock {\em Lab Chip}, 51:102--107, 2005.

\bibitem{Schlute2012BP}
D.~Schlüter, I.~Ramis-Conde, and M.~Chaplain.
\newblock Computational modeling of single-cell migration: The leading role of
  extracellular matrix fibers.
\newblock {\em Biophys. J.}, 103:1141--51, 2012.

\bibitem{Scianna_Preziosi.13.2}
M.~Scianna, L.~Preziosi, and K.~Wolf.
\newblock A cellular potts model simulating cell migration on and in matrix
  environments.
\newblock {\em Math. Biosci. Eng.}, 10:235--261, 2013.

\bibitem{Steeg2016NRC}
P.~Steeg.
\newblock Targeting metastasis.
\newblock {\em Nat. Rev. Cancer.}, 16:201--218, 2016.

\bibitem{Stroock}
D.~W. Stroock.
\newblock Some stochastic processes which arise from a model of the motion of a
  bacterium.
\newblock {\em Z. Wahrscheinlichkeit}, 28(4):305--315, 1974.

\bibitem{Sunda2013BB}
H.~Sundararaghavan, R.~Saunders, D.~Hammer, and J.~Burdick.
\newblock Fiber alignment directs cell motility over chemotactic gradients.
\newblock {\em Biotechnol. Bioeng.}, 110(4):1249--1254, 2013.

\bibitem{wagle2000}
M.~A. Wagle and R.~T. Tranquillo.
\newblock A self-consistent cell flux expression for simultaneous chemotaxis
  and contact guidance in tissues.
\newblock {\em J. Math. Biol.}, 41(4):315--330, 2000.

\bibitem{WILKINSON1983ECR}
P.~C. Wilkinson and J.~M. Lackie.
\newblock The influence of contact guidance on chemotaxis of human neutrophil
  leukocytes.
\newblock {\em Exp. Cell Res.}, 145(2):255--264, 1983.

\bibitem{Wolf2003JCB}
K.~Wolf, I.~Mazo, H.~Leung, K.~Engelke, U.~H. von Andrian, E.~I. Deryugina,
  A.~Y. Strongin, E.-B. Br\"ocker, and P.~Friedl.
\newblock {Compensation mechanism in tumor cell migration:
  mesenchymal–amoeboid transition after blocking of pericellular
  proteolysis}.
\newblock {\em Int. J. Cell Biol.}, 160(2):267--277, 2003.

\end{thebibliography}

\end{document}